\pdfoutput=1

\documentclass[11pt,twoside,a4paper,cmspaper,final,collab]{cms-tdr}
\def\svnVersion{177e650-D}\def\svnDate{2020/01/23}\def\cmsCernNoTag{CERN-EP-2019-022}\def\cmsCernDate{\today}\def\cmsMessage{"Published in the European Physical Journal C as \href{http://dx.doi.org/10.1140/epjc/s10052-019-7585-7}{\doi{10.1140/epjc/s10052-019-7585-7}.}"}

\begin{document}\cmsNoteHeader{SMP-17-011}

\hyphenation{had-ron-i-za-tion}
\hyphenation{cal-or-i-me-ter}
\hyphenation{de-vices}
\RCS$Revision$
\RCS$HeadURL$
\RCS$Id$
\newlength\cmsFigWidth
\ifthenelse{\boolean{cms@external}}{\setlength\cmsFigWidth{0.98\columnwidth}}{\setlength\cmsFigWidth{0.65\textwidth}}
\ifthenelse{\boolean{cms@external}}{\providecommand{\cmsLeft}{upper\xspace}}{\providecommand{\cmsLeft}{left\xspace}}
\ifthenelse{\boolean{cms@external}}{\providecommand{\cmsRight}{lower\xspace}}{\providecommand{\cmsRight}{right\xspace}}
\newlength\cmsTabSkip\setlength\cmsTabSkip{1.5ex}

\providecommand{\ll}{\ensuremath{\ell\ell}\xspace}
\providecommand{\lv}{\ensuremath{\ell\nu}\xspace}
\providecommand{\jj}{\ensuremath{\mathrm{jj}}\xspace}
\providecommand{\lljj}{\ensuremath{\ell\ell\mathrm{jj}}\xspace}
\providecommand{\lvjj}{\ensuremath{\ell\nu\mathrm{jj}}\xspace}
\providecommand{\Zjj}{\ensuremath{\cPZ\mathrm{jj}}\xspace}
\providecommand{\Wjj}{\ensuremath{\PW\mathrm{jj}}\xspace}
\providecommand{\ewklljj}{\ensuremath{\mathrm{EW}\,\ell\ell\mathrm{jj}}\xspace}
\providecommand{\ewkwjj}{\ensuremath{\mathrm{EW}\,\PW\mathrm{jj}}\xspace}
\providecommand{\ewkzjj}{\ensuremath{\mathrm{EW}\,\PZ\mathrm{jj}}\xspace}
\providecommand{\dywjj}{\ensuremath{\mathrm{DY}\,\PW\mathrm{jj}}\xspace}
\providecommand{\dyzjj}{\ensuremath{\mathrm{DY}\,\PZ\mathrm{jj}}\xspace}
\providecommand{\ptj}{\ensuremath{p_\mathrm{T j}}\xspace}
\providecommand{\ptjj}{\ensuremath{p_\mathrm{T jj}}\xspace}

\newcommand{\MT}{\ensuremath{{m}_{\mathrm{T}}}}
\newcommand{\MPT}{\ensuremath{\pt^{\text{miss}}}}
\newcommand{\MPTvec}{\ensuremath{\ptvec^{\text{miss}}}}
\cmsNoteHeader{SMP-17-011}
\title{Measurement of electroweak production of a $\PW$ boson in association with two jets in proton-proton collisions at \texorpdfstring{$\sqrt{s}=13\TeV$}{sqrt(s) = 13 TeV}}
\titlerunning{EW production of a W with two jets at 13 TeV}

\date{\today}

\abstract{
A measurement is presented of electroweak (EW) production of a $\PW$ boson in association with  two jets
in proton-proton collisions at $\sqrt{s}=13\TeV$. The data sample was
recorded by the CMS Collaboration at the LHC and  corresponds to an integrated luminosity
of 35.9\fbinv. The measurement is performed for the $\ell\nu$jj final state (with $\ell\nu$
indicating a lepton-neutrino pair, and j representing the quarks produced in the hard interaction) in a kinematic region defined by
invariant mass $m_\mathrm{jj} >120\GeV$ and transverse momenta $p_\mathrm{T j} > 25\GeV$.
The cross section of the process is measured in the electron and muon channels yielding
$\sigma_\mathrm{EW}(\PW\mathrm{jj})= 6.23 \pm 0.12 \stat\pm 0.61 \syst\unit{pb}$
per channel, in agreement with leading-order standard model predictions.
The additional hadronic activity of events in a signal-enriched region is studied,
and the measurements are compared with predictions.
The final state is also used to perform a search for anomalous trilinear gauge couplings.
Limits on anomalous trilinear gauge couplings associated with dimension-six
operators are given in the framework of an effective field theory.
The corresponding 95\% confidence level intervals are
$-2.3 <  c_{{\PW\PW\PW}}/\Lambda^2  < 2.5\TeV^{-2}$,
$-8.8 <  c_{\PW}/\Lambda^2  < 16\TeV^{-2}$, and
$-45 <  c_{\PB}/\Lambda^2  < 46\TeV^{-2}$.
These results are combined with the CMS EW \Zjj\ analysis, yielding the
constraint on the $c_{{\PW\PW\PW}}$ coupling :
$-1.8 <  c_{{\PW\PW\PW}}/\Lambda^2  < 2.0\TeV^{-2}$.
}

\hypersetup{%
pdfauthor={CMS Collaboration},%
pdftitle={Measurement of the electroweak production of a W boson in association with two jets in proton-proton collisions at sqrt(s)=13 TeV},%
pdfsubject={CMS},%
pdfkeywords={CMS, physics, electroweak production, W boson}}

\maketitle

\section{Introduction\label{sec:intro}}

In proton-proton (pp) collisions at the CERN LHC,
the pure electroweak (EW) production of a lepton-neutrino pair ($\ell\nu$)
in association with two jets (\jj)
includes production via vector boson fusion
(VBF).
This process has a distinctive signature of two jets with large energy and separation in pseudorapidity
($\eta$), produced in association with a lepton-neutrino pair.
This EW process is referred to as EW \Wjj, and the
two jets produced through the fragmentation of the outgoing quarks
are referred to as ``tagging jets''.

\begin{figure*}[htb] {
\centering
\includegraphics[width=0.32\textwidth]{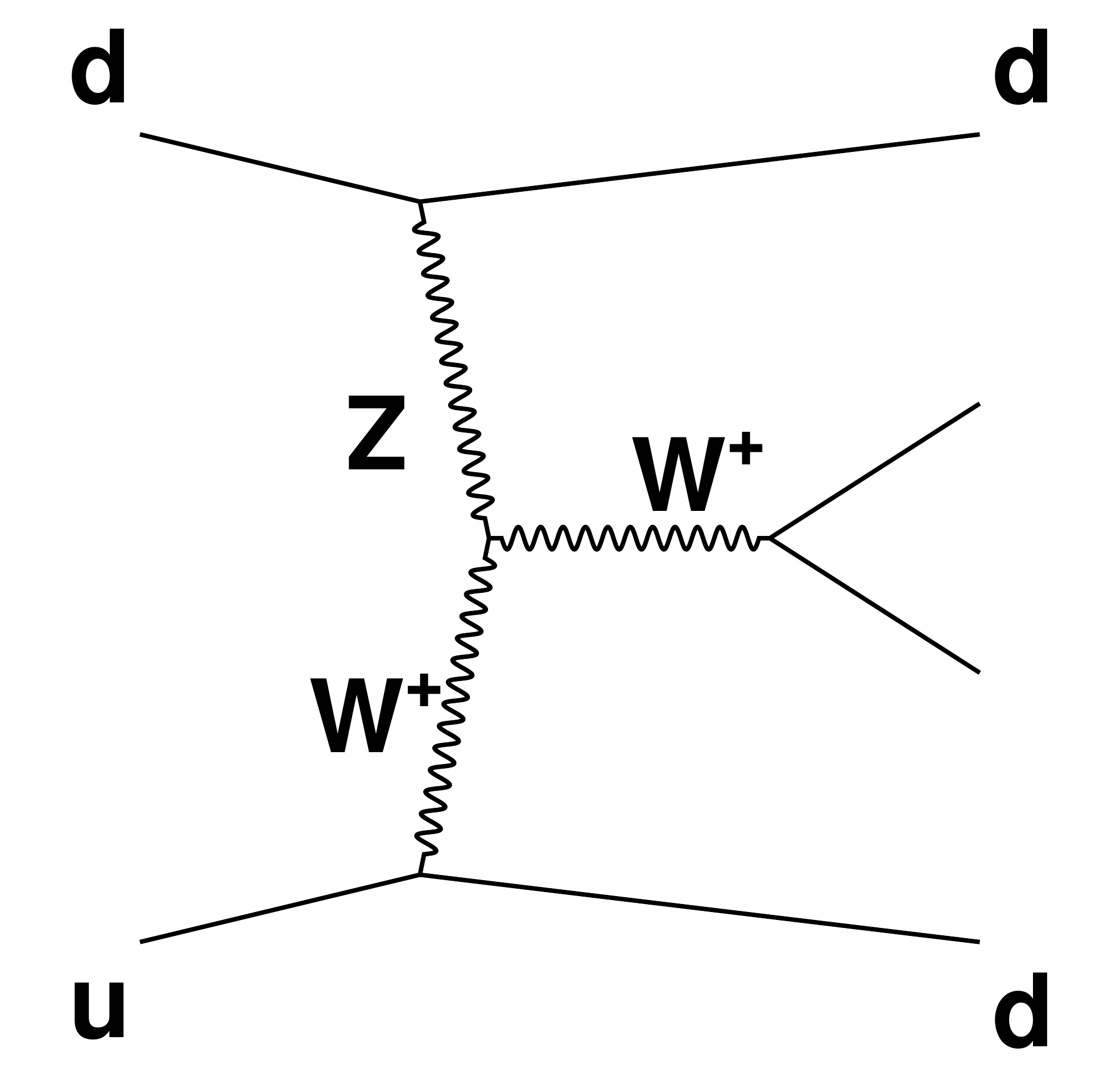}
\includegraphics[width=0.32\textwidth]{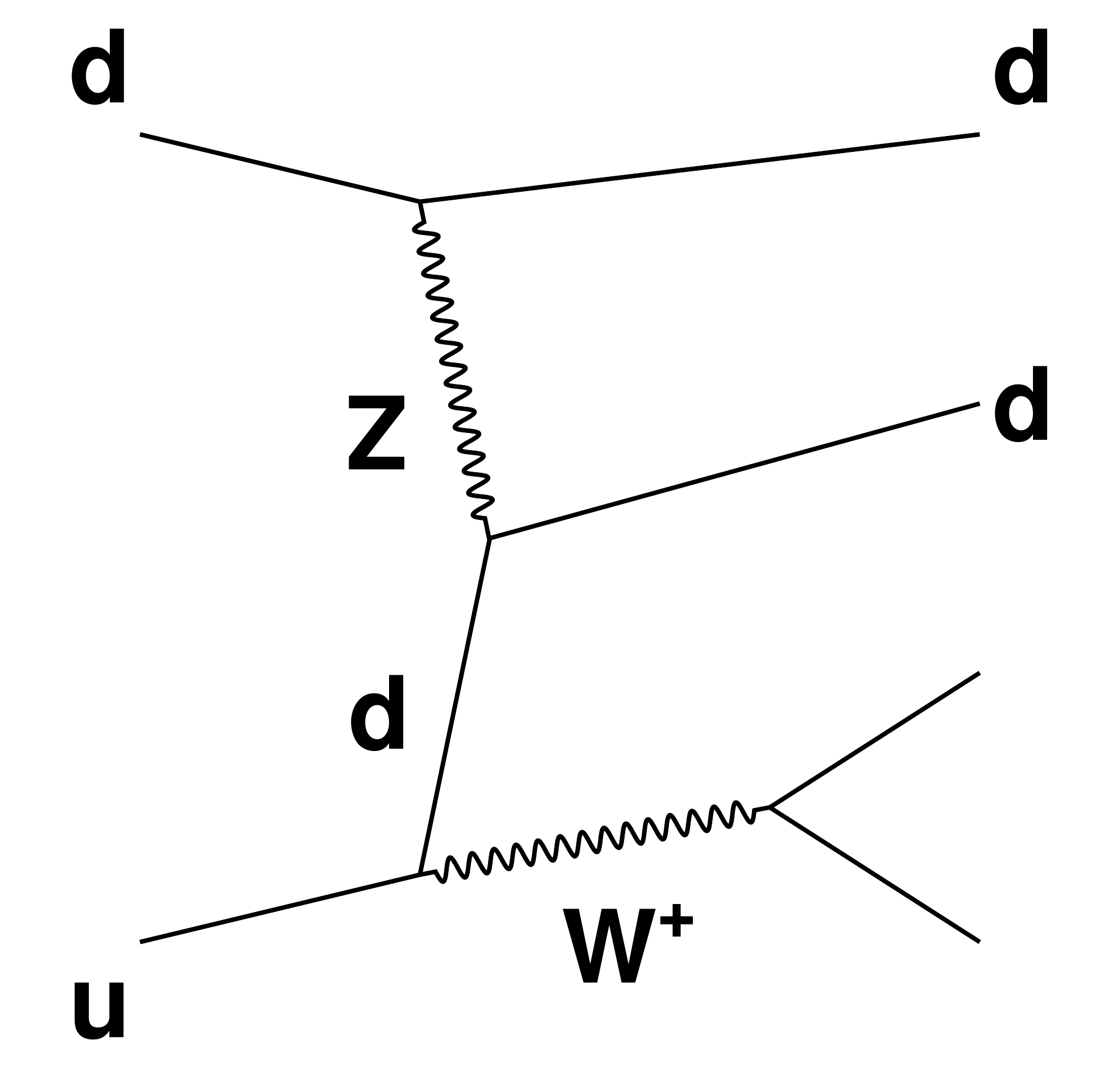}
\includegraphics[width=0.32\textwidth]{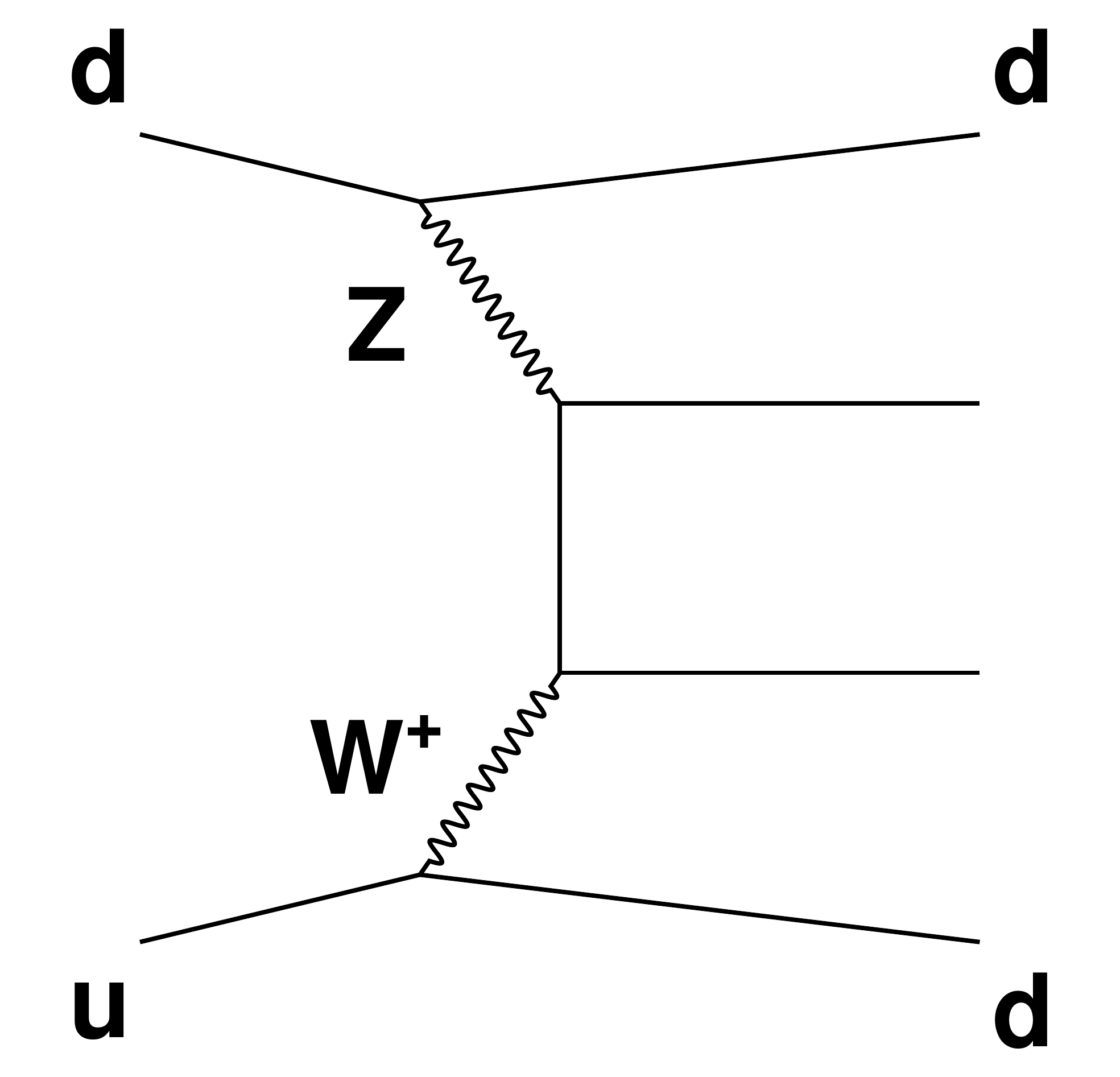}
\caption{
Representative Feynman diagrams for lepton-neutrino production in association
with two jets from purely electroweak amplitudes:
 vector boson fusion (left),
 bremsstrahlung-like (center),
and  multiperipheral (right) production.
\label{fig:sigdiagram}}}
\end{figure*}

Figure~\ref{fig:sigdiagram} shows representative Feynman diagrams for the EW \Wjj\
signal processes, namely VBF (Fig.~\ref{fig:sigdiagram}, left),
 bremsstrahlung-like (Fig.~\ref{fig:sigdiagram}, center), and
multiperipheral (Fig.~\ref{fig:sigdiagram}, right) production.
Gauge cancellations lead to a large negative interference between the VBF
diagram and the other two diagrams,
with the larger interference coming from bremsstrahlung-like production.
Interference with multiperipheral
production is limited to cases where the lepton-neutrino pair mass is close to
the $\PW$ boson mass.

\begin{figure*}[htb] {
\centering
\includegraphics[width=0.32\textwidth]{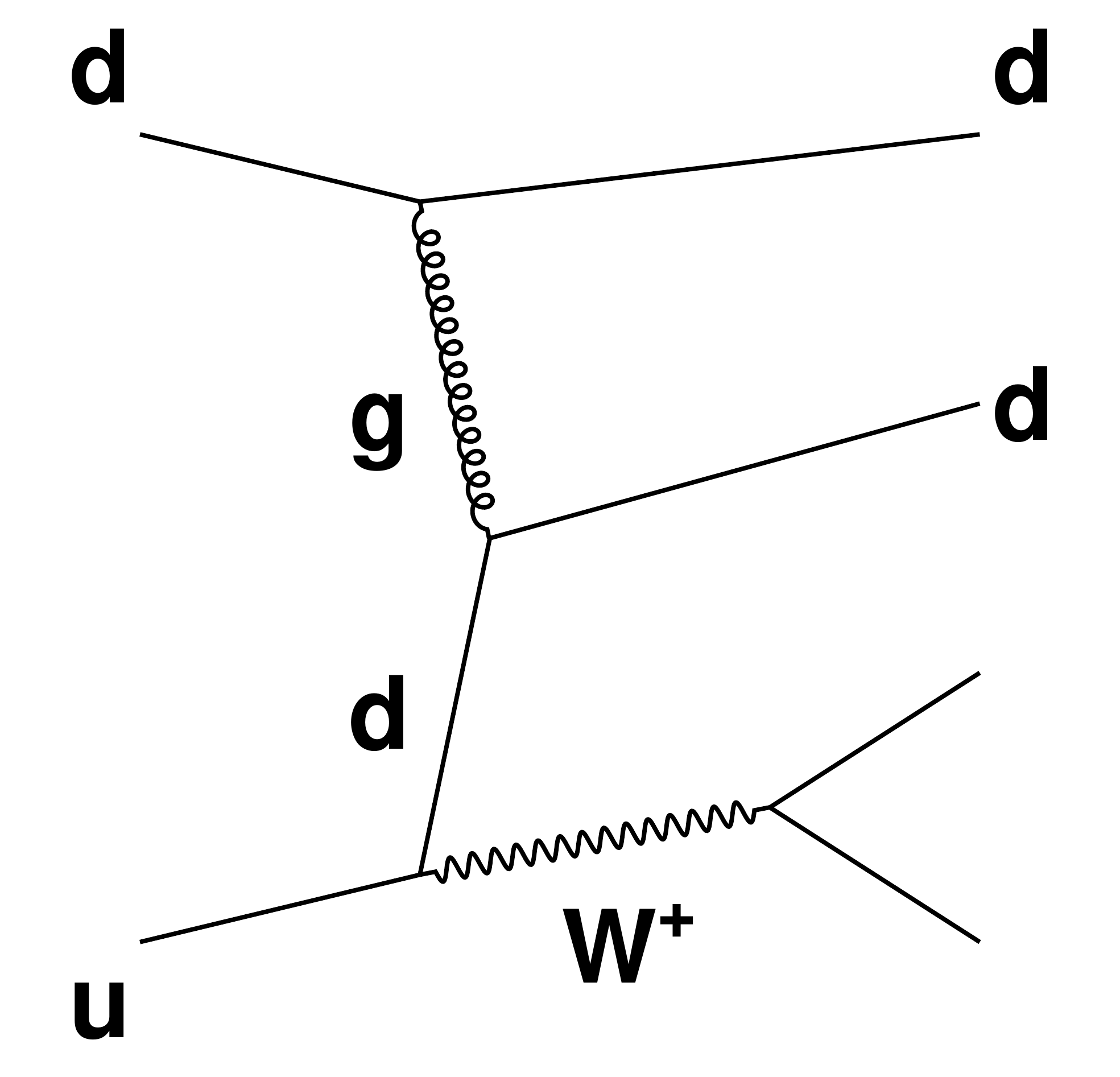}
\includegraphics[width=0.32\textwidth]{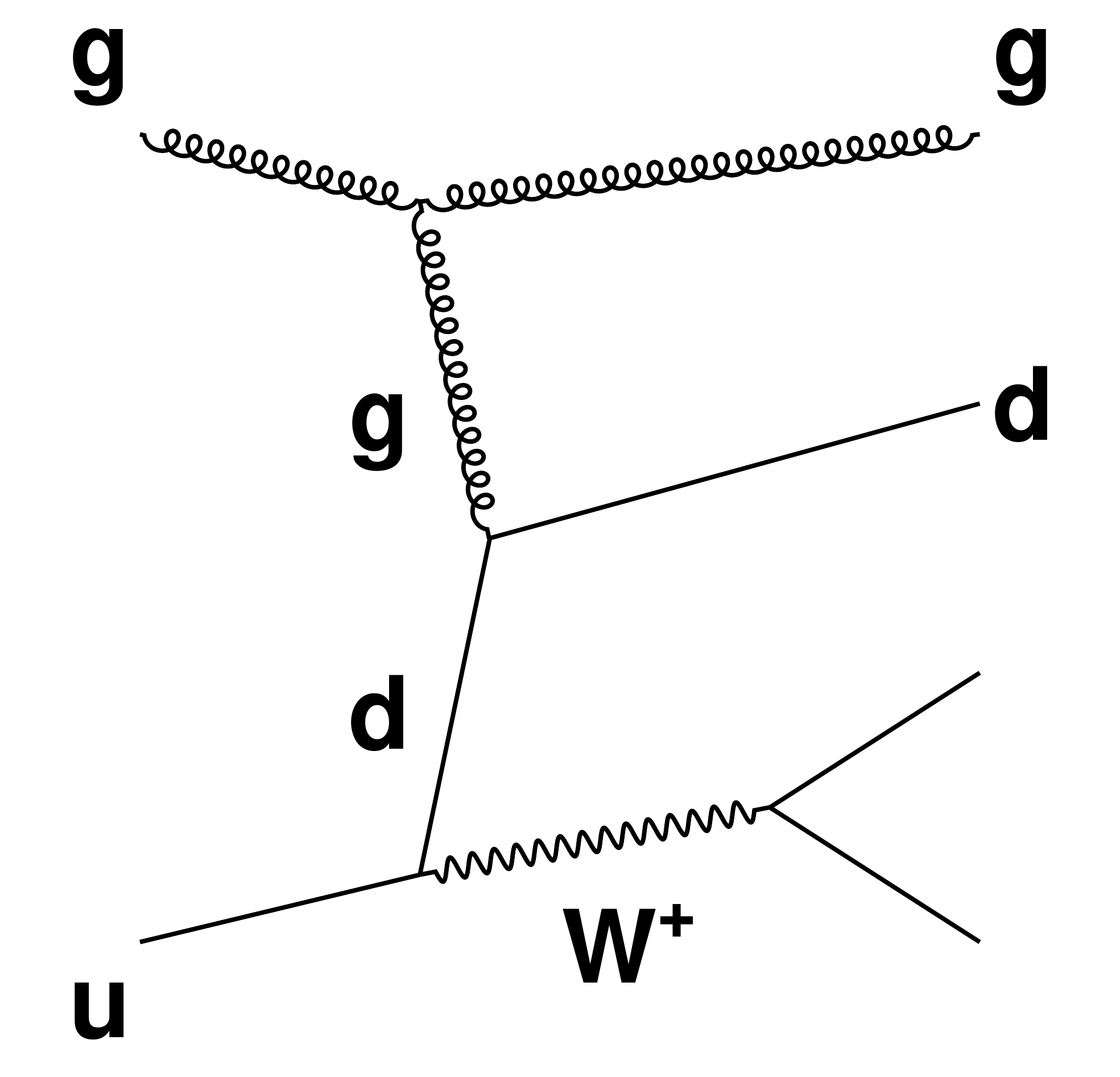}
\caption{Representative diagrams for
 $\PW$ boson production in association with two jets (\dywjj)
 that constitute the main background for the measurement.}
\label{fig:bkgdiagram}}
\end{figure*}

In addition to the purely EW signal diagrams described above, there are other, not purely EW processes, that lead
to the same \lvjj\ final states and can interfere with the signal diagrams
in Fig.~\ref{fig:sigdiagram}.
This interference effect between the signal production and the main Drell-Yan (DY) background processes (\dywjj)
is small compared to the interference effects among the EW production amplitudes,
but needs to be included when measuring the signal
contribution.
Figure~\ref{fig:bkgdiagram}~(left) shows
one example of $\PW$ boson
production in association with two jets
that has the same initial and final states as those in
Fig.~\ref{fig:sigdiagram}.
A process that does not interfere with the EW signal is shown in
Fig.~\ref{fig:bkgdiagram}~(right).

The study of EW \Wjj\ processes is part of a more general
investigation of standard model (SM) VBF and scattering processes that includes
the measurements of EW \Zjj\ processes,
Higgs boson production~\cite{Aad:2012tfa,Chatrchyan:2012ufa,Chatrchyan:2013lba}, and
searches for physics beyond the SM~\cite{Cho:2006sx}.
The properties of  EW \Wjj\ events that are isolated from the backgrounds can be compared with SM predictions.
Probing the additional hadronic activity in selected events
can shed light
on the modeling of the additional parton radiation~\cite{Bjorken:1992er,Schissler:2013nga},
which is important for signal selection and the vetoing of background events.

Higher-dimensional operators outside the SM can
generate anomalous trilinear gauge couplings (ATGCs)~\cite{Hagiwara:1993ck,Degrande:2012wf},
so the measurement of the coupling strengths provides
an indirect search for beyond-the-SM physics at mass scales not directly accessible at the LHC.

At the LHC, the EW \Wjj\ process was first measured
by the CMS Collaboration using pp collisions at $\sqrt{s}=8\TeV$~\cite{Khachatryan:2016qkk}
and then by the ATLAS Collaboration at both $\sqrt{s}=8\TeV$ and $\sqrt{s}=7\TeV$~\cite{Aaboud:2017fye}.
The closely related EW \Zjj\ process was first measured during Run 1
by the CMS Collaboration using pp collisions at
$\sqrt{s}=7\TeV$~\cite{Chatrchyan:2013jya},
and then at $\sqrt{s}=8\TeV$
by both the CMS~\cite{Khachatryan:2014dea} and ATLAS~\cite{Aad:2014dta} Collaborations.
The EW \Zjj\ measurements  using data samples of pp collisions at $\sqrt{s}=13\TeV$
have been performed by ATLAS~\cite{Aaboud:2017emo}
and by CMS~\cite{Sirunyan:2017jej}.
Considering leptonic final states in the same kinematic region the EW \Wjj\  cross section
is about a factor 10 larger than the EW \Zjj\  cross section.
All results so far agree with the expectations of the SM
within a precision of 10--20\%.

This paper presents measurements of the EW \Wjj process with the CMS detector
using pp collisions collected at $\sqrt{s}=$13\TeV
during 2016, corresponding to an integrated luminosity of 35.9\fbinv.
A multivariate analysis (BDT),
based on the methods developed  for the EW Zjj measurement~\cite{Chatrchyan:2013jya,Khachatryan:2014dea},
is used to separate signal events from the large \PW+jets background.
The analysis of the 13\TeV data
offers the opportunity to measure the cross section at a higher energy
than previously done  and to reduce
the uncertainties obtained with previous measurements,
given both the larger integrated luminosity and the larger predicted total
cross section.

This paper is organized as follows:
Section~\ref{sec:cmsexperiment} describes the experimental
apparatus
and Section~\ref{sec:simulation} the event simulations.
Event selection procedures are described in Section~\ref{sec:evsel},
together with the
selection efficiencies and background estimations using control regions (CRs).
Section~\ref{sec:qcd} describes
an estimation of the multijet background
from quantum chromodynamics (QCD), based on CRs in data.
Section~\ref{sec:mjj} discusses a correction applied to the simulation as a function of the invariant mass
$m_\mathrm{jj}$.
Section~\ref{sec:var} presents distributions of the main discriminating variables in data.
Section~\ref{sec:sigdisc} details the strategy adopted
to extract the signal from the data, and the corresponding
systematic uncertainties are summarized in Section~\ref{sec:systunc}.
The cross section and anomalous coupling results are presented in Sections~\ref{sec:results}
and \ref{sec:atgc}, respectively.
Section~\ref{sec:hadactivity} presents
a study of the additional hadronic activity
in an EW \Wjj enriched region.
Finally, a brief summary of the results is given in Section~\ref{sec:summary}.

\section{The CMS detector and physics objects}
\label{sec:cmsexperiment}
The central feature of the CMS apparatus is a
superconducting solenoid of 6\unit{m} internal diameter,
providing a magnetic field of 3.8\unit{T}.
Within the solenoid volume are a silicon pixel and strip tracker,
a lead tungstate crystal electromagnetic calorimeter (ECAL),
and a brass and scintillator hadron calorimeter,
each composed of a barrel and two endcap sections.
Forward calorimeters extend the $\eta$ coverage provided by the barrel and endcap detectors to $\abs{\eta}$ = 5.2.
Muons are measured in gas-ionization detectors embedded in the steel flux-return yoke outside the solenoid.

The tracker measures charged particles within the range $\abs{\eta} < 2.5$.
It consists of 1440 pixel and 15\,148 strip detector modules.
For nonisolated particles with transverse momenta  $1 < \pt < 10\GeV$ and $\abs{\eta} < 1.4$,
the track resolutions are typically 1.5\% in \pt and 25--90 (45--150)\mum
in the transverse (longitudinal) impact parameter~\cite{Chatrchyan:2014fea}.

The energy of electrons is measured after combining the information from
the ECAL and the tracker, whereas their direction is
measured by the tracker. The momentum resolution for electrons with
$\pt \approx$ 45\GeV from $\Z \to \Pe \Pe$ decays ranges from 1.7\% to 4.5\%.
It is generally better in the barrel region than in the endcaps,
and also depends on the bremsstrahlung energy emitted by the electron
as it traverses the material in front of the ECAL~\cite{Khachatryan:2015hwa}.

{\tolerance=1200
Muons are measured in the range $\abs{\eta} < 2.4$, with detection planes made using
three technologies:
drift tubes, cathode strip chambers, and resistive-plate chambers.
Matching muons to tracks measured in the silicon tracker results in a relative transverse momentum
resolution for muons with $20 <\pt < 100\GeV$ of 1.3--2.0\% in the barrel and better than 6\% in the endcaps.
The \pt resolution in the barrel is better than 10\% for muons with \pt up to 1\TeV~\cite{Sirunyan:2018fpa}.
\par}

Events of interest are selected using a two-tiered trigger system~\cite{Khachatryan:2016bia}.
The first level (L1), composed of custom hardware processors, uses information from the calorimeters and
muon detectors to select events at a rate of around 100\unit{kHz} within a time interval of less than 4\mus.
The second level, known as the high-level trigger (HLT), consists of a farm of processors running a version of
the full-event reconstruction software optimized for fast processing, and reduces the event rate to around 1\unit{kHz}
before data storage.

A more detailed description of the CMS detector, together with a
definition of the coordinate system used and the relevant kinematic variables,
can be found in Ref.~\cite{Chatrchyan:2008zzk}.

\section{Simulation of signal and background events}
\label{sec:simulation}
{\tolerance=4800
Signal events are simulated at leading order (LO) using
the \MGvATNLO
(v2.3.3) Monte Carlo (MC) generator~\cite{Alwall:2014hca},
interfaced with \PYTHIA (v8.212)~\cite{Sjostrand:2014zea}
for parton showering (PS) and hadronization.
The NNPDF30 \cite{Ball:2011uy} parton distribution functions (PDFs) are
used to generate the events.
The underlying event is modeled
using the CUETP8M1 tune~\cite{Khachatryan:2015pea}.
The simulation does not include extra partons at
matrix element (ME) level.
The signal is defined in the kinematic region with
parton transverse momentum $\ptj > 25\GeV$,
and diparton invariant mass $m_\mathrm{jj} >120\GeV$.
The simulated cross section for the $\ell\nu$jj final state (with $\ell$ = e, $\mu$ or $\tau$),
applying the above requirements, is
$\sigma_\mathrm{LO}(\mathrm{EW}~\ell\nu\mathrm{jj})=
6.81 ^{+0.03}_{-0.06} \,\text{(scale)}\pm 0.26\,\text{(PDFs)}\unit{pb}$,
where the first uncertainty is obtained by changing simultaneously
the factorization ($\mu_\mathrm{F}$) and renormalization ($\mu_\mathrm{R}$) scales
by factors of $2$ and $1/2$, and the second one reflects the uncertainties
in the NNPDF30 PDFs.
The LO signal cross section and relevant kinematic distributions
estimated with \MGvATNLO are in agreement within 2--5\%
with the next-to-leading-order (NLO)
predictions of the \textsc{vbfnlo} generator (v2.6.3)~\cite{Arnold:2008rz,Arnold:2011wj,Arnold:2012xn},
which include QCD NLO corrections to the LO ME-level diagrams evaluated with \MGvATNLO.
For additional comparisons, signal events produced with \MGvATNLO
are also processed with
the \HERWIGpp (v2.7.1)~\cite{Bahr:2008pv} PS, using the EE5C~\cite{Seymour:2013qka}  tune.
\par}

An additional signal sample that includes NLO QCD corrections but does not include
the s-channel contributions to the final state has been generated with
\POWHEG (v2.0)~\cite{Nason:2004rx,Frixione:2007vw,Alioli:2010xd}, based on the \textsc{vbfnlo} ME
calculations~\cite{Oleari:2003tc,Jager:2012xk}.
In the \POWHEG sample the $m_\mathrm{jj} >120\GeV$ condition is applied on the two \pt-leading
parton-level jets, after clustering the ME final state partons with
the \kt-algorithm~\cite{Catani:1992zp,Catani:1993hr,Ellis:1993tq},
with a distance parameter $D=0.8$, as done in Ref.~\cite{Oleari:2003tc}.
The \POWHEG sample has also been processed alternatively with \PYTHIA and \HERWIGpp
parton showering (PS) and hadronization programs, as done for the \MGvATNLO  samples.
In the following, results obtained with the \POWHEG signal samples are given
as a cross check of the main results obtained with the \MGvATNLO  signal samples.

{\tolerance=1200
Events coming from processes including ATGCs are generated with the same settings
as the SM sample, but include additional information for reweighting
in the three-dimensional effective field theory (EFT) parameter space,
which is described in more detail in Section~\ref{sec:atgc}.
The 'EWdim6NLO' model~\cite{Alwall:2014hca,Degrande:2012wf} is used for the generation of anomalous couplings.
\par}

Background $\PW$ boson events are also simulated
with \MGvATNLO using (i) an NLO ME calculation with up to three final-state
partons generated from QCD interactions, and
(ii) an LO ME calculation with up to four partons from QCD interactions. The ME-PS matching is
performed following the FxFx prescription~\cite{Frederix:2012ps} for the NLO case,
and the MLM prescription~\cite{Mangano:2006rw,Alwall:2007fs} for the LO case.
The NLO background simulation is used to extract the final results, while the
independent LO samples are used to perform the multivariate discriminant training.
The inclusive $\PW$ boson production is normalized
to $\sigma_\text{th}(\PW)=61.5 \unit{nb}$, as computed
at next-to-next-to-leading order (NNLO)
with \FEWZ (v3.1)~\cite{Melnikov:2006kv}.

The evaluation of the interference between \ewkwjj\ and \dywjj\ processes
relies on the predictions obtained with \MGvATNLO.
A dedicated sample of  events arising from the interference terms is generated directly
by selecting the contributions of order $\alpha_\mathrm{s}\alpha_\mathrm{EW}^3$,
and passed through the full detector simulation to estimate
the expected interference contribution.

Other backgrounds are expected from events with
one electron or muon and missing transverse momentum
together with jets in the final state.
Events from top quark pair production are
generated with \POWHEG (v2.0)~\cite{Nason:2004rx,Frixione:2007vw,Alioli:2010xd},
and normalized to the inclusive cross section
calculated at NNLO, including next-to-next-to-leading logarithmic corrections,
of 832\unit{pb}~\cite{Kidonakis:2012db,Czakon:2013goa}.
Single top quark processes are modeled at NLO
with \POWHEG~\cite{Alioli:2010xd,Nason:2004rx,Frixione:2007vw,Alioli:2009je}
and normalized to cross sections of $71.7\pm2.0\unit{pb}$,
$217\pm3\unit{pb}$, and $10.32\pm0.20\unit{pb}$, respectively, for
the tW (\POWHEG v1)~\cite{Re:2010bp}, $t$-, and $s$-channel production~\cite{Kidonakis:2012db,Kidonakis:2013zqa}.
The diboson (VV) production processes ($\PW\PW$,
$\PW\cPZ$, and $\cPZ\cPZ$) are generated with \PYTHIA
and normalized to NNLO cross section computations obtained with
\MCFM (v8.0)~\cite{Campbell:2010ff}.

The contribution from QCD multijet processes is derived via an extrapolation from a QCD data CR with the
lepton relative isolation selection inverted.
All background simulations make use of the \PYTHIA
PS model with the CUETP8M1 tune.

A detector simulation based on \GEANTfour (v9.4p03)~\cite{Agostinelli:2002hh,Allison:2006ve}
is applied to all the generated signal and background samples.
The presence of multiple pp interactions
is incorporated by simulating additional interactions (pileup),
both in-time and out-of-time with respect to the hard
interaction, with a multiplicity
that matches the distribution observed in data.
The average pileup is measured to be about 23
additional interactions per bunch crossing.

\section{Reconstruction and selection of events}
\label{sec:evsel}

Events containing exactly one isolated, high-\pt lepton and at least two high-\pt jets are selected.
Isolated single-lepton triggers are used to acquire the data,
where the lepton is required to have $\pt>27\GeV$ for the electron trigger
and $\pt>24\GeV$ for the muon trigger.

The offline analysis uses candidates reconstructed by the particle-flow (PF)
algorithm~\cite{Sirunyan:2017ulk}.
In the PF event reconstruction, all stable particles in the event --- \ie, electrons, muons, photons,
charged and neutral hadrons ---
are reconstructed as PF candidates using information from all subdetectors to obtain an
optimal determination of their direction, energy, and type.
The PF candidates are used to reconstruct the jets and the missing transverse momentum.

The reconstructed primary vertex (PV) with the largest value of summed physics-object $\pt^2$ is
the primary $\Pp\Pp$ interaction vertex.
The physics objects are the objects returned by a jet finding algorithm~\cite{Cacciari:2008gp,Cacciari:2011ma}
applied to all charged particle tracks associated with the vertex, along with the corresponding
associated missing transverse momentum.
Charged tracks identified as hadrons from pileup vertices are omitted in the subsequent PF event
reconstruction~\cite{Sirunyan:2017ulk}.

{\tolerance=1200
Offline electrons are reconstructed from clusters of energy
deposits in the ECAL that match tracks extrapolated from the
silicon tracker~\cite{Khachatryan:2015hwa}.
Offline muons are reconstructed by fitting trajectories based on hits in the silicon
tracker and in the muon system~\cite{Chatrchyan:2012xi}.
Reconstructed electron or muon candidates
are required to have $\pt>20\GeV$.
Electron candidates are required to be reconstructed
within $\abs{\eta}\leq 2.4$, excluding the barrel-to-endcap transitional region
$1.444 < \abs{\eta} < 1.566$
of the ECAL~\cite{Chatrchyan:2008zzk}.
Muon candidates are required to be reconstructed in the fiducial
region $\abs{\eta}\leq 2.4$.
The track associated with a lepton candidate is required
to have both its transverse and longitudinal impact parameters
compatible with the position of the PV of the event.
\par}

The leptons are required to be isolated;
the isolation ($I$) variable is calculated from PF candidates
and is corrected for pileup on an event-by-event basis~\cite{Cacciari:2007fd}.
The scalar \pt sum of all PF candidates
reconstructed in an isolation cone
with radius $\Delta R=\sqrt{\smash[b]{(\Delta \eta)^{2}+(\Delta \phi)^{2}}}=0.4$
around the lepton's momentum vector, excluding the lepton itself, is required
to be less than 6\% of the electron or muon \pt value.
For additional offline analysis, the isolated lepton is required to have
$\pt>25$\GeV for the muon channel and $\pt>30$\GeV for the electron channel.
Events with more than one lepton satisfying the above requirements are rejected.
The lepton flavor samples are exclusive and precedence is given to the selection of muons.

The missing transverse momentum vector, \MPTvec, is calculated offline as the negative
of the vector sum of transverse momenta of all PF objects identified in the
event~\cite{CMS-PAS-JME-16-004}, and the magnitude of this vector is denoted $\MPT$.
Events are required to have $\MPT$ in excess of 20\GeV in the muon channel
and 40\GeV in the electron channel. The tighter requirement for the electron channel
reduces the corresponding higher background of QCD multijet events.
The transverse mass ($\MT$) of the lepton and $\MPTvec$ four-vector sum is then
required to exceed 40\GeV in both channels.

Jets are reconstructed by clustering PF candidates with the anti-\kt
algorithm~\cite{Cacciari:2005hq,Cacciari:2008gp} using a distance parameter of 0.4.
The jet momentum is the vector sum of all particle momenta in the jet
and is typically within 5 to 10\% of the true momentum over the
whole \pt spectrum and detector acceptance.

An offset correction is applied to jet
energies because of the contribution  from pileup. Jet energy corrections are derived from simulation,
and are confirmed with in situ measurements of the energy balance in dijet, multijet,
photon+jet, and Z+jets events with leptonic Z boson decays~\cite{Chatrchyan:2011ds}.
Loose jet identification criteria are applied to reject misreconstructed jets resulting
from detector noise~\cite{CMS-PAS-JME-16-003}.
Loose criteria are also applied to remove jets heavily contaminated with pileup energy
(clustering of energy deposits not associated with a parton from the
primary $\Pp\Pp$ interaction)~\cite{CMS-PAS-JME-13-005,CMS-PAS-JME-16-003}.
The efficiency of the jet identification is greater than 99\%, with a rejection of 90\%
of background pileup jets with $\pt\simeq50\GeV$ and $\abs{\eta}\leq 2.5$. For jets with
$\abs{\eta} > 2.5$ and $30 < \pt < 50\GeV$, the efficiency is approximately 90\% and the
pileup jet rejection is approximately 50\%.
The jet energy resolution (JER) is typically ${\approx}$15\% at 10\GeV,
8\% at 100\GeV, and 4\% at 1\TeV for jets with $\abs{\eta}\leq 1$~\cite{Chatrchyan:2011ds}.
Jets reconstructed with $\pt\geq 15$\GeV and $\abs{\eta}\leq 4.7$ are used in the analysis.

The two highest \pt jets are defined as the tagging jets,
and are required  to have $\pt>50$\GeV and $\pt>30$\GeV
for the leading and subleading (in \pt) jet, respectively.
The invariant mass of the two tagging jets is required to satisfy $m_\mathrm{jj}>200\GeV$.

The transverse momentum of the $\PW$ boson ($\vec{p}_{\mathrm{T} \PW}$) is evaluated as the vector sum
of the lepton \pt and \MPTvec.
The event $\pt$ balance ($R(\pt$)) is then defined as

\begin{equation}
R(\pt)=
\frac
{\abs{ \vec{p}_{\mathrm{T} \mathrm{j}_1}+\vec{p}_{\mathrm{T} \mathrm{j}_2}+\vec{p}_{\mathrm{T} \PW}}}
{ \abs{\vec{p}_{\mathrm{T} \mathrm{j}_1}} +\abs{\vec{p}_{\mathrm{T} \mathrm{j}_2}} + \abs{\vec{p}_{\mathrm{T} \PW}} }
\end{equation}
where $\vec{p}_{\mathrm{T} \mathrm{j}_1}$ and $\vec{p}_{\mathrm{T} \mathrm{j}_2}$ are the transverse momenta of the two tagging jets.

Finally, events are required to have $R(\pt) < 0.2$.
This has a negligible effect on the analysis sensitivity and allows the definition
of a nonoverlapping control sample with $R(\pt) > 0.2$ that is used to derive
a correction to the invariant mass based on a CR in data, as described in Section~\ref{sec:mjj}.

A multivariate analysis technique, described in Section~\ref{sec:sigdisc}, is used
to provide an optimal separation of the \dywjj\ and \ewkwjj\ components of the inclusive
\lvjj\ spectrum.
The main discriminating variables are the
dijet invariant mass $m_\mathrm{jj}$ and pseudorapidity separation $\Delta\eta_\mathrm{jj}$.

Angular variables useful for signal discrimination include
 the $y^*$ Zeppenfeld variable~\cite{Schissler:2013nga}, defined as
the difference between the rapidity of the
$\PW$ boson $y_{\PW}$ and the average rapidity of the two tagging jets, \ie,
\begin{equation}
y^*=y_{\PW}-\frac{1}{2}(y_{\mathrm{j}_1}+y_{\mathrm{j}_2}),
\end{equation}
and the $z^*$ Zeppenfeld variable~\cite{Schissler:2013nga}
defined as \begin{equation}
z^*=\frac{ y^* } { \Delta y_\mathrm{jj} },
\end{equation}
where $\Delta y_\mathrm{jj}$ is the dijet rapidity separation.

Table~\ref{tab:yields} reports the expected and observed
event yields after the initial  selection and after imposing a minimum
value for the final multivariate discriminant output applied to define the signal-enriched
region used for the studies of additional hadronic activity described in Section~\ref{sec:hadactivity}.

\begin{table*}[htbp]
\centering
\topcaption{Event yields expected for background and signal
  processes using the initial selections
and with a selection on the multivariate analysis output (BDT) that provides
similar signal and background yields. The yields are
  compared to the data observed in the different channels and categories.
  The total uncertainties quoted for signal, \dywjj
and diboson backgrounds, and processes with top
  quarks (\ttbar and single top quarks)
include the systematic uncertainties.\label{tab:yields}}
\begin{tabular}{l{c}@{\hspace*{5pt}}cc{c}@{\hspace*{5pt}}cc}\hline
{Sample}
 &&  \multicolumn{2}{c}{Initial} &&  \multicolumn{2}{c}{$\mathrm{BDT} > 0.95 $ } \\
\cline{1-1} \cline{3-4}\cline{6-7}
  &&  $\mu$ & $\Pe$ &&  $\mu$ & $\Pe$ \\
VV && 20300 $\pm$ 2000   & 9820 $\pm$ 980   && 11.0 $\pm$ 2.5    & 9.6 $\pm$ 2.8    \\
\dyzjj&& 102000 $\pm$ 10000 & 29900 $\pm$ 3000   && 9.4 $\pm$ 5.9   & 7.7 $\pm$ 3.0   \\[\cmsTabSkip]
$\ttbar$ &&  298000 $\pm$ 28000  & 164000 $\pm$ 15000   && 146 $\pm$ 17   & 102 $\pm$ 12    \\
Single top quark && 96000 $\pm$ 14000   & 45800 $\pm$ 6900  && 35.5 $\pm$ 5.6  & 25.7 $\pm$ 4.2   \\
QCD multijet && 100000 $\pm$ 39000   & 65000 $\pm$ 21000  && 98 $\pm$ 39  & 17.0 $\pm$ 5.6   \\
\dywjj&& 1720000 $\pm$ 120000 & 715000 $\pm$ 51000   && 356 $\pm$ 65   &  240 $\pm$ 41   \\[\cmsTabSkip]
Interference && 7000 $\pm$ 2100   &  3400 $\pm$ 1000   && 18.2 $\pm$ 8.1     & 9.8 $\pm$ 5.5    \\
Total backgrounds&& 2340000 $\pm$ 170000  & 1032000 $\pm$ 58000   && 674 $\pm$ 78   &412 $\pm$ 44    \\
\ewkwjj signal && 43100 $\pm$ 4300   &  20700 $\pm$ 2100   && 503 $\pm$ 54     & 308 $\pm$ 34    \\
\ewkzjj signal && 1330 $\pm$ 130   &  407 $\pm$ 41   && 11.2 $\pm$ 1.3     & 6.6 $\pm$ 0.9    \\
Total prediction&& 2390000  $\pm$ 170000   & 1054000 $\pm$ 58000    && 1186 $\pm$ 95   & 726 $\pm$ 56    \\[\cmsTabSkip]
Data &&  $2381901$ &  $1051285$ &&  $1138$ &  $686$  \\
\end{tabular}
\end{table*}

\subsection{Discriminating quarks from gluons}
\label{sec:gtag}
Jets in signal events are expected to originate from quarks,
whereas for background events it is more probable that jets are
initiated by a gluon.
A quark-gluon likelihood (QGL) discriminant~\cite{Chatrchyan:2013jya}
is evaluated for the two tagging jets with the intent of distinguishing
the nature of each jet.

The QGL discriminant exploits differences in the showering and fragmentation of quarks and gluons,
making use of the following internal jet composition observables:
(i) the particle multiplicity of the jet, (ii) the  minor root-mean-square of distance between
the jet constituents
in the $\eta$--$\phi$ plane, and (iii) the  \pt
distribution function of the jet constituents, as defined in Ref.~\cite{CMS-PAS-JME-13-002}.

The variables are used as inputs to a likelihood
discriminant on gluon and quark jets constructed from simulated dijet events.
The performance of the QGL discriminant is
evaluated and validated using independent,
exclusive samples of \cPZ+jet and dijet data~\cite{CMS-PAS-JME-13-002}.
Corrections to the simulated QGL distributions and related systematic uncertainties
are derived from a comparison of simulation and data distributions.

\section{The QCD multijet background}
\label{sec:qcd}
The QCD multijet contribution is estimated by defining a multijet-enriched CR with inverted
lepton isolation criteria for both the muon and electron channels.
In the nominal selection both lepton types are required to pass the relative isolation
requirement $I<0.06$, whereas the multijet-enriched CRs are defined
with the same event selection but with isolation
requirements $0.06<I<0.12$ and  $0.06<I<0.15$,  for the muon and electron channel respectively.
It is then assumed that the $\MPT$ distribution of QCD events has the same shape in both the
nominal and the multijet-enriched CR.

The various components, with floating \PW+jets and QCD multijet background scale factors,
are simultaneously fitted to the $\MPT$ data distributions,
independently in the muon and electron channels, and the expected QCD multijet yields
in the nominal regions are derived.

The contribution of QCD multijet processes in any other observable ($x$) used in the analysis
is then normalized to the yields obtained above from the fit to the $\MPT$ distribution, and the
shape for the distribution $x$ is taken as the difference between data and all simulated background contributions
in the $x$ distribution in the multijet-enriched CR.

The estimation of the QCD multijet contribution based on a CR in data is validated by checking the modeling
of other variables that discriminate QCD multijets from \PW+jets such as the W transverse mass and the minimum difference in $\phi$ between
the missing transerse energy and the jets. Good agreement with the data is observed in all distributions.
The stability of the \PW+jets fitted normalization is checked by varying the selection requirements
for the fitted region and repeating the QCD extraction fit.
The observed variations in fitted normalization when varying the $m_\mathrm{T}$(W) and $\MPT$ selection
requirements with respect to the fit region definition are much smaller than systematic uncertainties.

Although {\cPqb} tagging is not used in this analysis, a
{\cPqb}-tagging discriminant output~\cite{BTV-16-002}
is used to check the fitted \PW+jets background normalization as well as the $\ttbar$ normalization from simulation, and they
agree with data within the uncertainties.
Finally, the selections on $m_\jj$, $\MPT$, and $m_\mathrm{T}$(W) are also loosened in order to verify that the \PW+jets
background scale factor is not biased by these requirements.

\section{The \texorpdfstring{$m_\jj$}{mjj} correction} \label{sec:mjj}
A systematic overestimation of the simulation yields
is caused by a partial mistiming of the signals in the forward region of the ECAL endcaps
($2.5<\vert \eta \vert<3.0$).
This effect, which increases with increasing $m_\jj$, is observed in both electron and muon channels.
A correction for this effect is derived in the nonoverlapping
signal-depleted CR obtained by requiring that the event transverse momentum balance $R(\pt)$,
defined in Section~\ref{sec:evsel}, exceeds 0.2.

A third-order polynomial correction is first applied to the \PW+jets simulation separately in the muon and electron channels
in order to match the $R(\pt)$ distribution in data. The magnitude of the applied $R(\pt)$ corrections is about 10\%.
The uncertainty in this correction due to the limited statistical
precision of the simulation as well as data is propagated to the fitted \PW+jets templates.

A correction to the $m_\jj$ prediction from simulation is derived in the signal-depleted $R(\pt)>0.2$ CR
via a third-order polynomial fit to the ratio of data to the overall prediction
from simulation for signal and background as a function of $\ln(m_\jj/\GeV)$.
The electron and muon channels are combined when deriving the $m_\jj$ correction. The uncertainty in the correction includes
the data statistical component as well as the systematic uncertainty due to the limited statistical precision of the simulation.

\begin{figure}[htb!]
  \centering
      \includegraphics[width=\cmsFigWidth]{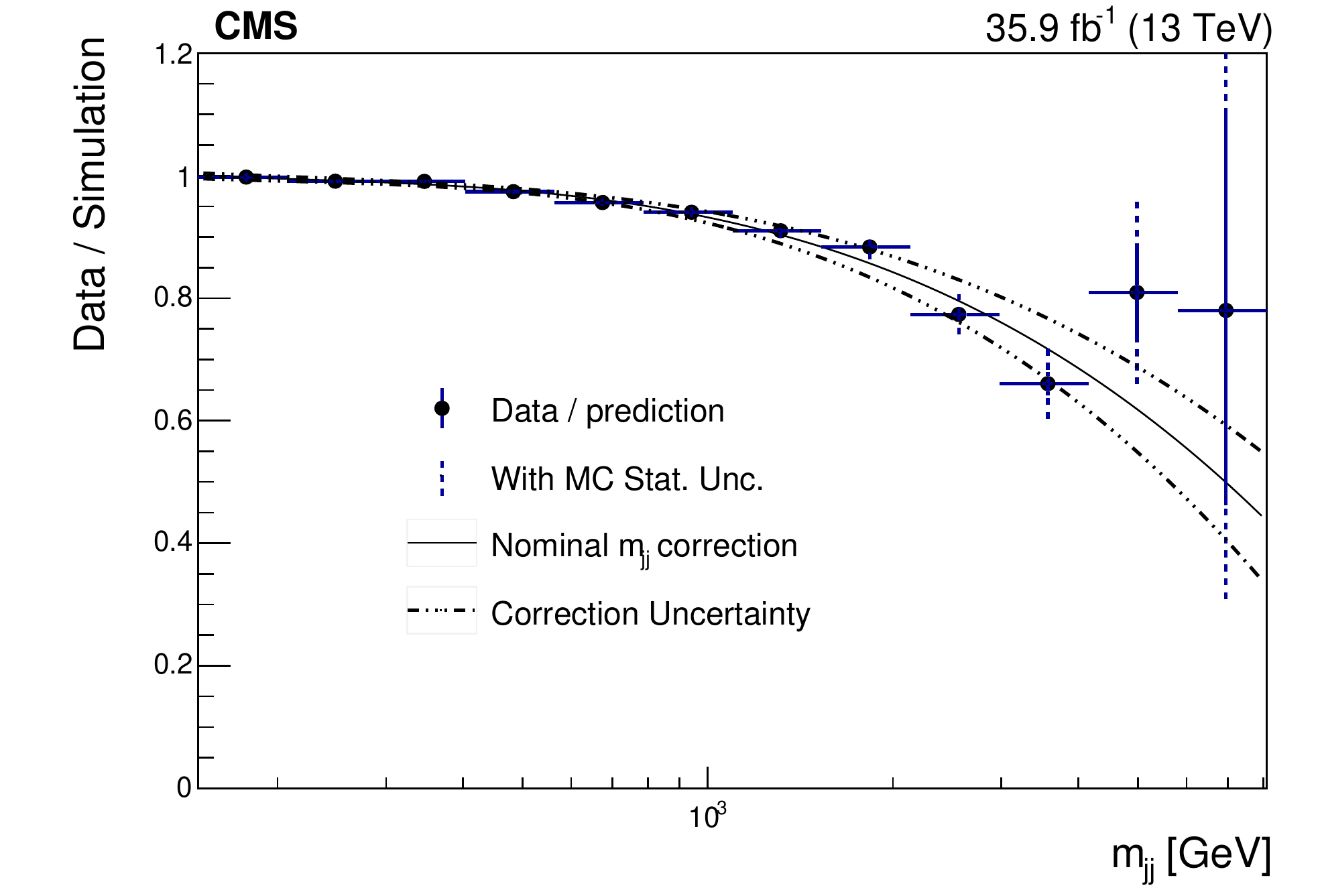}
    \caption{ Data divided by simulation as a function of $\ln(m_\jj/\GeV)$ in a signal-depleted control sample with $R(\pt)>0.2$.
     This distribution is fit by a third-order polynomial (solid black line) in order
    to derive a correction on the simulation $m_\jj$ prediction.
    The points are varied by the uncertainty, including the effect of the limited number of simulated events
    and refitted in order to derive the systematic variations
    on the correction (dashed lines) corresponding to a standard deviation (s.d.).
    }
  \label{fig:mjjfit}

\end{figure}

Figure \ref{fig:mjjfit} shows the fitted correction including the uncertainty.
This correction is applied to all simulated results, including the signal, and the
corresponding uncertainty is propagated to the signal extraction fits.

\section{Distributions of discriminating variables}
\label{sec:var}

\begin{figure*}[htb!]
  \centering
      \includegraphics[width=0.98\textwidth]{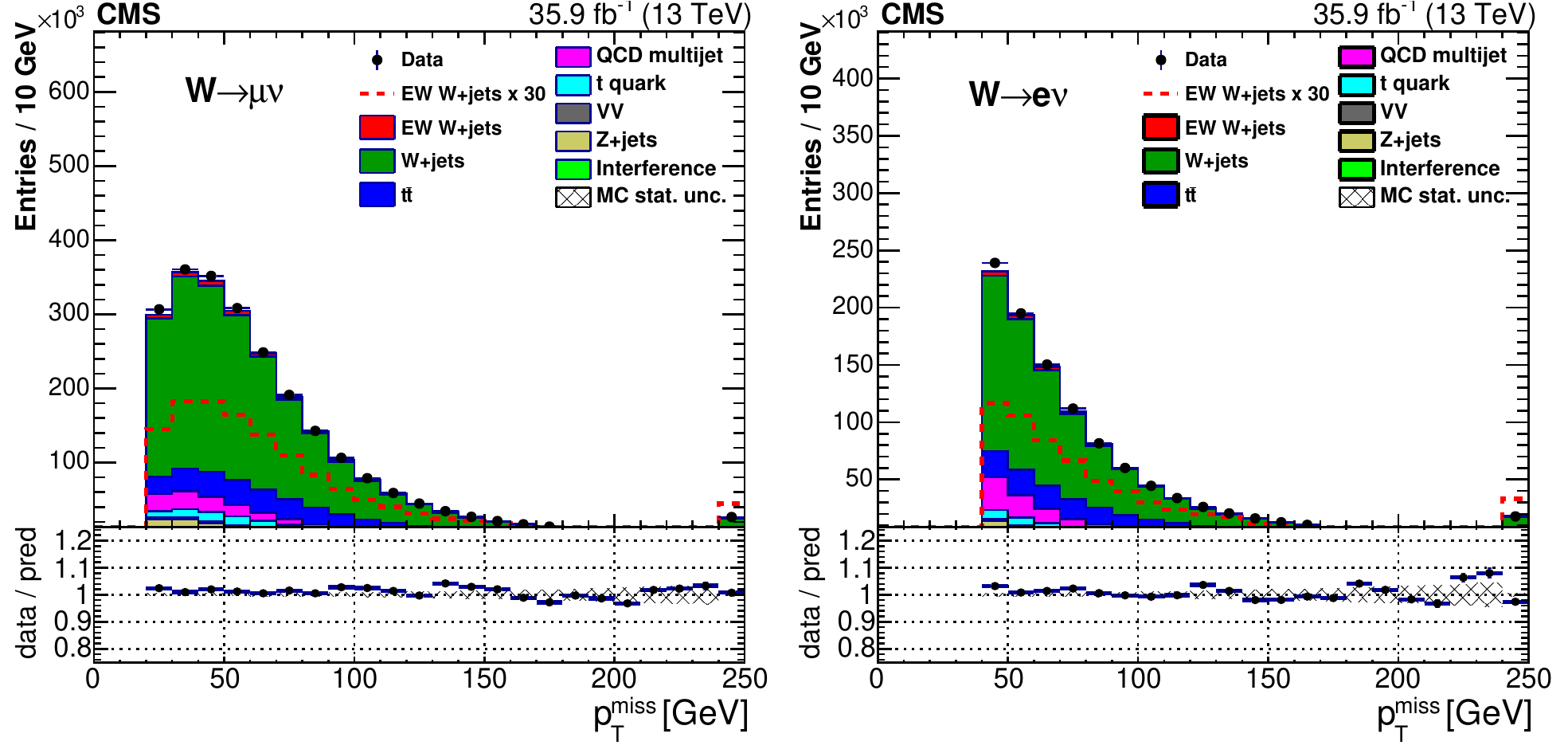}
      \includegraphics[width=0.98\textwidth]{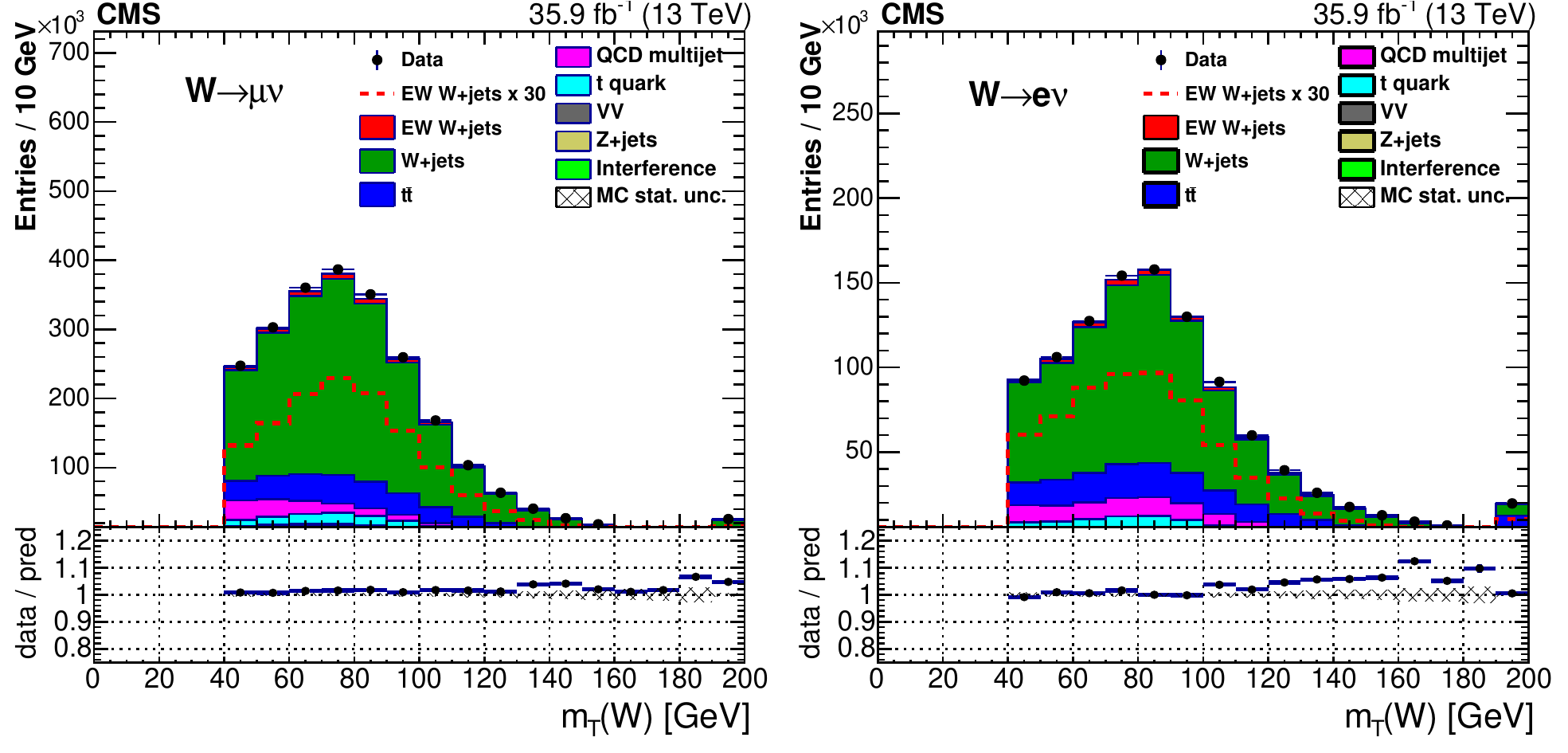}
    \caption{Distribution of the missing transverse momentum (upper) and the lepton-\MPT
    system transverse mass (lower)
    after the event preselection for the selected leading lepton in the event,
      in the muon (left) and electron (right) channels.
      In all plots the last bin contains overflow events.}
    \label{fig:met}

\end{figure*}

\begin{figure*}[htb!]
  \centering
      \includegraphics[width=0.98\textwidth]{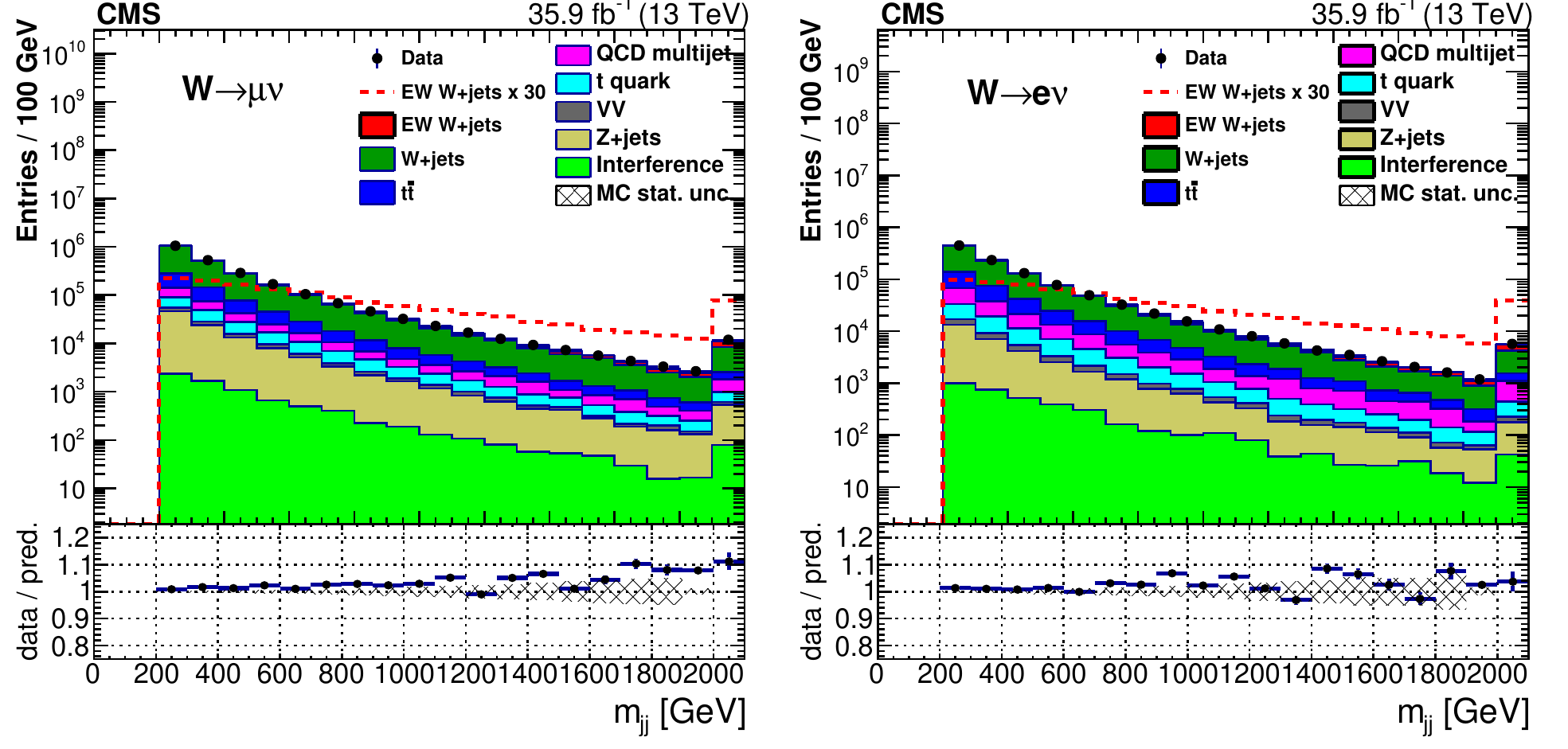}
      \includegraphics[width=0.98\textwidth]{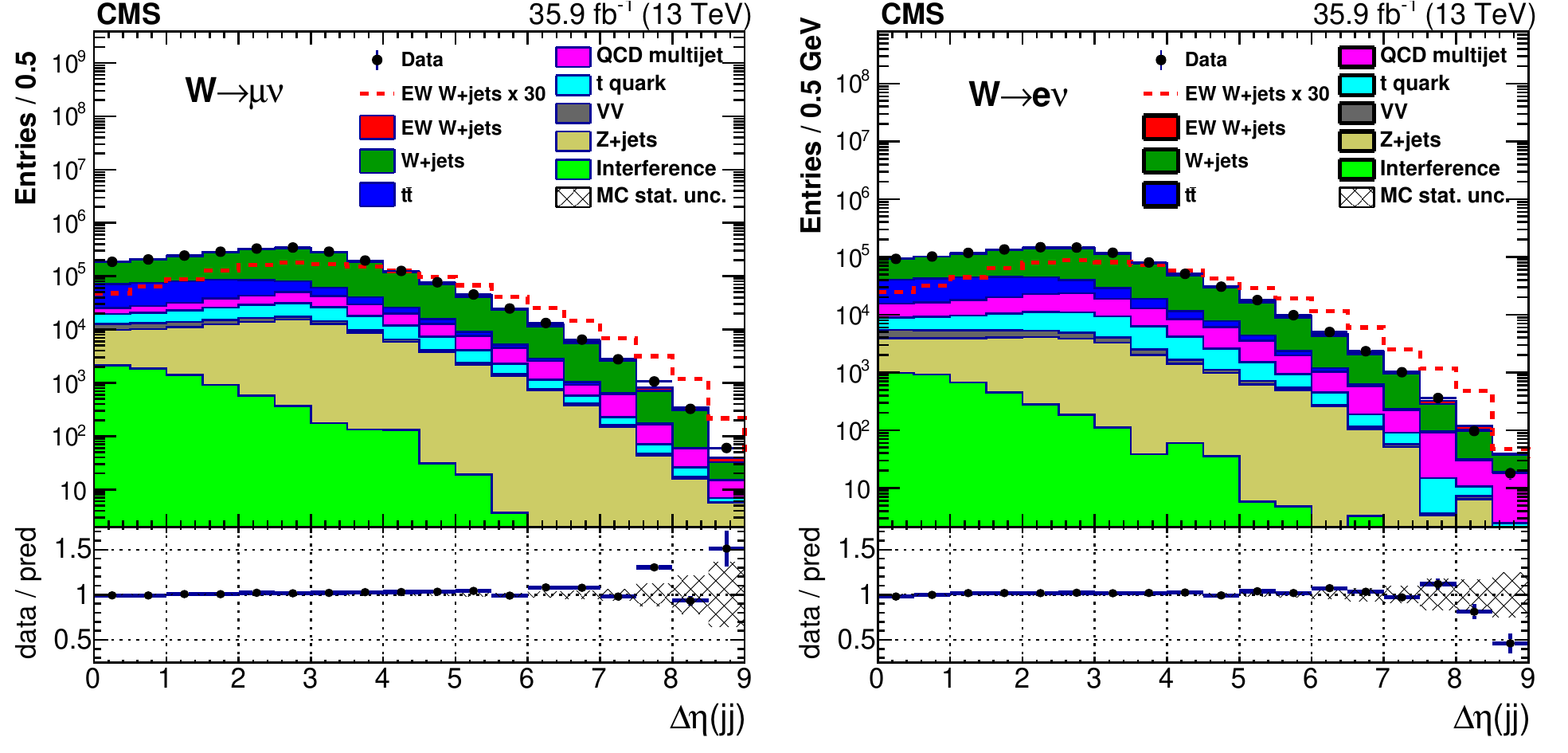}
      \caption{Dijet invariant mass (upper) and pseudorapidity difference (lower)
      distributions after the event preselection,
      in the muon (left) and electron (right) channels.
      In all plots the last bin contains overflow events.}
    \label{fig:dijet}

\end{figure*}

\begin{figure*}[htb!]
  \centering
  \includegraphics[width=0.98\textwidth]{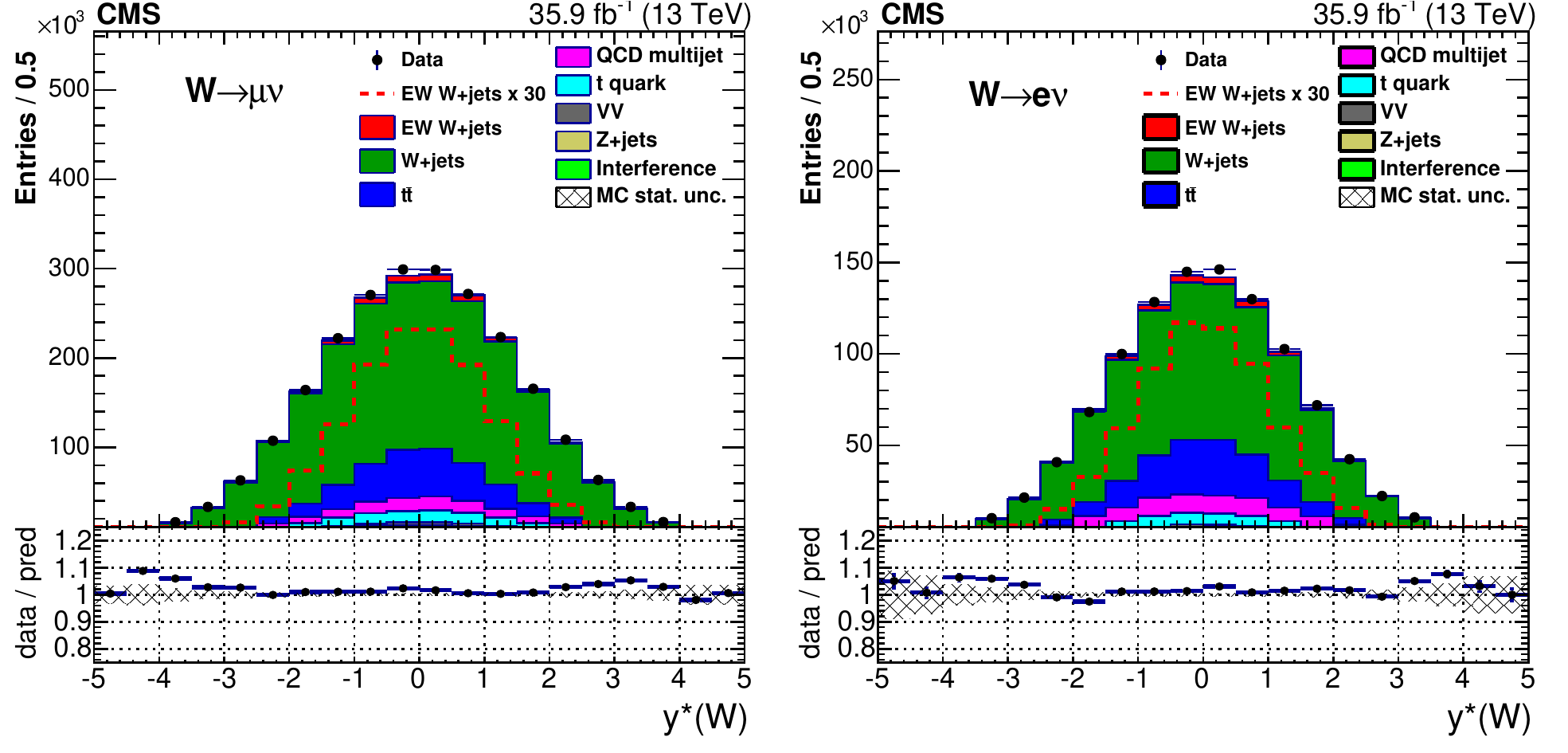}
  \includegraphics[width=0.98\textwidth]{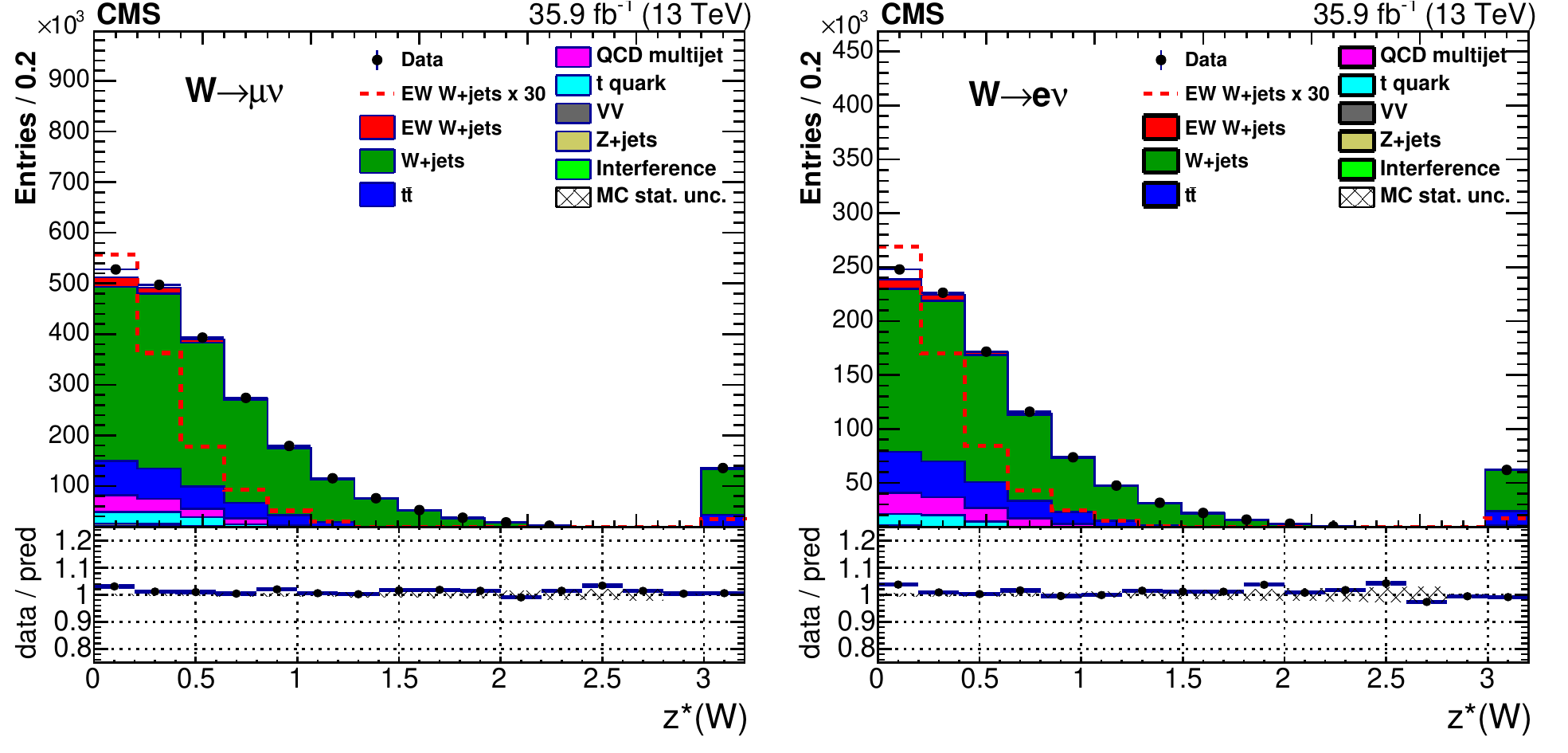}
    \caption{Distributions of the ``Zeppenfeld'' variables $y^\star$(W) (upper)
    and $z^\star$(W) (lower) after event preselection
      in the muon (left) and electron (right) channels.
      In all plots the first and last bins contain overflow events.}
    \label{fig:zep}

\end{figure*}

\begin{figure*}[htb!]
  \centering
    \includegraphics[width=0.98\textwidth]{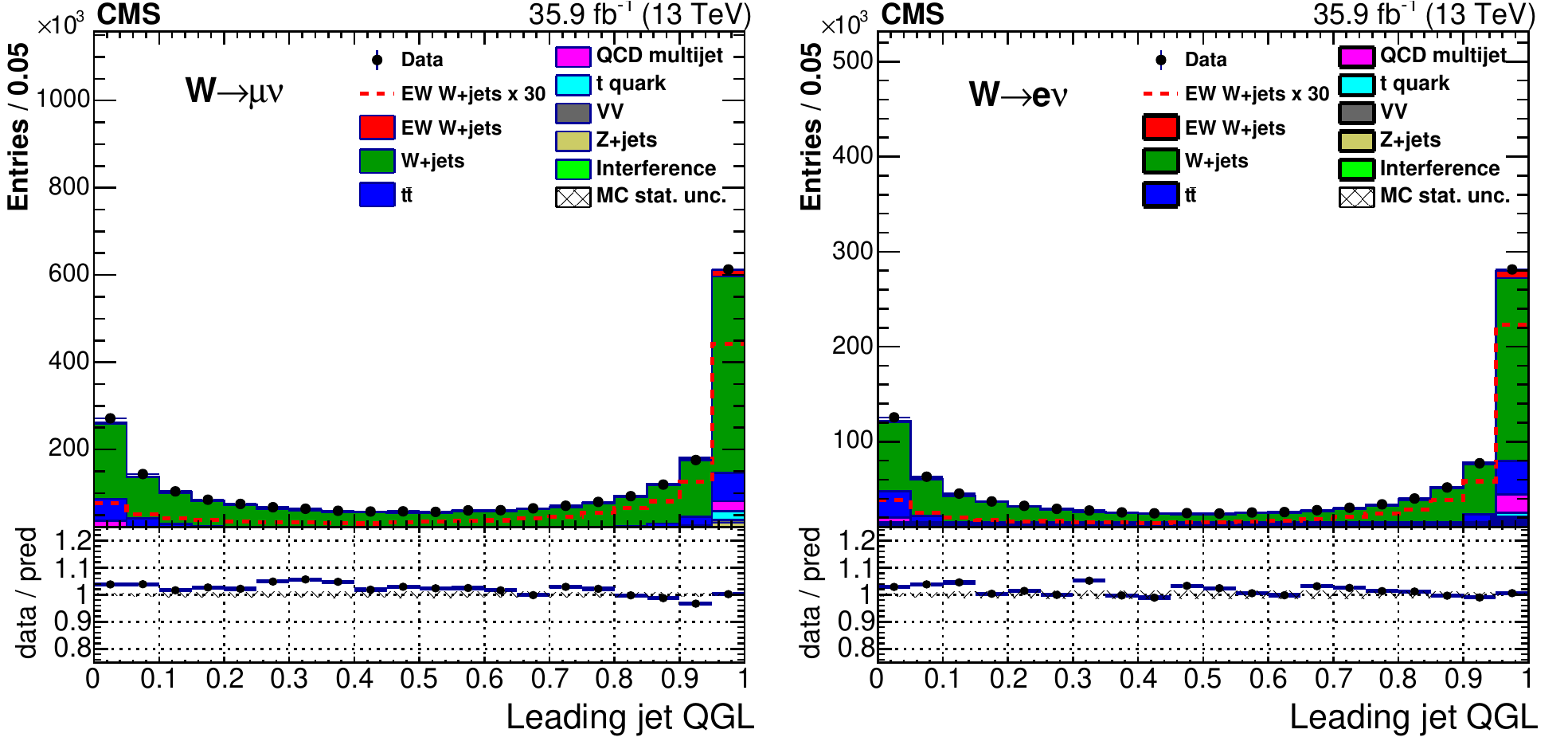}
    \includegraphics[width=0.98\textwidth]{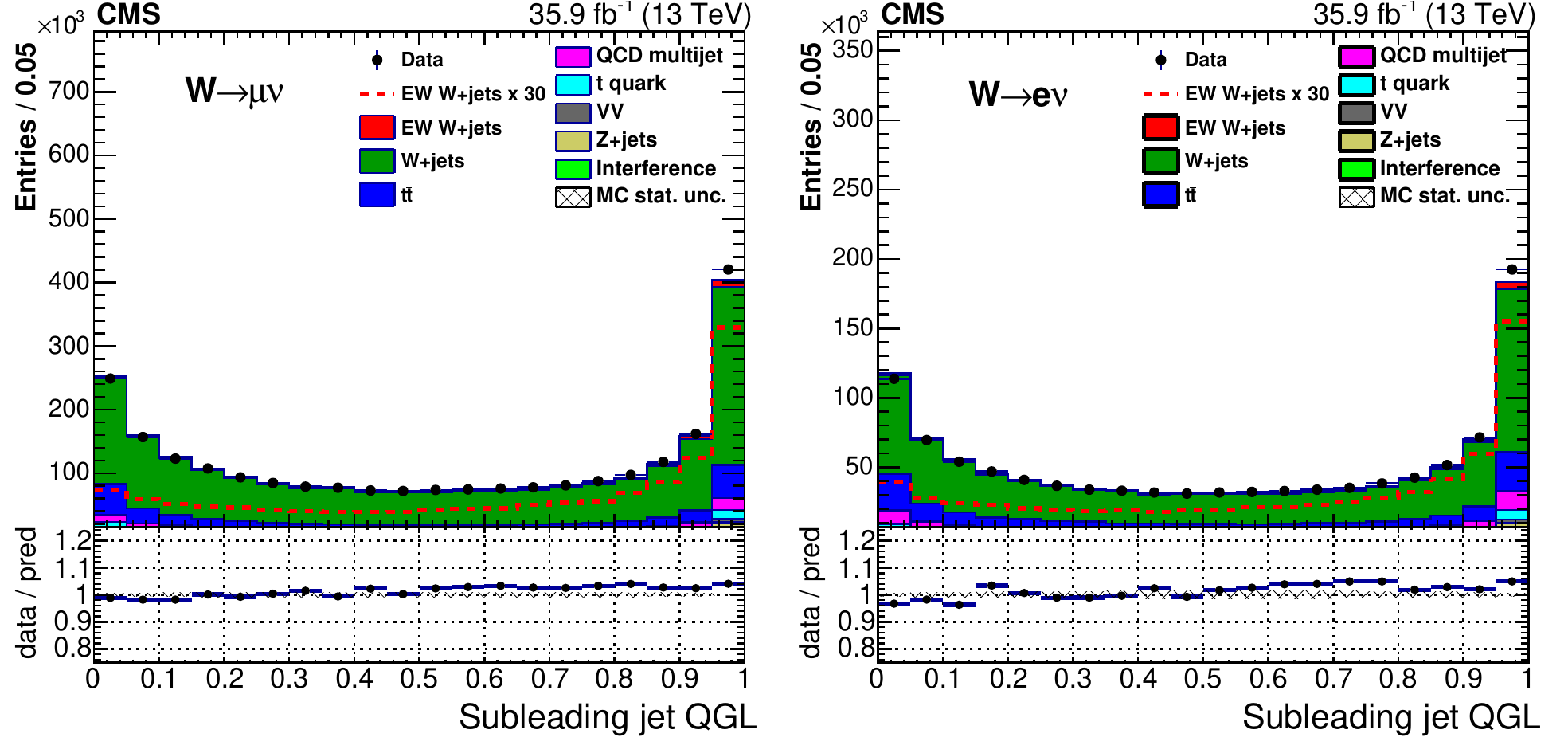}
    \caption{The QGL output for the leading (upper) and subleading (lower) quark jet candidates
     in the preselected muon (left) and electron (right) samples.}
     \label{fig:JetQgl}

\end{figure*}

Figure \ref{fig:met}
shows the $\MPT$ and $\MT$(W) distributions after the
event preselection.
The dijet invariant mass and pseudorapidity difference ($\Delta\eta_\jj$)
after preselection are presented in Fig.~\ref{fig:dijet}, and
Fig.~\ref{fig:zep} shows the $y^\star$ and $z^\star$
distributions after the event preselection.
The distributions of the QGL likelihood output values in data and simulation for
the two tagging jets are shown in Fig.~\ref{fig:JetQgl}.
The prediction from simulated events and the data agree within total uncertainties for all
discriminating variables.

\section{Signal discriminants and extraction procedure}
\label{sec:sigdisc}
The \ewkwjj\ signal is characterized
by a large
pseudorapidity separation between the tagging jets,
due to the small-angle scattering of the two initial partons.
Because of both the topological configuration and the large energy
of the outgoing partons, $m_\mathrm{jj}$ is also
expected to be large, and can be used to distinguish the \ewkwjj\ and \dywjj\ processes.
The correlation between $\Delta\eta_\mathrm{jj}$ and $m_\mathrm{jj}$ is expected
to be different in signal and background events, therefore
these characteristics are expected to yield a high separation power
between \ewkwjj\ and \dywjj\ production.
In addition, in signal events
it is expected that the $\PW$ boson candidate is produced
centrally
in the rapidity region defined by the two tagging jets.
As a consequence, signal events are expected to yield lower values
of $z^*$
compared to the DY background.
Other variables that are used to enhance the signal-to-background separation
are related to the kinematics of the event
or to the properties of the jets that are expected to be initiated by
quarks.
The variables that are used in the
multivariate analysis are:
(i)~$m_\mathrm{jj}$,
(ii)~$\Delta\eta_\mathrm{jj}$,
(iii)~$z^*$, and
(iv)~the QGL values of the two tagging jets.

The output is built by training a boosted decision tree
(BDT) discriminator
with the \textsc{tmva} package~\cite{Hocker:2007ht}
to achieve an optimal separation between the
\ewkwjj\ and \dywjj\ processes.
The simulated events that are used for the BDT training
are not used for the signal extraction.

To improve the sensitivity for the extraction of the signal
component, the transformation that originally projects the BDT output
value in the [$-$1,$+$1] interval is changed to  $\mathrm{BDT'} = \tanh^{-1}((\mathrm{BDT}+1)/2)$.
This allows the purest signal region of the BDT output to be better
sampled while keeping an equal-width binning of the BDT variable.

\begin{figure}[htb!]
  \centering
    \includegraphics[width=0.48\textwidth]{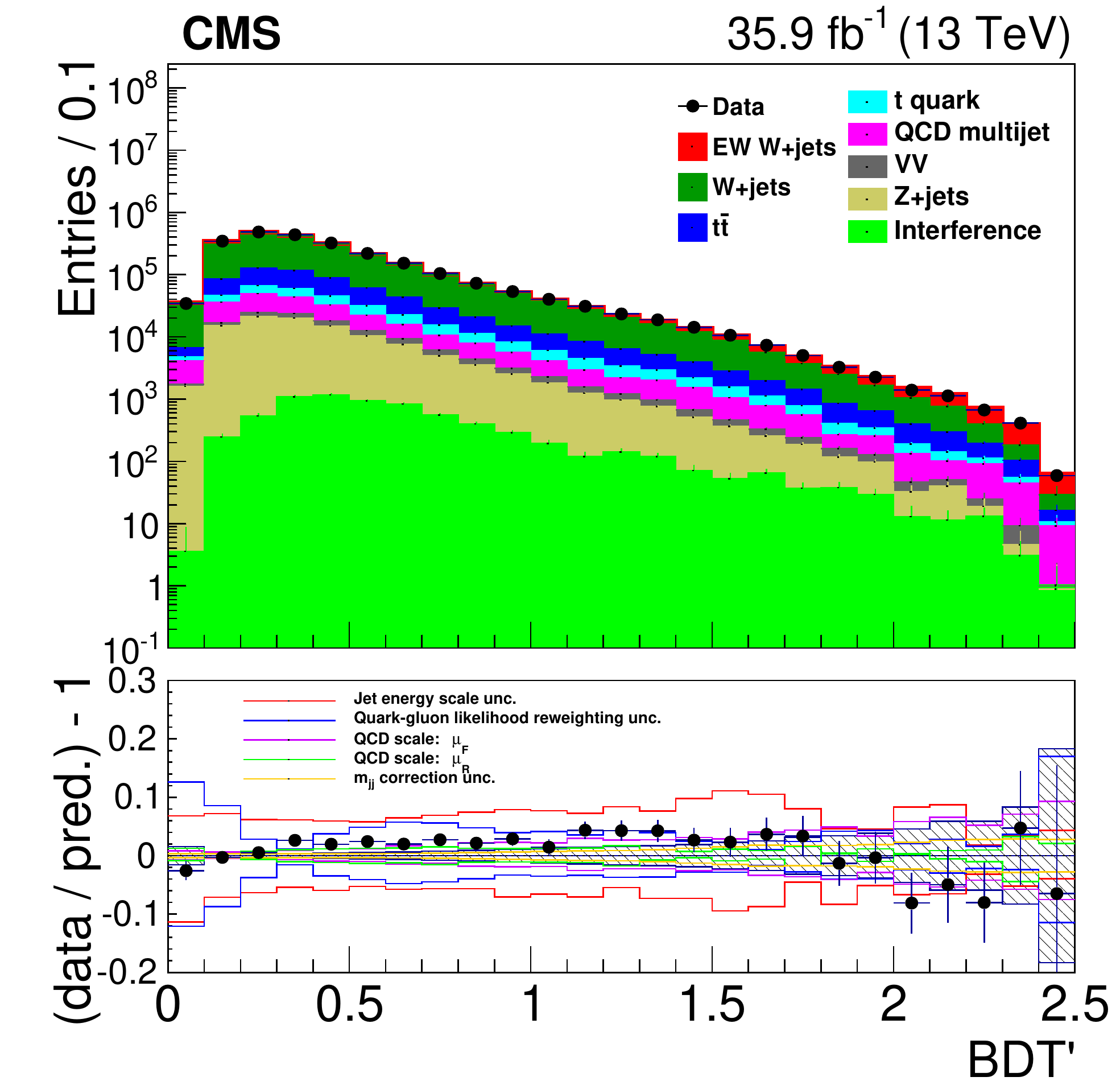}
    \includegraphics[width=0.48\textwidth]{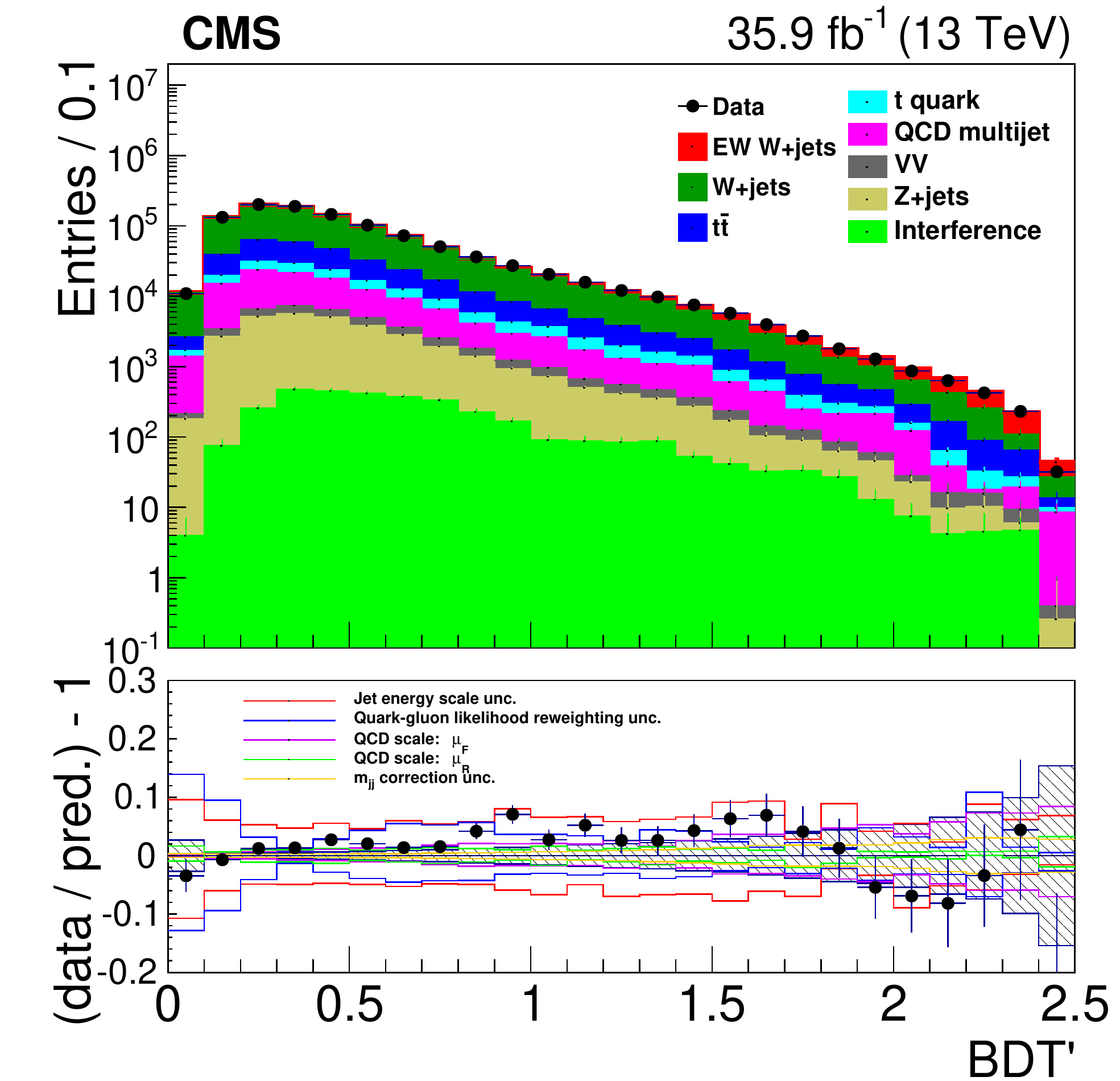}
    \caption{Data and MC simulation BDT' output distributions for the muon (\cmsLeft) and electron (\cmsRight) channels,
      using the BDT output transformed
      with the $\tanh^{-1}$ function to enhance the purest signal region. The ratio panel shows the statistical uncertainty
      from the simulation as well as the independent systematic uncertainties front the leading sources.}
     \label{fig:bdt}

\end{figure}

Figure~\ref{fig:bdt} shows
the distributions of the discriminants for the two leptonic channels.
Good overall agreement between simulation and data is observed in all distributions,
and the signal presence is visible at high BDT' values.

A binned maximum likelihood is built from the expected
rates for each process, as a function of the value of the discriminant,
which is fit to extract
the strength modifiers for the \ewkwjj\ and \dywjj\ processes,
$\mu = \sigma({\mathrm{EW}~\PW\mathrm{jj}}) / \sigma_\mathrm{LO}({\mathrm{EW}~\ell\nu\mathrm{jj}})$
and
$\upsilon = \sigma({\PW})/\sigma_\text{NNLO}({\PW})$.
Nuisance parameters are added to modify the
expected rates and shapes according to the estimate of the systematic
uncertainties affecting the measurement.

The interference between the \ewkwjj\ and
\dywjj\ processes is included in the fit procedure,
and its strength scales as $\sqrt{\mu\upsilon}$.
The interference model is derived from
the \MGvATNLO simulation described
in Section~\ref{sec:simulation}.

The parameters of the model ($\mu$ and $\upsilon$) are determined by
maximizing the likelihood. 
The statistical methodology follows the one used in other
analyses~\cite{CMS-NOTE-2011-005}
using asymptotic formulas~\cite{Cowan:2010js}.
In this procedure the systematic uncertainties affecting the
measurement of the signal and background strengths ($\mu$ and $\upsilon$)
are constrained with log-normal probability distributions.

\section{Systematic uncertainties}
\label{sec:systunc}

The main systematic uncertainties affecting the measurement
are classified into experimental and theoretical according to their sources.
Some uncertainties affect only
normalizations, whereas others affect both the normalization and shape
of the BDT output distribution.

\subsection{Experimental uncertainties}
\label{subsec:expunc}

The following experimental uncertainties are considered.

\begin{description}
\item[Integrated luminosity.] A 2.5\% uncertainty is assigned to the value of
the integrated luminosity~\cite{CMS-PAS-LUM-17-001}.
\item[Trigger and selection efficiencies.]
Uncertainties in the efficiency corrections based on control samples in data for
the leptonic trigger and
offline selections are included and amount to a total of 2--3\%
  depending on the lepton $\pt$ and $\eta$,
  for both the e and $\mu$ channels. These uncertainties are
  estimated by comparing the lepton efficiencies expected in
  simulation and measured in data with
  a ``tag-and-probe'' method~\cite{Khachatryan:2010xn}.
\item[Jet energy scale and resolution.]
 The uncertainty in the energy of the jets affects the event selection and the
  computation of the kinematic variables used to calculate the discriminants.
  Therefore, the uncertainty in the jet energy scale (JES) affects both the
  expected event yields and the final shapes.
  The effect of the JES uncertainty is studied by
rescaling up and down the reconstructed jet energy
  by \pt- and $\eta$-dependent scale factors~\cite{Chatrchyan:2011ds}.
  An analogous approach is used for the JER.
\item[QGL discriminator.] The uncertainty in the performance of the
  QGL discriminator is measured
using independent \cPZ+jet and dijet data,
after comparing with the corresponding simulation predictions~\cite{CMS-PAS-JME-13-002}.
Shape variations corresponding to
the full differences between the data and the simulation are used as estimates of the uncertainty.
\item[Pileup.] Pileup can affect the identification and
  isolation of the leptons or the corrected energy of the jets. When
  the jet clustering algorithm is run, pileup can distort the reconstructed
  dijet system because of the contamination
  of tracks and calorimetric deposits.
  This uncertainty is evaluated by generating
  alternative distributions of the
number of pileup interactions, corresponding
  to a 4.6\% uncertainty in the total
  inelastic pp cross section at $\sqrt{s}=13\TeV$~\cite{Sirunyan:2018nqx}.
\item[Limited number of simulated events.] For each signal and background simulation,
  shape variations for the distributions are considered by shifting the content of each bin
  up or down by its statistical uncertainty~\cite{Barlow:1993dm}.
  This generates alternatives to the nominal shape that are used to describe the uncertainty
from the limited number of simulated events.
\item[$m_\jj $ correction.] As described in Section \ref{sec:mjj}, the $m_\jj $ prediction from
  simulation is corrected to match the distribution in data in a signal-depleted $R(\pt)>0.2$ control region.
  The uncertainty in this correction is derived by varying the fitted points within the statistical uncertainty from
  data and simulation combined and refitting the correction.
\item[QCD multijet background template.] As described in Section \ref{sec:qcd},
the QCD multijet prediction is extrapolated from the data in  a nonoverlapping CR.
  The uncertainty in the QCD multijet background template shape is derived by taking the envelope of the
  shape obtained when varying the lepton isolation requirement used to define the multijet-enriched CR.
  A 50\% uncertainty in the QCD multijet background normalization is also included.
\end{description}

\subsection{Theoretical uncertainties}
\label{subsec:thunc}

The following theoretical uncertainties are considered in the analysis.

{\tolerance=99999
\begin{description}
\item[PDF.] The PDF uncertainties are evaluated
by comparing the nominal distributions to those obtained when using the alternative PDFs of the NNPDF set,
including $\alpha_\mathrm{s}$ variations.
\item[Factorization and renormalization scales.] To account for theoretical uncertainties, signal and background shape variations are built by changing the values of $\mu_\mathrm{F}$ and $\mu_\mathrm{R}$ from their defaults by factors of 2 or 1/2  in the ME calculation, simultaneously for $\mu_\mathrm{F}$ and $\mu_\mathrm{R}$, but independently for each simulated sample.
\item[Signal acceptance.] A 5\% uncertainty on the signal yield is assigned to account for differences between the prediction for the LO signal with respect to the NLO
predictions of the \textsc{vbfnlo} generator (v2.6.3).
\item[Normalization of top quark and diboson backgrounds.] Diboson and top quark
production  processes are modeled with MC simulations.
An uncertainty in the normalization of these backgrounds
is assigned based on the PDF and $\mu_\mathrm{F}$, $\mu_\mathrm{R}$
uncertainties, following
calculations in Refs.~\cite{Campbell:2010ff,Czakon:2013goa,Kidonakis:2012db}.
\item[Interference between {{\ewkwjj}} and {{\dywjj}}.] An overall normalization
 and a shape uncertainty
are  assigned to the interference term in the fit, based on an envelope of predictions with
different $\mu_\mathrm{F}$, $\mu_\mathrm{R}$ scales.
\item[Parton showering model.] The uncertainty in the PS
model and the event tune is assessed as the full difference
of the acceptance and shape predictions using \PYTHIA and \HERWIGpp.
\item[$R(\pt)$ correction.] As described in Section \ref{sec:mjj}, the $R(\pt)$ prediction from \PW+jets
  simulation is corrected to match the distribution in data with all expected contributions other than \PW+jets subtracted.
  The uncertainty in this correction is derived by varying the fitted points within the statistical uncertainty from
  data and simulation combined and refitting the correction.
\end{description}
\par}

\section{Measurement of the \texorpdfstring{\ewkwjj}{(EW) Wjj} production cross section}
\label{sec:results}

\begin{figure}[htb!]
  \centering
    \includegraphics[width=0.48\textwidth]{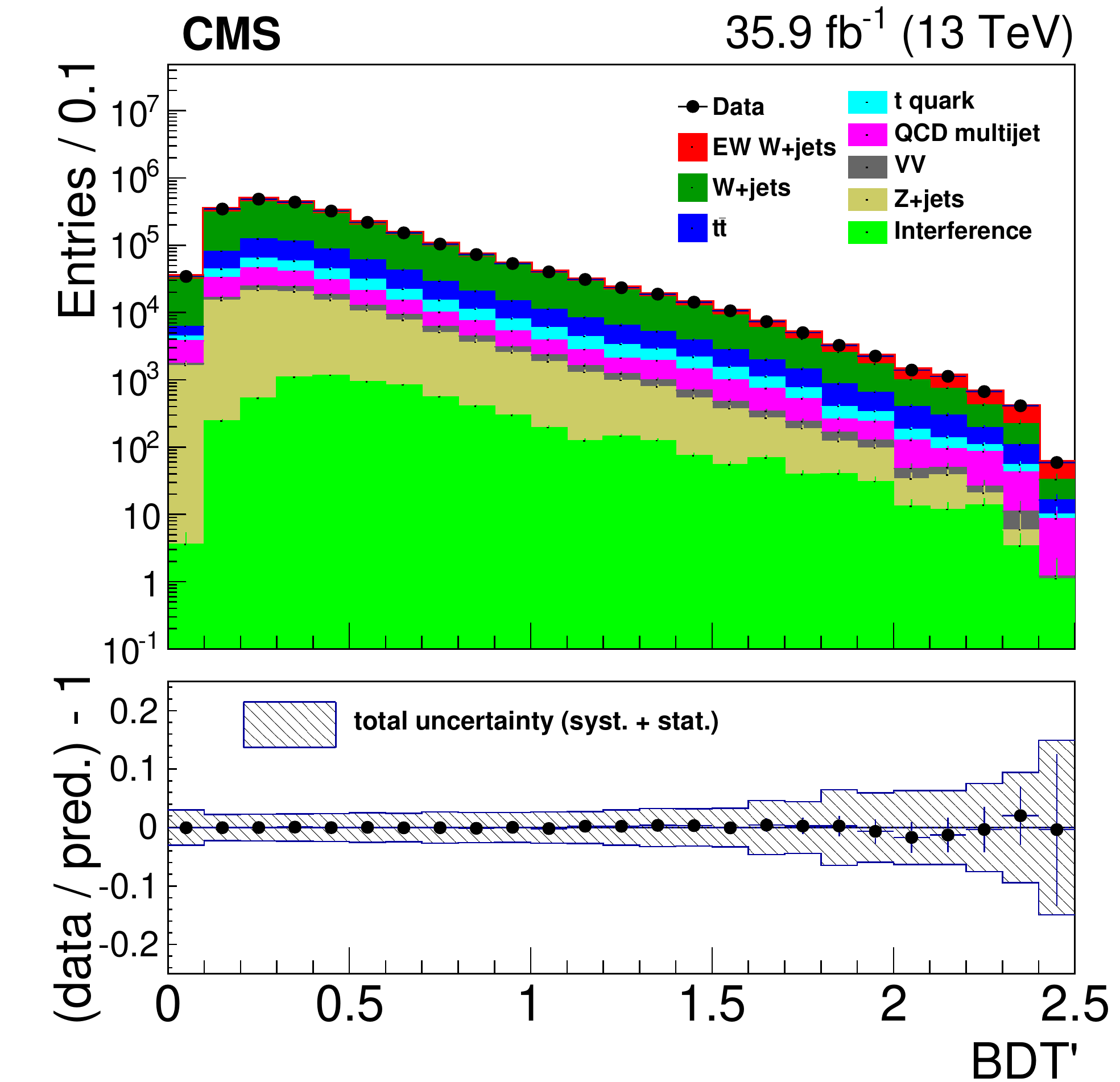}
    \includegraphics[width=0.48\textwidth]{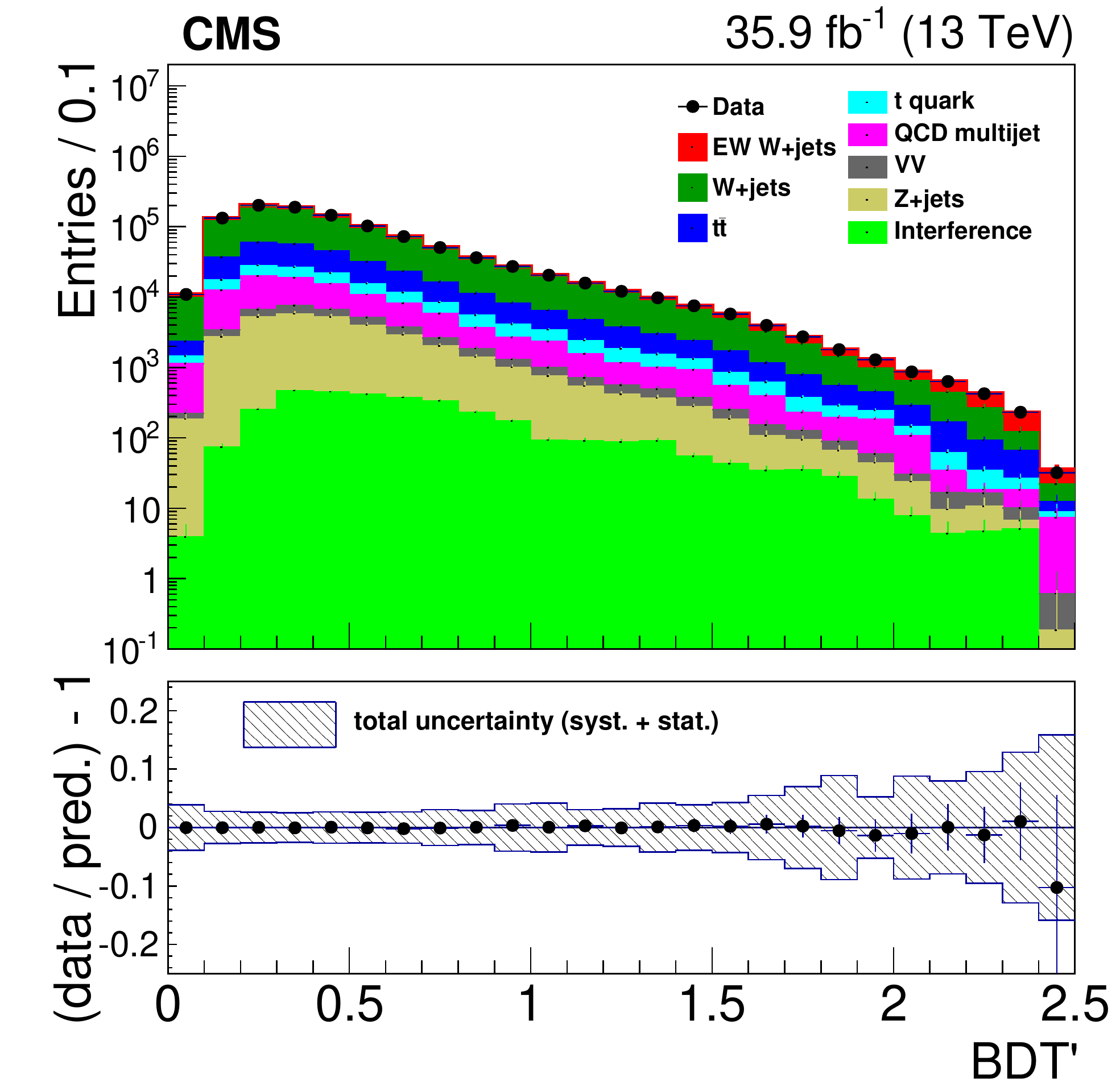}
    \caption{Data compared with simulation for the BDT' output distribution for the muon (\cmsLeft) and electron (\cmsRight) channels,
    after the fit. The grey uncertainty band in the ratio panel includes all systematic uncertainties.
      }
     \label{fig:bdt_postfit}

\end{figure}

The signal strength, defined
with the $\ell\nu$jj final state
in the kinematic region described in Section~\ref{sec:simulation},
is extracted from the fit to the BDT output distribution
as discussed in Section~\ref{sec:sigdisc}. Figure \ref{fig:bdt_postfit}
shows the BDT distribution in the muon and electron
channels for data and simulation after the fit, where the grey uncertainty
band includes all systematic uncertainties. Good agreement is observed between the
data and simulation within the uncertainties.

In the muon channel, the signal strength is measured to be
\begin{equation*}
\mu=0.91 \pm 0.02\stat \pm0.12\syst=0.91\pm 0.12\,\text{(total)},
\end{equation*}
corresponding to a measured signal cross section
\ifthenelse{\boolean{cms@external}}{
\begin{equation*}
\begin{split}
\sigma({\mathrm{EW}~\ell\nu\mathrm{jj}})&=
6.22 \pm 0.12\stat \pm 0.74\syst\unit{pb}\\
&=6.22\pm 0.75 \,\text{(total)}\unit{pb}.
\end{split}\end{equation*}
}{
\begin{equation*}
\sigma({\mathrm{EW}~\ell\nu\mathrm{jj}})=
6.22 \pm 0.12\stat \pm 0.74\syst\unit{pb}=6.22\pm 0.75 \,\text{(total)}\unit{pb}.
\end{equation*}
}

In the electron channel, the signal strength is measured to be
\begin{equation*}
\mu=0.92\pm 0.03\stat \pm0.13\syst=0.92\pm 0.13 \,\text{(total)},
\end{equation*}
corresponding to a measured signal cross section
\ifthenelse{\boolean{cms@external}}{
\begin{equation*}
\begin{split}
\sigma({\mathrm{EW}~\ell\nu\mathrm{jj}})&=
6.27 \pm 0.19\stat \pm 0.80\syst\unit{pb}\\
&=6.27\pm 0.82\,\text{(total)}\unit{pb}.
\end{split}\end{equation*}
}{
\begin{equation*}
\sigma({\mathrm{EW}~\ell\nu\mathrm{jj}})=
6.27 \pm 0.19\stat \pm 0.80\syst\unit{pb}=6.27\pm 0.82\,\text{(total)}\unit{pb}.
\end{equation*}
}

The results obtained for the different lepton channels are compatible with each other, and in
agreement with the SM predictions.

From the combined fit of the two channels, the signal strength is measured to be
\begin{equation*}
\mu=0.91\pm 0.02\stat \pm0.10\syst=0.91\pm 0.10\,\text{(total)},
\end{equation*}
corresponding to a measured signal cross section
\ifthenelse{\boolean{cms@external}}{
\begin{equation*}
\begin{split}
\sigma({\mathrm{EW}~\ell\nu\mathrm{jj}})&=
6.23 \pm 0.12\stat \pm 0.61\syst\unit{pb}\\&=6.23\pm 0.62\,\text{(total)}\unit{pb},
\end{split}\end{equation*}
}{
\begin{equation*}
\sigma({\mathrm{EW}~\ell\nu\mathrm{jj}})=
6.23 \pm 0.12\stat \pm 0.61\syst\unit{pb}=6.23\pm 0.62\,\text{(total)}\unit{pb},
\end{equation*}
}

in agreement with the \MGvATNLO LO prediction
$\sigma_\mathrm{LO}(\mathrm{EW}\,\ell\nu\mathrm{jj})=6.81^{+0.03}_{-0.06}\,\text{(scale)}\pm 0.26\,\text{(PDF)} \unit{pb}$.
In the combined fit, the DY strength is $\nu=0.88\pm 0.07$.
Using the statistical methodology described in Section~\ref{sec:sigdisc},
the background-only hypotheses in the electron, muon, and combined
channels are all excluded with significance above five standard deviations.
Table~\ref{tab:unc_breakdown} lists the major sources of uncertainty and
their impact on the measured precision of $\mu$.
The largest sources of experimental uncertainty are the $m_\jj $ correction, the JES, and the limited number of simulated events,
while the largest sources of theoretical uncertainty are the
$\mu_\mathrm{F}$, $\mu_\mathrm{R}$ scale uncertainties and the uncertainty in the signal acceptance, derived by comparing
the LO signal prediction with the prediction from the \textsc{vbfnlo} generator.

\begin{table}[htb]
\centering
\caption{Major sources of uncertainty in the measurement of the signal strength $\mu$, and their
impact. The total uncertainty is separated into four components: statistical, number of simulated events,
experimental, and theory. The experimental and theory components are further decomposed into their primary
individual uncertainty sources. }
\label{tab:unc_breakdown}
\begin{tabular}{ l l l }
  \noalign{\vskip 2mm}
 \hline \hline
Uncertainty source & \multicolumn{2}{c}{$\Delta\mu$} \\
\hline
  Statistical   & $+0.02$ & $-0.02$ \\
  \noalign{\vskip 2mm}
  Size of simulated samples                                 & $+0.05$ & $-0.05$ \\
  \noalign{\vskip 2mm}
  Experimental                                              & $+0.07$ & $-0.07$ \\
  \qquad Jet energy scale and resolution                   & $+0.03$ & $-0.01$ \\
  \qquad QCD multijet estimation                           & $+0.03$ & $-0.03$ \\
  \qquad $m_{jj}$ correction                              & $+0.05$ & $-0.05$ \\
  \qquad Background normalization                          & $+0.02$ & $-0.02$ \\
  \qquad Other experimental uncertainties                & $<0.01$ & $ $ \\
  \noalign{\vskip 2mm}
  Theory                                                    & $+0.07$ & $-0.07$ \\
  \qquad QCD scale and PDF                                       & $+0.05$ & $-0.05$\\
  \qquad Interference                                         & $+0.02$ & $-0.02$\\
  \qquad Signal acceptance                                         & $+0.05$ & $-0.05$\\
  \qquad Other theory uncertainties                            & $+0.01$ & $-0.01$ \\
  \noalign{\vskip 2mm}
  Total   & $+0.10$ & $-0.10$ \\
\hline
\end{tabular}
\end{table}

The signal strength is also measured with respect to the NLO signal prediction, as described
in Section~\ref{sec:simulation}. In the muon channel, the signal strength is measured to be

\begin{equation*}
\mu_{\mathrm{NLO}}=0.91 \pm 0.02\stat \pm0.12\syst=0.91\pm 0.12\,\text{(total)}.
\end{equation*}

In the electron channel, the signal strength is measured to be

\begin{equation*}
\mu_{\mathrm{NLO}}=0.89 \pm 0.03\stat \pm0.12\syst=0.89\pm 0.12\,\text{(total)}.
\end{equation*}

From the combined fit of the two channels, the signal strength is measured to be

\begin{equation*}
\mu_{\mathrm{NLO}}=0.90 \pm 0.02\stat \pm0.10\syst=0.90\pm 0.10\,\text{(total)},
\end{equation*}

corresponding to a measured signal cross section
\ifthenelse{\boolean{cms@external}}{
\begin{equation*}
\begin{split}
\sigma({\mathrm{EW}~\ell\nu\mathrm{jj}})&=
6.07 \pm 0.12\stat \pm 0.57\syst\unit{pb}\\&=6.07\pm 0.58\,\text{(total)}\unit{pb},
\end{split}\end{equation*}
}{
\begin{equation*}
\sigma({\mathrm{EW}~\ell\nu\mathrm{jj}})=
6.07 \pm 0.12\stat \pm 0.57\syst\unit{pb}=6.07\pm 0.58\,\text{(total)}\unit{pb},
\end{equation*}
}

in agreement with the \POWHEG NLO prediction
$\sigma_\mathrm{NLO}(\mathrm{EW}\,\ell\nu\mathrm{jj})=6.74^{+0.02}_{-0.04}\,\text{(scale)}\pm 0.26\,\text{(PDF)} \unit{pb}$.

\section{Limits on anomalous gauge couplings}
\label{sec:atgc}

It is useful to look for signs of new physics via a model-independent EFT framework.
In the framework of EFT, new physics can be
described as an infinite series of new interaction terms organized as an
expansion in the mass dimension of the operators.

In the EW sector of the SM, the first higher-dimensional
operators containing bosons are six-dimensional~\cite{Degrande:2012wf}:
\begin{equation}\label{atgc:eq1}
\begin{split}
\mathcal{O}_{WWW} & = \frac{c_{WWW}}{\Lambda^2}W_{\mu\nu}W^{\nu\rho}W_{\rho}^{\mu},\\
\mathcal{O}_{W} & = \frac{c_{W}}{\Lambda^2}(D^{\mu}\Phi)^{\dagger}W_{\mu\nu}(D^{\nu}\Phi),\\
\mathcal{O}_{B} & = \frac{c_{B}}{\Lambda^2}(D^{\mu}\Phi)^{\dagger}B_{\mu\nu}(D^{\nu}\Phi),\\
\widetilde{\mathcal{O}}_{WWW} & = \frac{\widetilde{c}_{WWW}}{\Lambda^2}\widetilde{W}_{\mu\nu}W^{\nu\rho}W_{\rho}^{\mu},\\
\widetilde{\mathcal{O}}_{W} & = \frac{\widetilde{c}_{W}}{\Lambda^2}(D^{\mu}\Phi)^{\dagger}\widetilde{W}_{\mu\nu}(D^{\nu}\Phi),
\end{split}
\end{equation}
where, as is customary, group indices are suppressed and the mass scale
$\Lambda$ is factorized
from the coupling constants $c$. In Eq.~\eqref{atgc:eq1}, $W_{\mu\nu}$
is the SU(2) field strength, $B_{\mu\nu}$ is the U(1) field strength, $\Phi$
is the Higgs doublet, and operators with a tilde are the magnetic duals of the
field strengths. The first three operators are charge and parity conserving, whereas the last
two are not.
Models with operators that preserve charge conjugation and parity symmetries
can be included in the calculation either individually or in pairs.
With these assumptions,
the values of coupling constants divided by the mass scale $c/\Lambda^2$ are measured.

These operators have a rich phenomenology since they contribute to many
multiboson scattering processes at tree level.
The operator $\mathcal{O}_{WWW}$ modifies
vertices with three or six vector bosons, whereas the operators $\mathcal{O}_{W}$ and
$\mathcal{O}_{B}$ modify both the HVV vertices and vertices with three or four vector
bosons. A more detailed description of the phenomenology of these operators
can be found in Ref.~\cite{Degrande:2013yda}.
Modifications to the ZWW and $\gamma$WW vertices are investigated in this analysis, since these modify
the $ \Pp\Pp \to \PW \mathrm{jj} $ cross section.

{\tolerance=1200
Previously, modifications to these vertices have been studied using
anomalous trilinear gauge couplings~\cite{Nakamura:2010zzi}.
The relationship between the dimension-six operators in Eq.~\eqref{atgc:eq1}
and ATGCs can be found in Ref.~\cite{Degrande:2012wf}.
Most stringent limits on ATGC parameters were previously set by LEP~\cite{Schael:2013ita},
CDF~\cite{Aaltonen:2012vu}, D0~\cite{Abazov:2012ze}, ATLAS~\cite{ATLAS:WV8TeV,Aad:2016ett}, and CMS~\cite{CMS:WV8TeV,Khachatryan:2016poo}.
\par}

\subsection{Statistical analysis}\label{atgc:sec5}

The measurement of the coupling constants uses templates in the \pt of the lepton from
the $W\to\ell\nu$ decay. Because this is well measured and longitudinally
Lorentz invariant, this variable is robust against mismodeling and ideal for this purpose.
An additional requirement of $\mathrm{BDT} >0.5$ has been applied, which is optimized based on the expected sensitivity
to the ATGC signal. The expected limits are subsequently improved by 20--25\% with respect to the expected
limits without a BDT selection.
In each channel, four bins from $ 0 < \pt^\ell < 1.2 \TeV $ are used,
where the last bin contains overflow and its lower bin edge boundary has been optimized separately for each channel.

{\tolerance=9600
For each signal MC event, 125 weights are assigned that correspond to a $5{\times} 5{\times} 5$ grid in $(c_{WWW}/\Lambda^2) \, (c_{W}/\Lambda^2) \, (c_{B}/\Lambda^2)$.
Equal bins are used in the interval $[-15, 15]\TeV^{-2}$ for $c_{WWW}/\Lambda^2$,  $[-40, 40]\TeV^{-2}$ for $c_{W}/\Lambda^2$,
and equal bins in the interval $[-175, 175]\TeV^{-2}$ for $c_{B}/\Lambda^2$.
\par}

\begin{figure*}[htbp]
\centering
\includegraphics[width=0.48\textwidth]{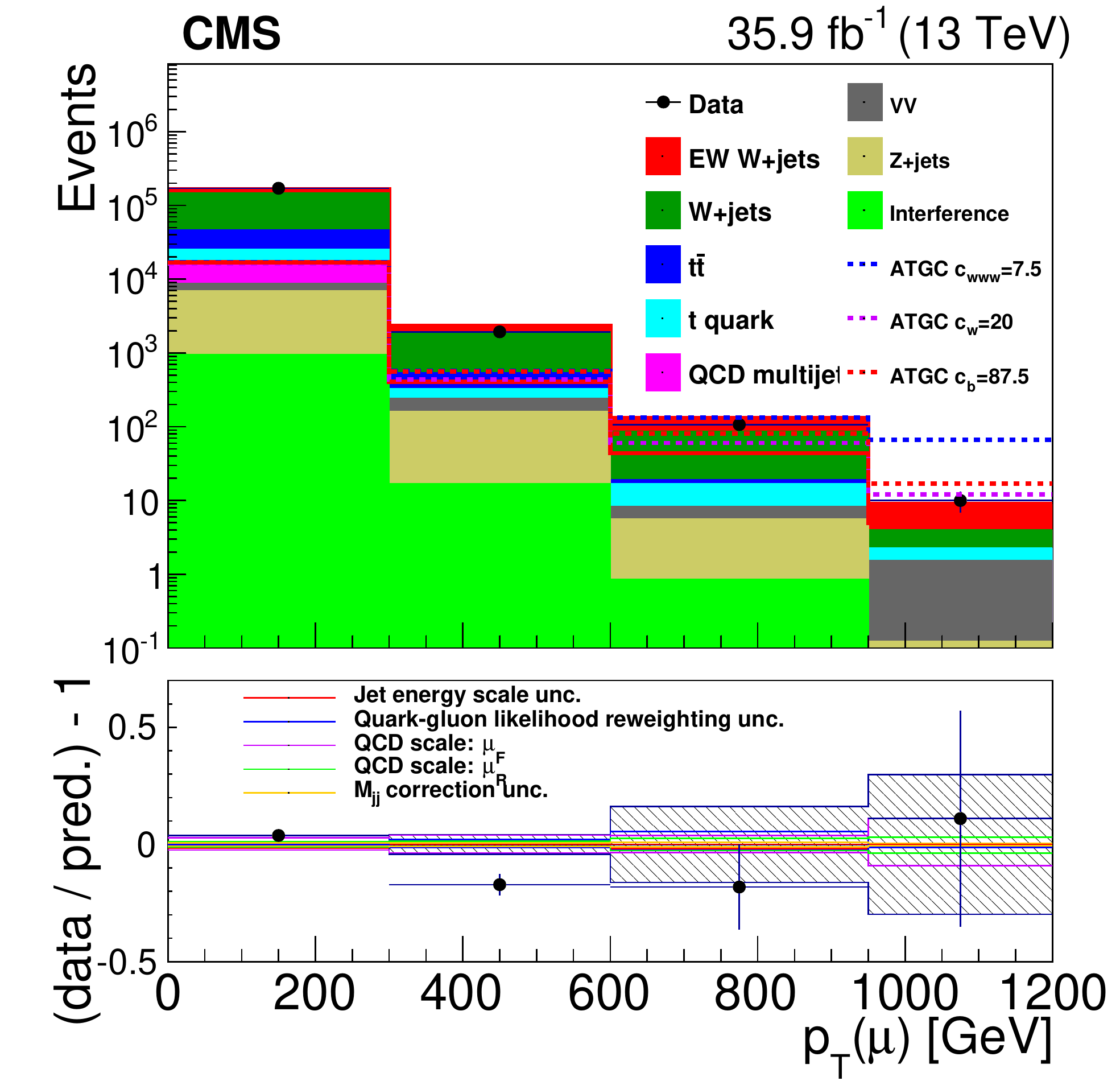}
\includegraphics[width=0.48\textwidth]{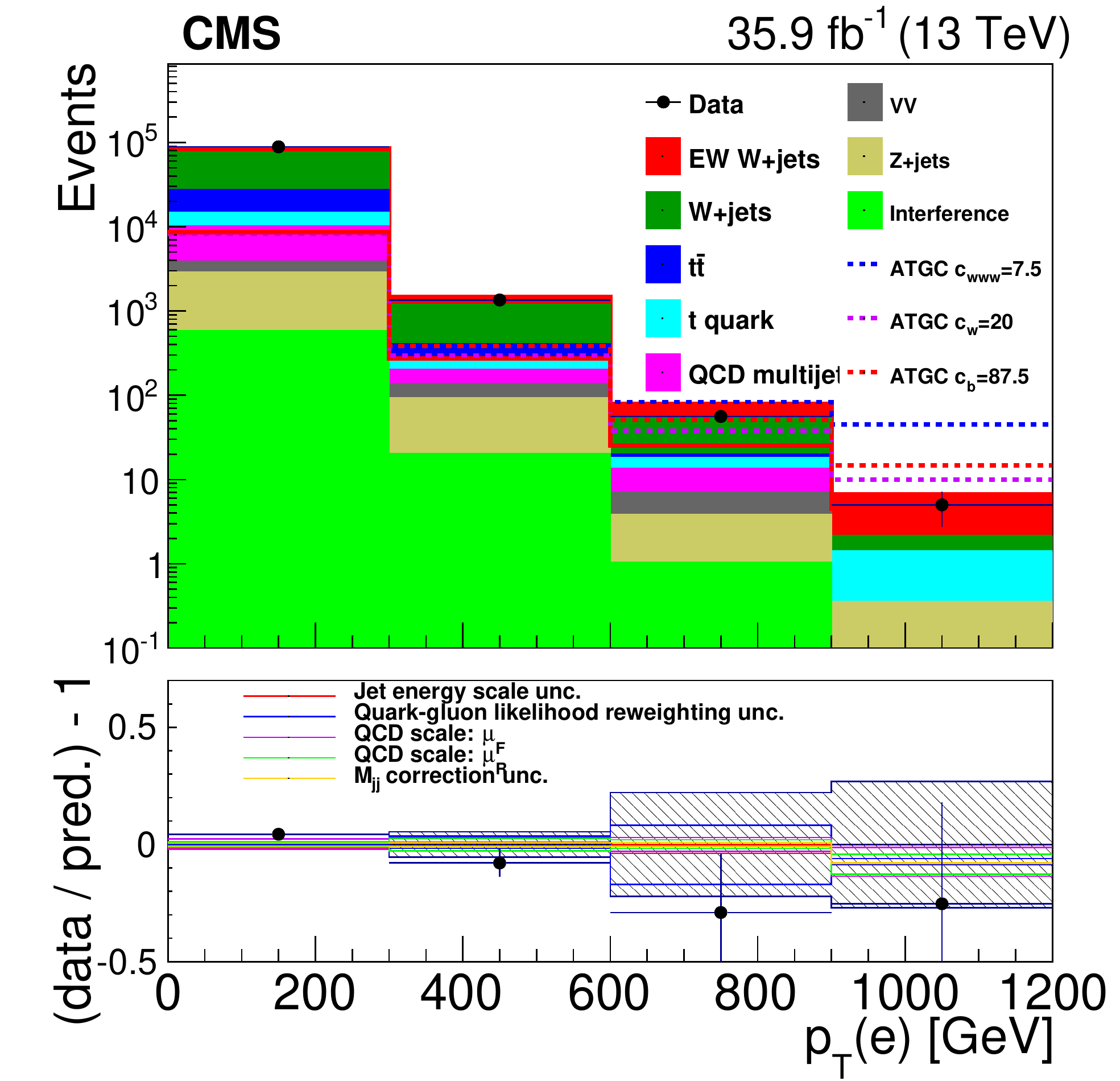}
\caption{\label{atgc:fig5} Distributions of $\pt^\ell$ in data and SM backgrounds, and various
ATGC scenarios in the muon (left) and electron (right) channels, before the fit. For each ATGC scenario plotted a particular
parameter is varied while the other ATGC parameters are fixed to zero. The lower panels show the ratio between
data and prediction minus one with the statistical uncertainty from simulation (grey hatched band) as well as
the leading systematic uncertainties in the shape of the $\pt^\ell$ distribution.}
\end{figure*}

{\tolerance=4800
To construct the $\pt^\ell$ templates, the associated weights calculated
for each event are used to construct a parametrized model
of the expected yield in each bin as a function of the values of the dimension-six operators' coupling constants.
For each bin,
the ratios of the expected signal yield with dimension-six operators to the
one without (leaving only the SM contribution) are fitted at each point of the grid to a quadratic polynomial.
The highest $\pt^\ell$  bin has the largest statistical
power to detect the presence of higher-dimensional operators.
Figure~\ref{atgc:fig5} shows examples of the final templates, with the expected signal overlaid on the background expectation,
for three different hypotheses of dimension-six operators. The SM distribution is normalized to the expected cross section.
\par}

A simultaneous binned fit for the values of the ATGCs is performed in the two lepton channels.
A profile likelihood method, the Wald Gaussian approximation, and Wilks' theorem~\cite{Khachatryan:2014jba}
are used to derive confidence intervals at 95\% confidence level (CL).
One-dimensional and two-dimensional limits are derived
on each of the three ATGC parameters and each combination of two ATGC parameters
while all other parameters are set to their SM values.
Systematic and theoretical uncertainties are represented by the individual nuisance
parameters with log-normal distributions and are profiled in the fit.

\subsection{Results}\label{atgc:sec6}

{\tolerance=600
No significant deviation from the SM expectation is observed.
Limits on the EFT parameters are reported and also translated into the equivalent parameters defined
in an effective Lagrangian (LEP parametrization) in Ref.~\cite{hagiwara1}, without form factors:
$\lambda^{\gamma} = \lambda^{\PZ} = \lambda$, $\Delta{\kappa^{\PZ}} = \Delta{g_1^{\PZ}}-\Delta{\kappa^\gamma} \, \tan^2\theta_{{\PW}}$.
The parameters $\lambda$, $\Delta{\kappa^{Z}}$, and $\Delta{g_1^{\PZ}}$ are considered,
where the $\Delta$ symbols represent deviations from their respective SM values.
\par}

\begin{figure}
\centering
\includegraphics[width=0.42\textwidth]{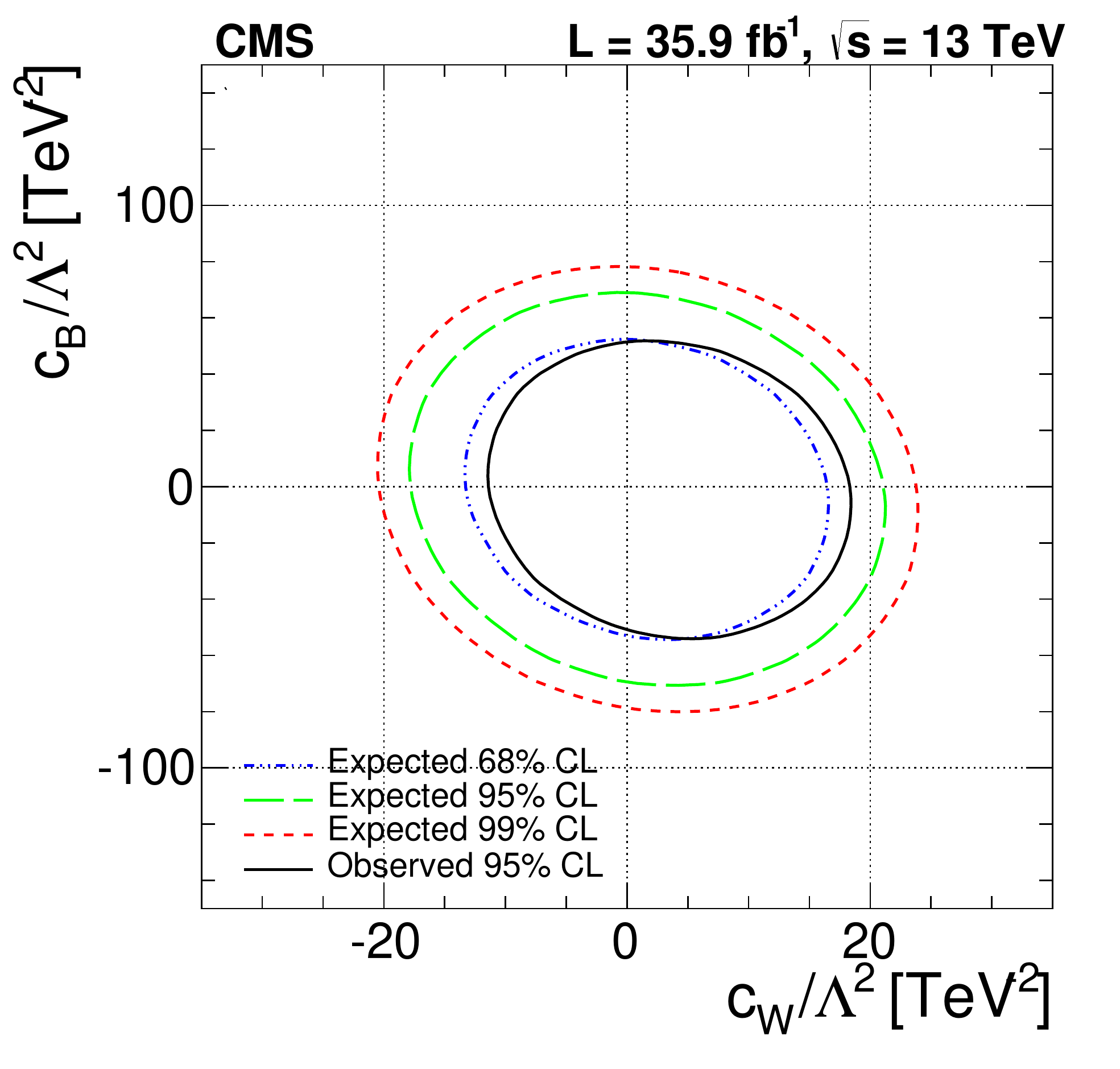}
\includegraphics[width=0.42\textwidth]{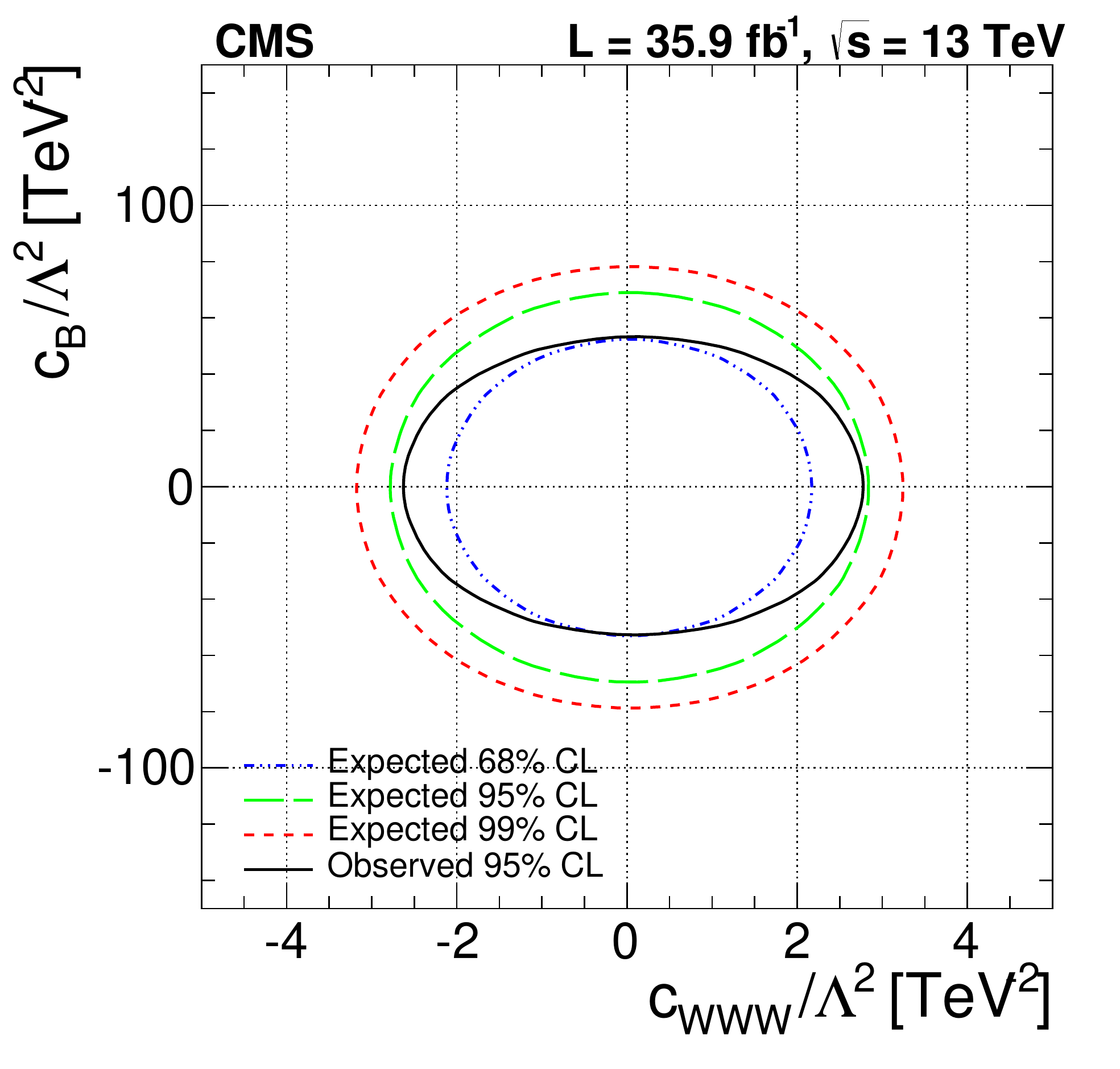}
\includegraphics[width=0.42\textwidth]{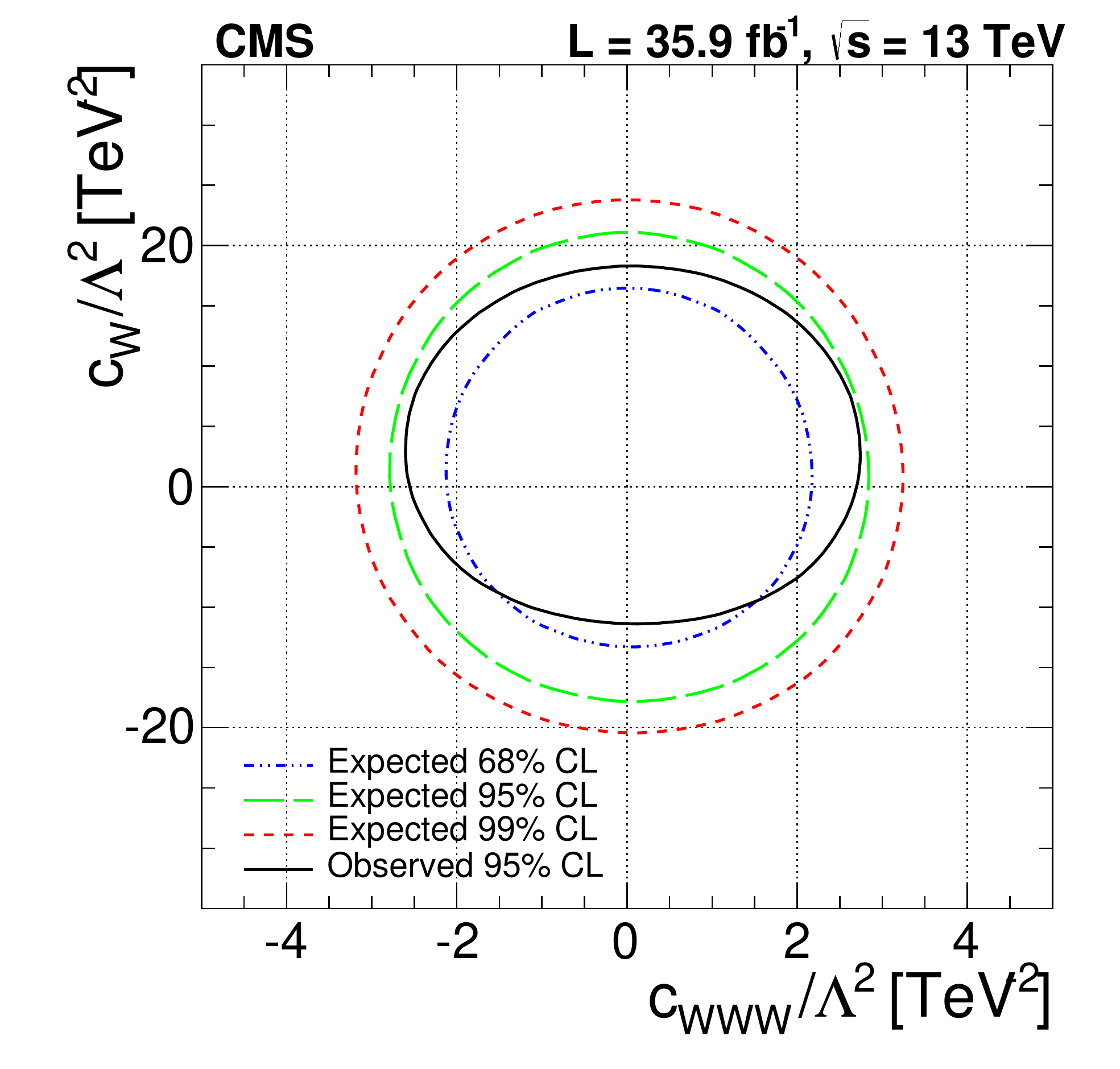}
\caption{\label{atgc:2dlimits_eft}
Expected and observed two-dimensional limits on the EFT parameters at 95\% CL.
}
\end{figure}

\begin{figure}
\centering
\includegraphics[width=0.42\textwidth]{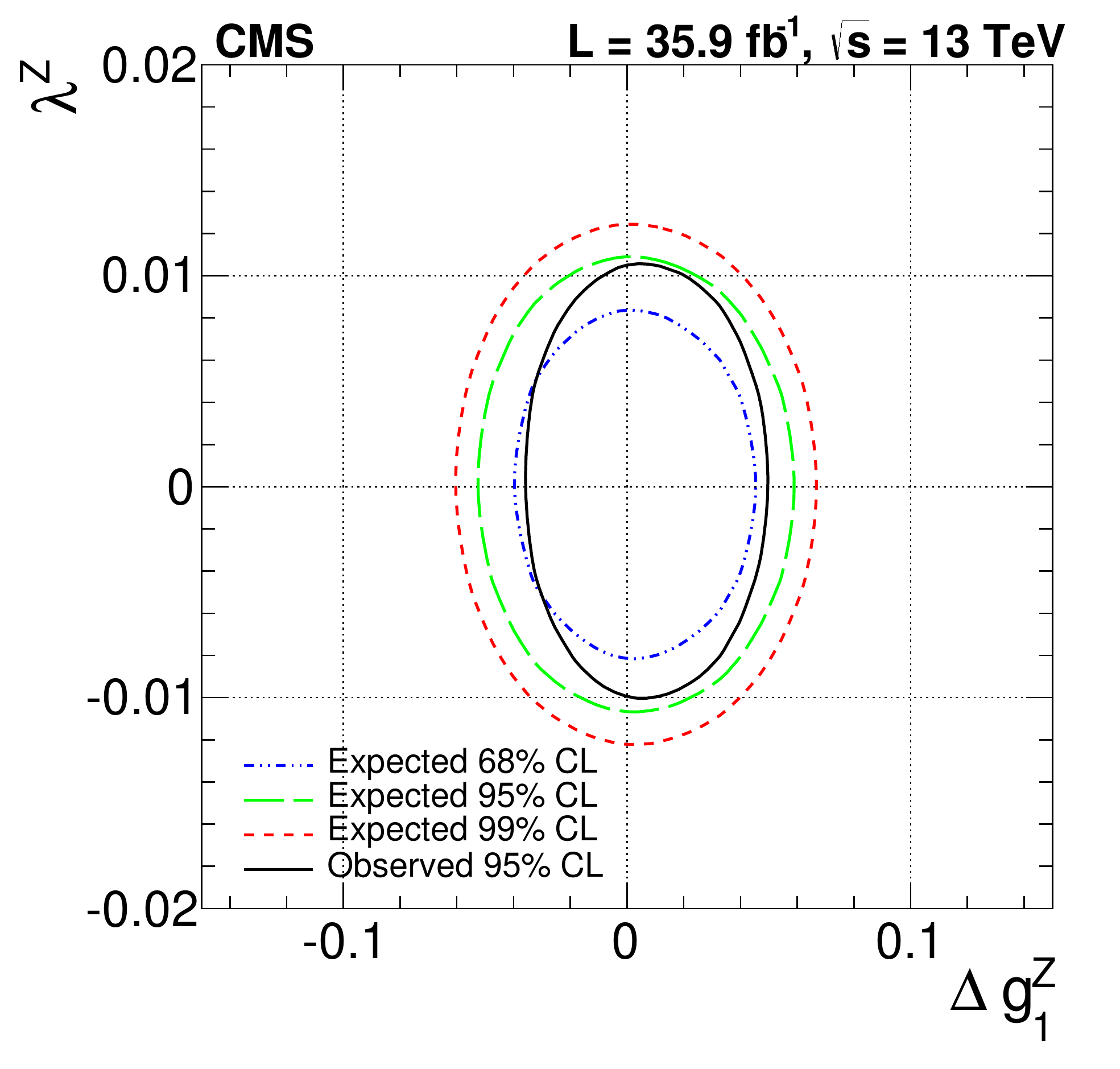}
\includegraphics[width=0.42\textwidth]{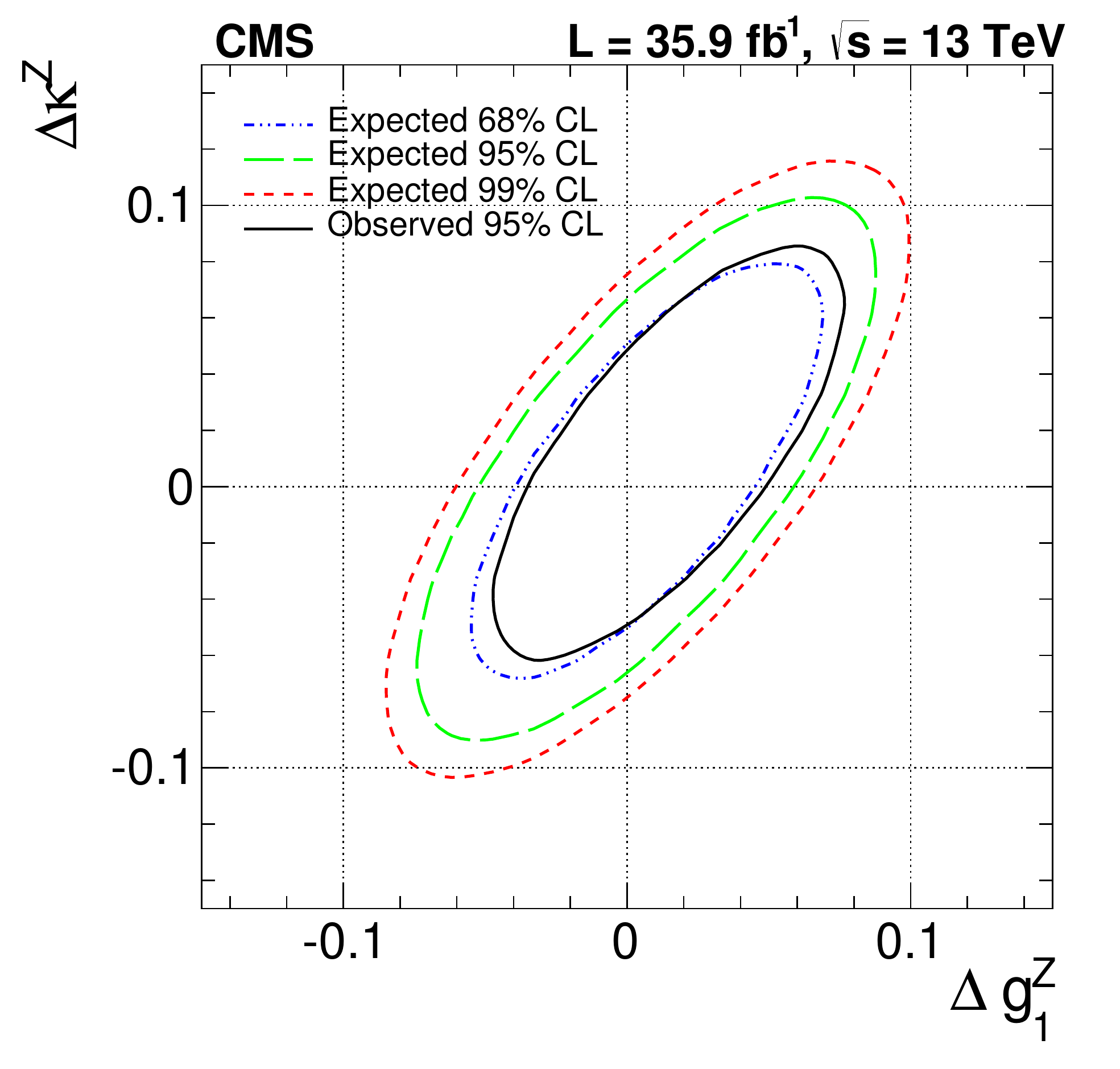}
\includegraphics[width=0.42\textwidth]{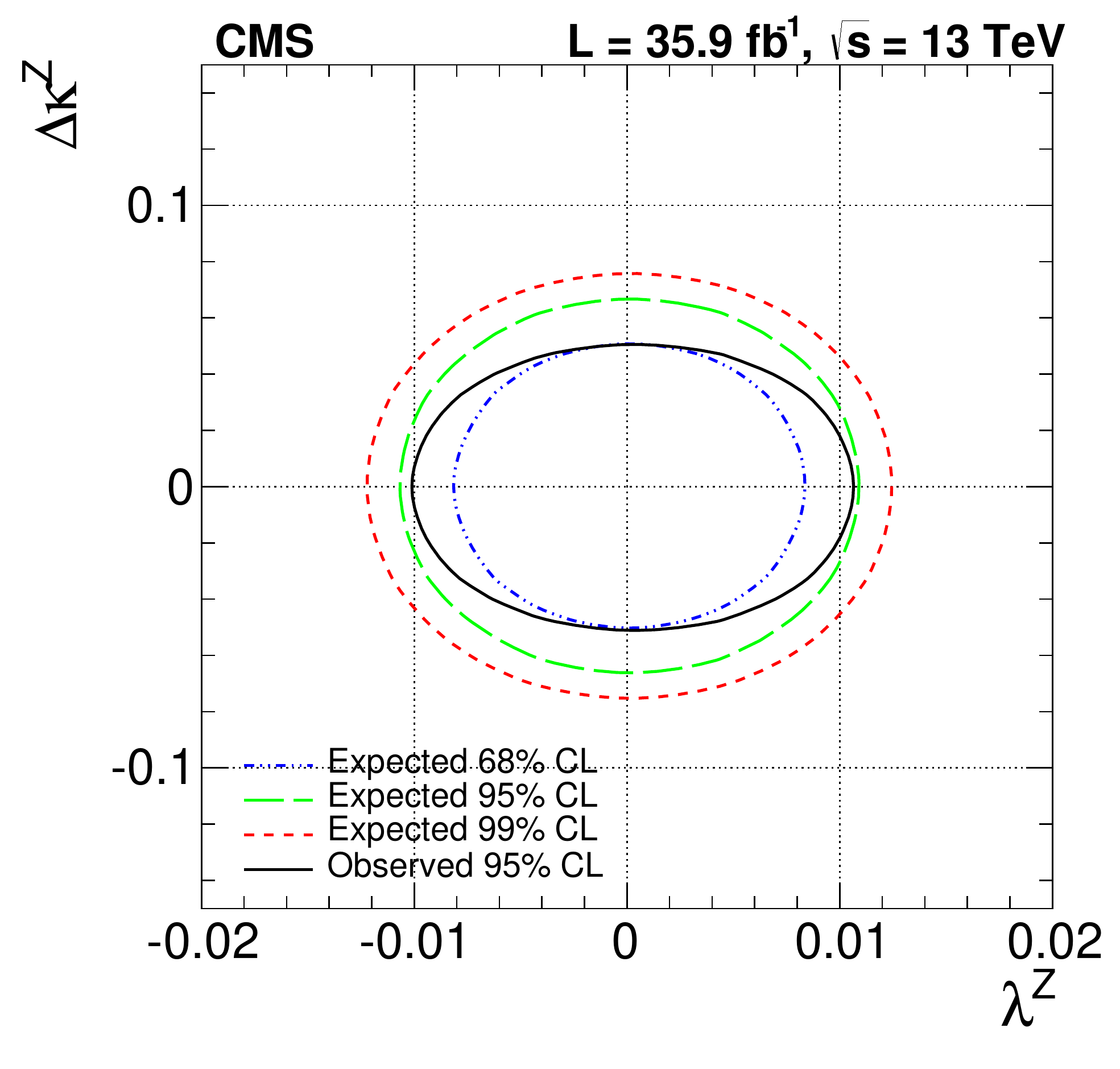}
\caption{\label{atgc:2dlimits_eft_lep}
Expected and observed two-dimensional limits on the ATGC effective Lagrangian (LEP parametrization) parameters at 95\% CL.
}
\end{figure}

Results for the one-dimensional limits are listed in Table~\ref{atgc:tbl3} for  $c_{WWW}$,  $c_W$ and $c_B$,
and in Table~\ref{atgc:limits_atgc} for $\lambda$, $\Delta g_{1}^{\PZ}$ and $\Delta \kappa_{1}^{\PZ}$;
two-dimensions limits are shown in Figs.~\ref{atgc:2dlimits_eft} and ~\ref{atgc:2dlimits_eft_lep}.
The results are dominated by the sensitivity in the muon channel
due to the larger acceptance for muons.
An ATGC signal is not included in the interference between EW and DY production.
The effect on the limits is small (${<}$3\%).
The LHC semileptonic WZ analysis using 13\TeV data currently sets the most stringent limits on
$c_{WWW}/\Lambda^2$ and $c_W/\Lambda^2$, while the WW analysis using 8\TeV data currently sets the tightest
limits on $c_B/\Lambda^2$.
This analysis is most sensitive to $c_{WWW}/\Lambda^2$, where the limit is slightly less restrictive but
comparable.

\begin{table*}[htbp]
\centering
\topcaption{\label{atgc:tbl3}
One-dimensional limits on the ATGC EFT parameters at 95\% CL.
}
\renewcommand{\arraystretch}{1.2}
\begin{tabular}{ccc}
Coupling constant & Expected 95\% CL interval ($\TeVns{}^{-2}$) & Observed 95\% CL interval ($\TeVns{}^{-2}$)\\ \hline
$c_{WWW}/\Lambda^{2}$ & $[-2.5, 2.5]$ & $[-2.3, 2.5]$\\
$c_{W}/\Lambda^{2}$ & $[-16, 19]$ & $[-8.8, 16]$\\
$c_{B}/\Lambda^{2}$ & $[-62, 61]$ & $[-45,46]$
\end{tabular}
\end{table*}

\begin{table*}[htbp]
\centering
\topcaption{\label{atgc:limits_atgc}
One-dimensional limits on the ATGC effective Lagrangian (LEP parametrization) parameters at 95\% CL.
}
\renewcommand{\arraystretch}{1.2}
\begin{tabular}{ccc}
Coupling constant & Expected 95\% CL interval & Observed 95\% CL interval \\ \hline
$\lambda^{Z}$ & $[-0.0094, 0.0097]$ & $[-0.0088, 0.0095]$\\
$\Delta g_{1}^{Z}$ & $[-0.046, 0.053]$ & $[-0.029, 0.044]$\\
$\Delta \kappa_{1}^{Z}$ & $[-0.059, 0.059]$ & $[-0.044, 0.044]$
\end{tabular}
\end{table*}

\subsection{Combination with the VBF Z boson production analysis}\label{atgc:sec7}
\begin{figure}
\centering
\includegraphics[width=0.42\textwidth]{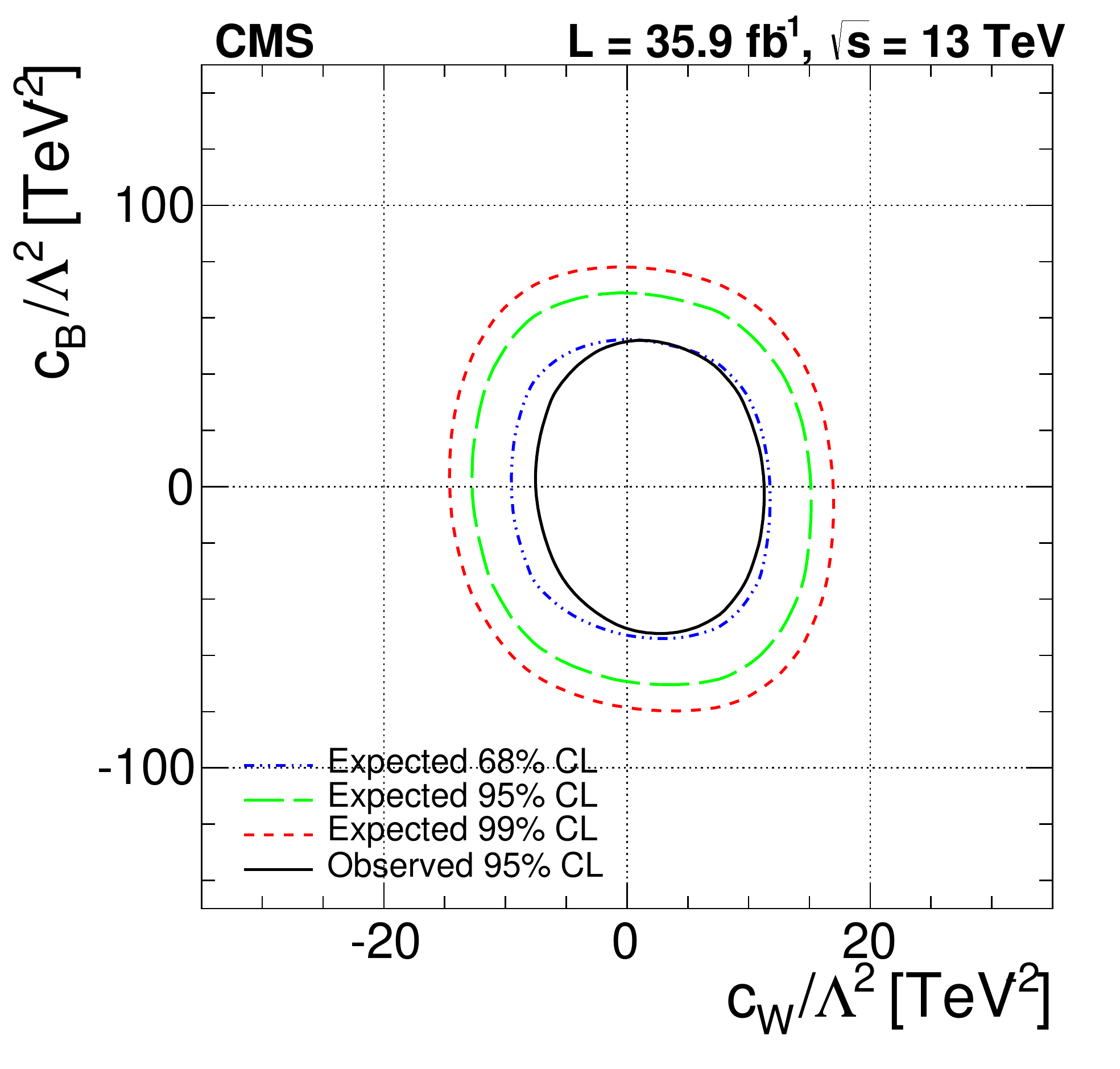}
\includegraphics[width=0.42\textwidth]{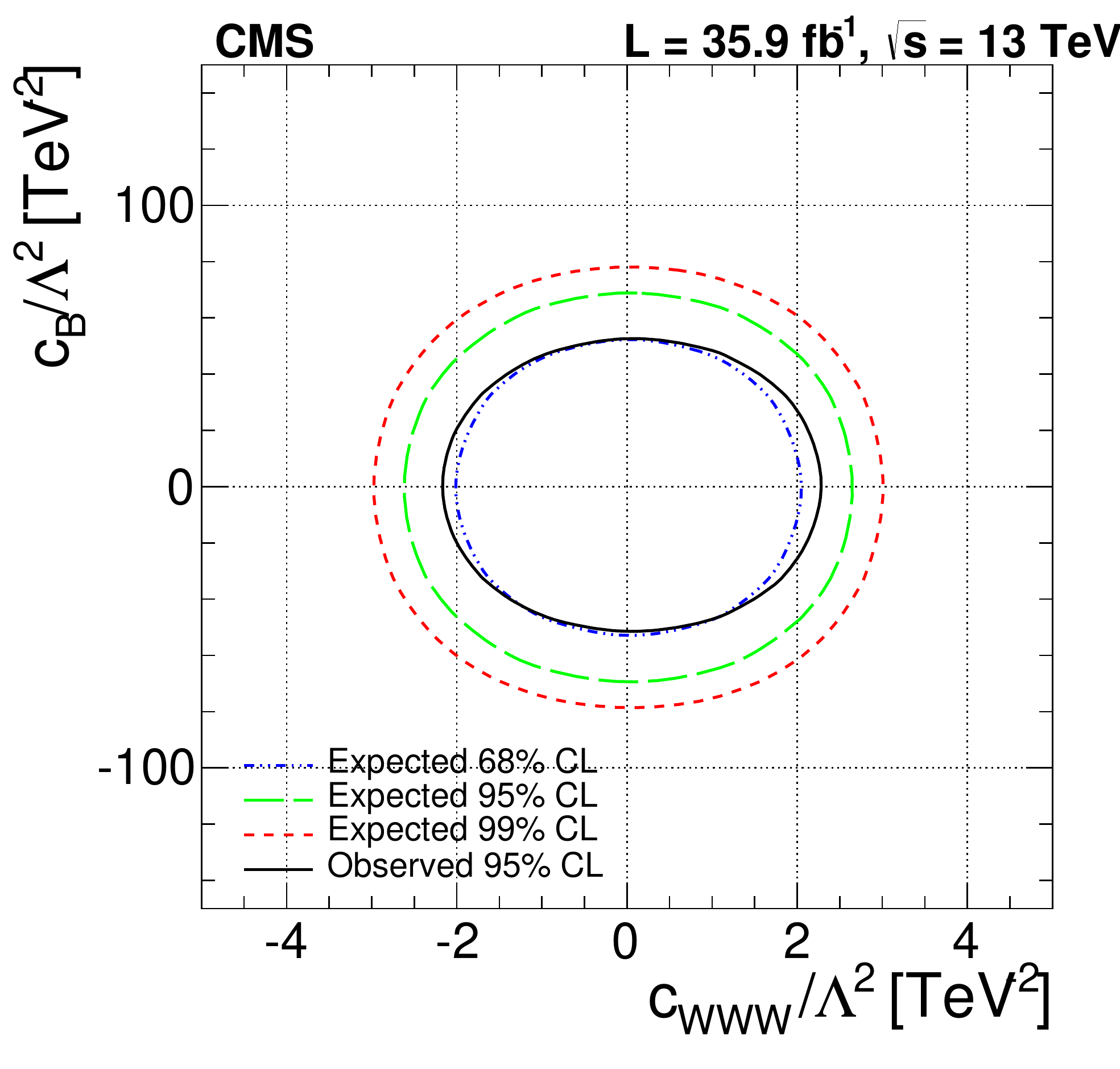}
\includegraphics[width=0.42\textwidth]{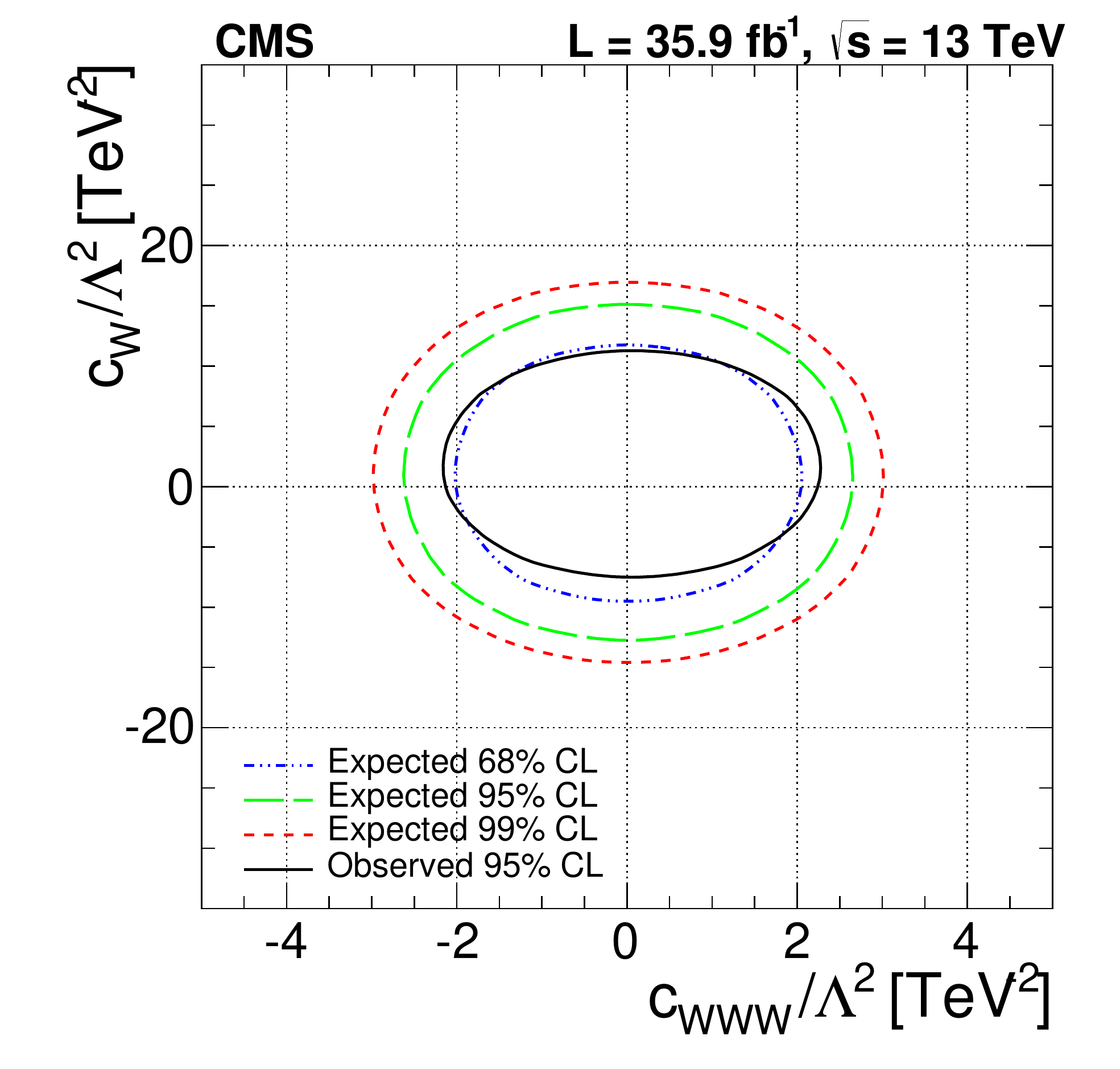}
\caption{\label{atgc:2dlimits_eft_comb}
Expected and observed two-dimensional limits on the EFT parameters at 95\% CL from the combination of EW \Wjj\ and EW \Zjj\ analyses.
}
\end{figure}

\begin{figure}
\centering
\includegraphics[width=0.42\textwidth]{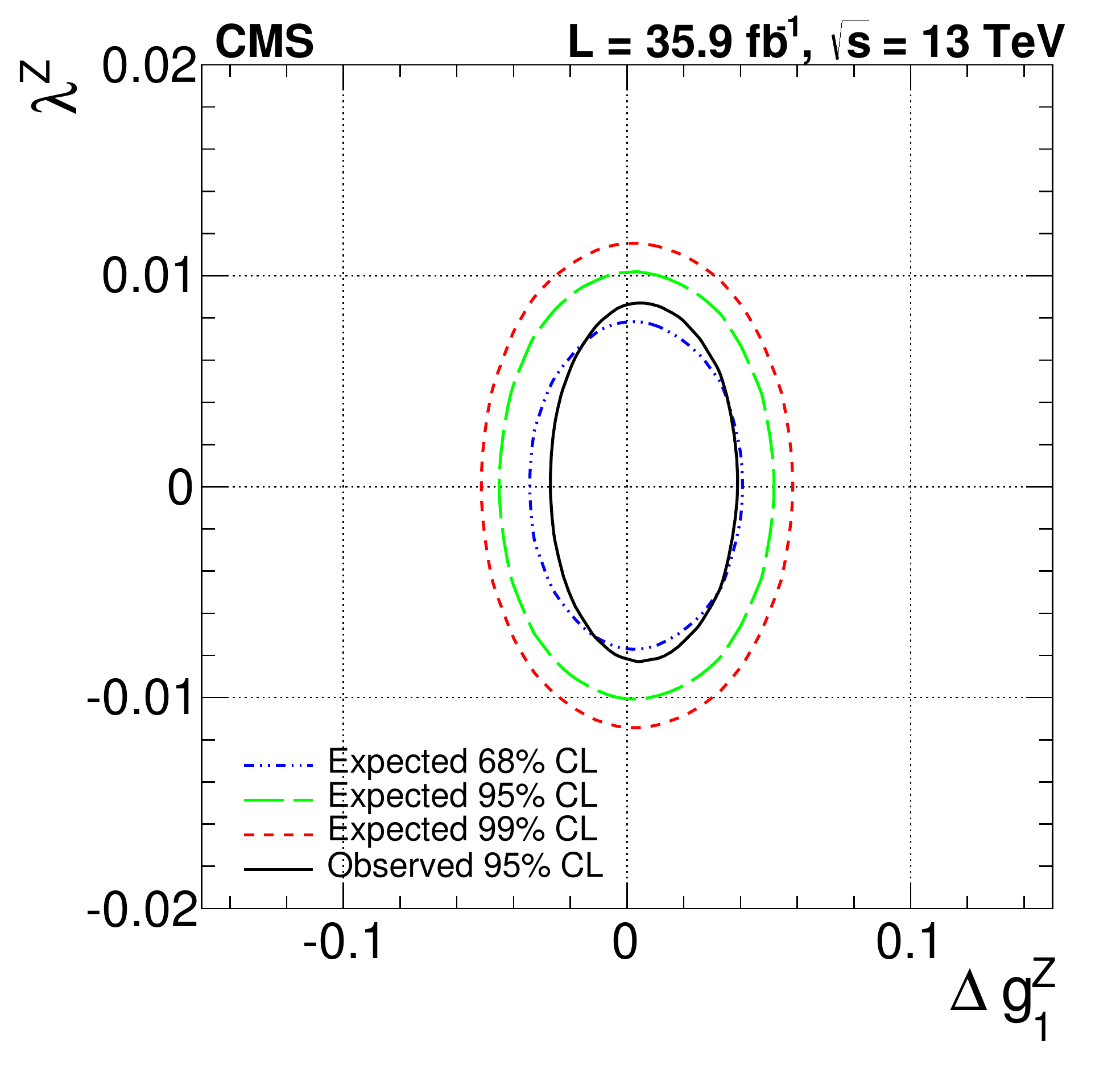}
\includegraphics[width=0.42\textwidth]{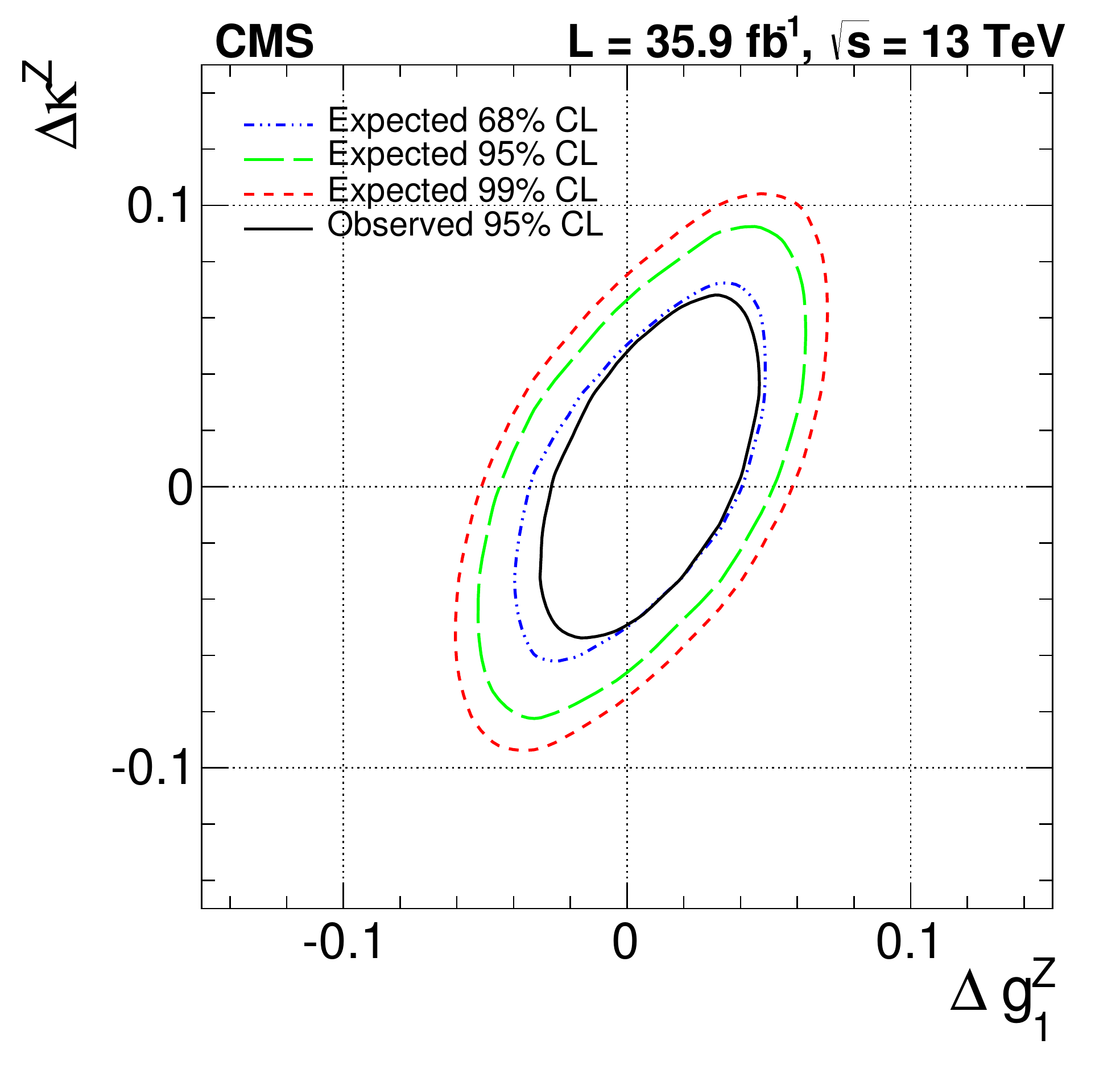}
\includegraphics[width=0.42\textwidth]{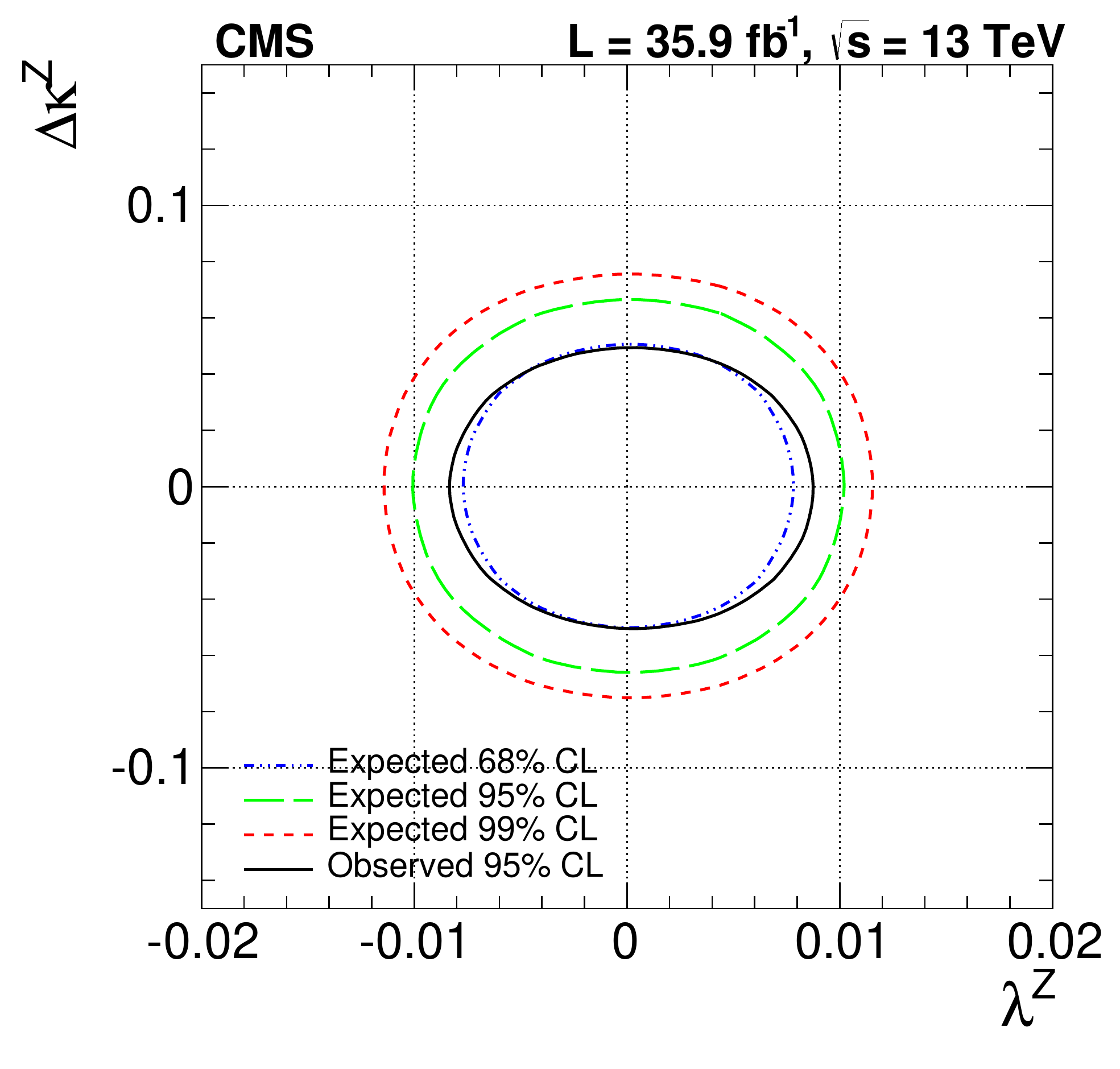}
\caption{\label{atgc:2dlimits_eft_lep_comb}
Expected and observed two-dimensional limits
on the ATGC effective Lagrangian (LEP parametrization) parameters at 95\% CL from the combination of EW \Wjj\ and EW \Zjj\ analyses.
}
\end{figure}

As mentioned in Section~\ref{sec:intro}, the closely-related EW \Zjj\ process has been measured by CMS at
$\sqrt{s}=13\TeV$~\cite{Sirunyan:2017jej}. This result included constraints on ATGC EFT parameters obtained via a fit
to the $\pt$(Z) distribution, an experimentally clean observable sensitive to deviations from zero in the ATGC parameters.
Both the EW \Zjj\ and EW \Wjj\ analyses are sensitive to anomalous couplings related to the WWZ vertex.
A simultaneous binned likelihood fit for the ATGC parameters is performed to the $\pt$(Z) distribution in the EW Zjj production and
and $\pt^\ell$   in the EW \Wjj\ production.
In the combined fit, the primary uncertainty sources are correlated including the JES and JER uncertainties.
Results for the one-dimensional limits are listed in Table~\ref{atgc:tbl3_comb} for
$c_{WWW}$,  $c_W$ and $c_B$, and in Table~\ref{atgc:table_comb} for $\lambda$, $\Delta g_{1}^{\PZ}$, and $\Delta \kappa_{1}^{\PZ}$;
two-dimensions limits are shown in Figs.~\ref{atgc:2dlimits_eft_comb} and ~\ref{atgc:2dlimits_eft_lep_comb}.

\begin{table*}[htbp]
\centering
\topcaption{\label{atgc:tbl3_comb}
One-dimensional limits on the ATGC EFT parameters at 95\% CL from the combination of EW \Wjj\ and EW \Zjj\ analyses.
}
\renewcommand{\arraystretch}{1.2}
\begin{tabular}{ccc}
Coupling constant & Expected 95\% CL interval ($\TeVns{}^{-2}$) & Observed 95\% CL interval ($\TeVns{}^{-2}$)\\ \hline
$c_{WWW}/\Lambda^{2}$ & $[-2.3, 2.4]$ & $[-1.8, 2.0]$\\
$c_{W}/\Lambda^{2}$ & $[-11, 14]$ & $[-5.8, 10.0]$\\
$c_{B}/\Lambda^{2}$ & $[-61, 61]$ & $[-43,45]$
\end{tabular}
\end{table*}

\begin{table*}[htbp]
\centering
\topcaption{\label{atgc:table_comb}
One-dimensional limits on the ATGC effective Lagrangian (LEP parametrization) parameters at 95\% CL from the combination of EW \Wjj\ and EW \Zjj\ analyses.
}
\begin{tabular}{ccc}
Coupling constant & Expected 95\% CL interval & Observed 95\% CL interval \\ \hline
$\lambda^{Z}$ & $[-0.0089, 0.0091]$ & $[-0.0071, 0.0076]$\\
$\Delta g_{1}^{Z}$ & $[-0.040, 0.047]$ & $[-0.021, 0.034]$\\
$\Delta \kappa_{1}^{Z}$ & $[-0.058, 0.059]$ & $[-0.043, 0.042]$
\end{tabular}
\end{table*}

\section{Study of the hadronic and jet activity in \texorpdfstring{\PW+jet}{W+jet} events}
\label{sec:hadactivity}

Having established the presence of the SM signal,
the properties of the hadronic
activity in the selected events can be examined,
in particular in the the region in rapidity between the two tagging jets,
with low expected hadron activity (rapidity gap).
The production of additional jets in the rapidity gap, in a region with a
larger contribution of \ewkwjj\ processes
is explored in Section~\ref{subsec:highpur}.
Studies of the rapidity gap hadronic activity
using track-only observables, are
presented in Section~\ref{subsec:soft}.
Finally, a study of hadronic activity vetoes, using both
PF jets and track-only observables, is presented in Section~\ref{subsec:gapveto}.
A significant suppression of the hadronic activity
in signal events is expected because the final-state objects originate from EW
interactions, in contrast with the radiative QCD
production of jets in \dywjj\ events.

In all these studies,
event distributions are shown with a selection on the output value at
BDT~$>0.95$, which allows a signal-enriched region to be selected
with a similar
fraction of signal and background events.
None of the BDT input observables listed in Section~\ref{sec:sigdisc}
are related to additional hadronic activity
observables, as a consequence there is no bias on the additional
 hadronic activity observables due to the BDT output cut.
The reconstructed distributions are compared directly to the
prediction obtained with a full simulation of the CMS detector.
In the BDT~$>0.95$ region, the dominant uncertainty on the prediction
from simulation is due to the limited number of generated events.

\subsection{Jet activity studies in a high-purity region}
\label{subsec:highpur}

For this study, aside from the two tagging jets used in the preselection, all PF jets with
$\pt>15\GeV$ found
within the pseudorapidity gap of the tagging jets,
$\eta^\text{tag jet}_\text{min} < \eta < \eta^\text{tag jet}_\text{max}$,
are used.
For the estimation of the background contributions, the normalizations obtained from the fit discussed in
Section~\ref{sec:results} are used.

\begin{figure*}[htb!]
  \centering
    \includegraphics[width=0.98\textwidth]{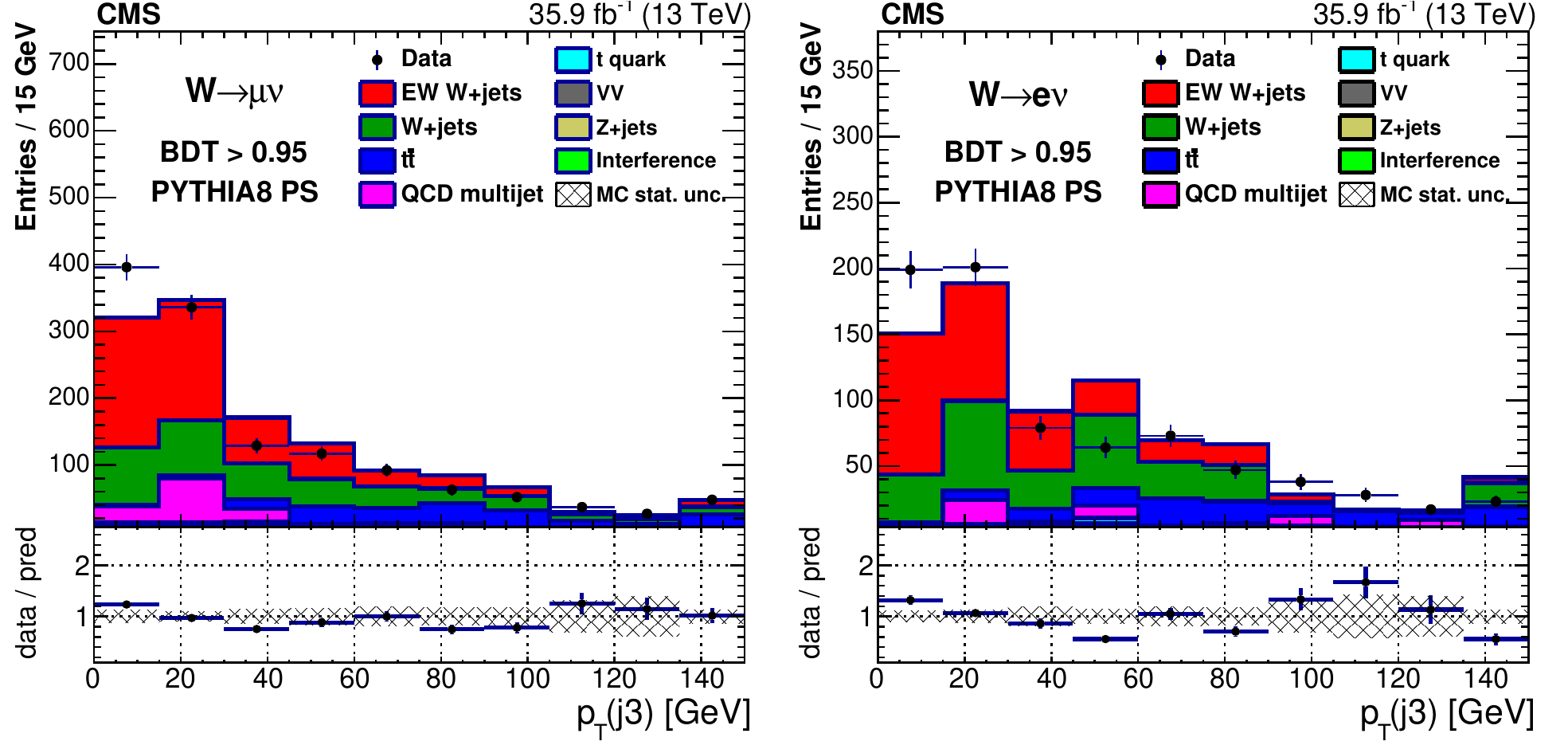}
    \includegraphics[width=0.98\textwidth]{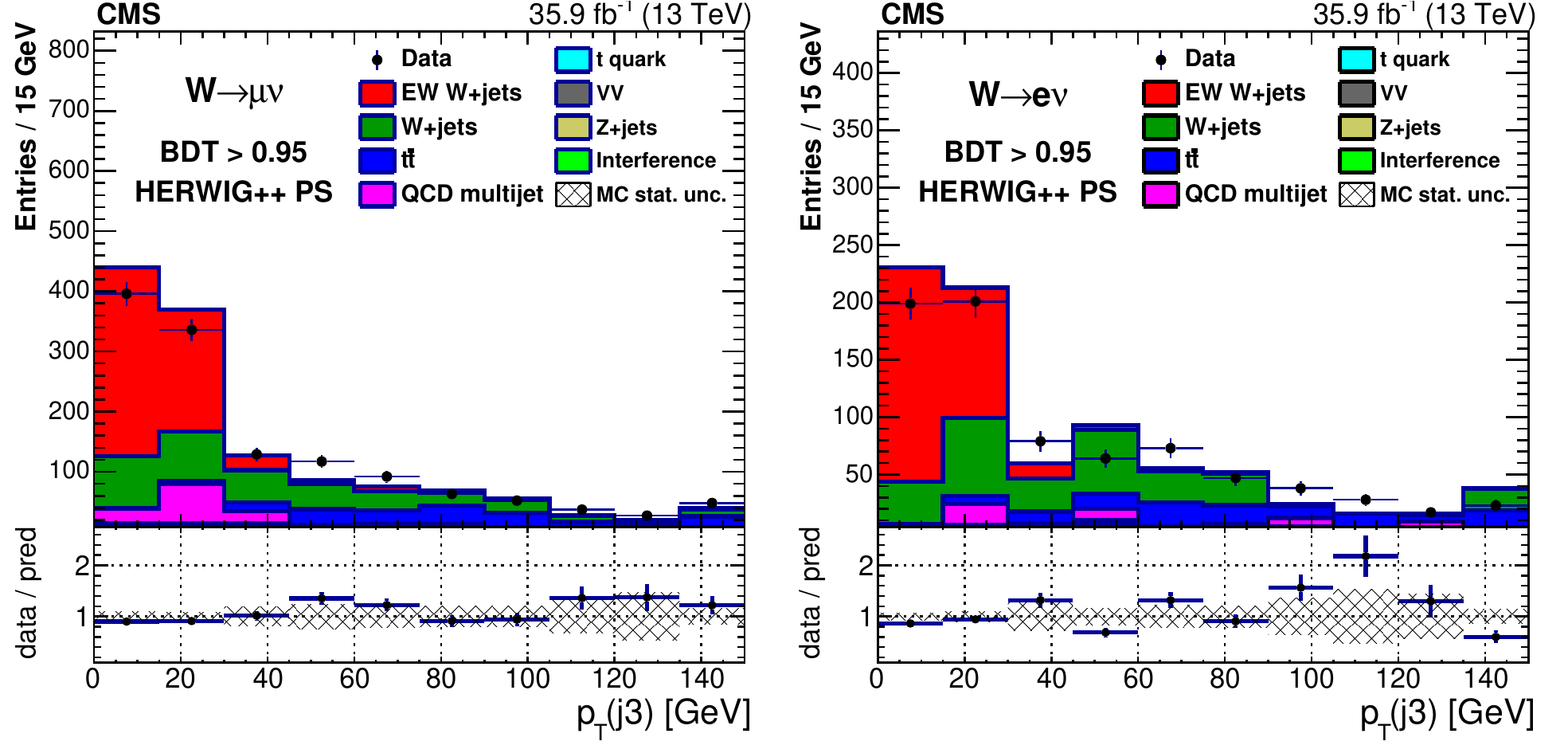}
    \caption{Leading additional jet \pt (\pt(j3)) for $\mathrm{BDT} > 0.95 $ in the muon (left) and electron (right)
    channels including the signal prediction from \MGvATNLO interfaced with \PYTHIA parton showering (upper) and \HERWIGpp
    parton showering (lower). In all plots the last bin contains overflow events, and the first bin
    contains events where no additional jet with \pt$>15$\GeV is present.}
     \label{fig:jet3pt_bdtcut}

\end{figure*}

\begin{figure*}[htb!]
  \centering
    \includegraphics[width=0.98\textwidth]{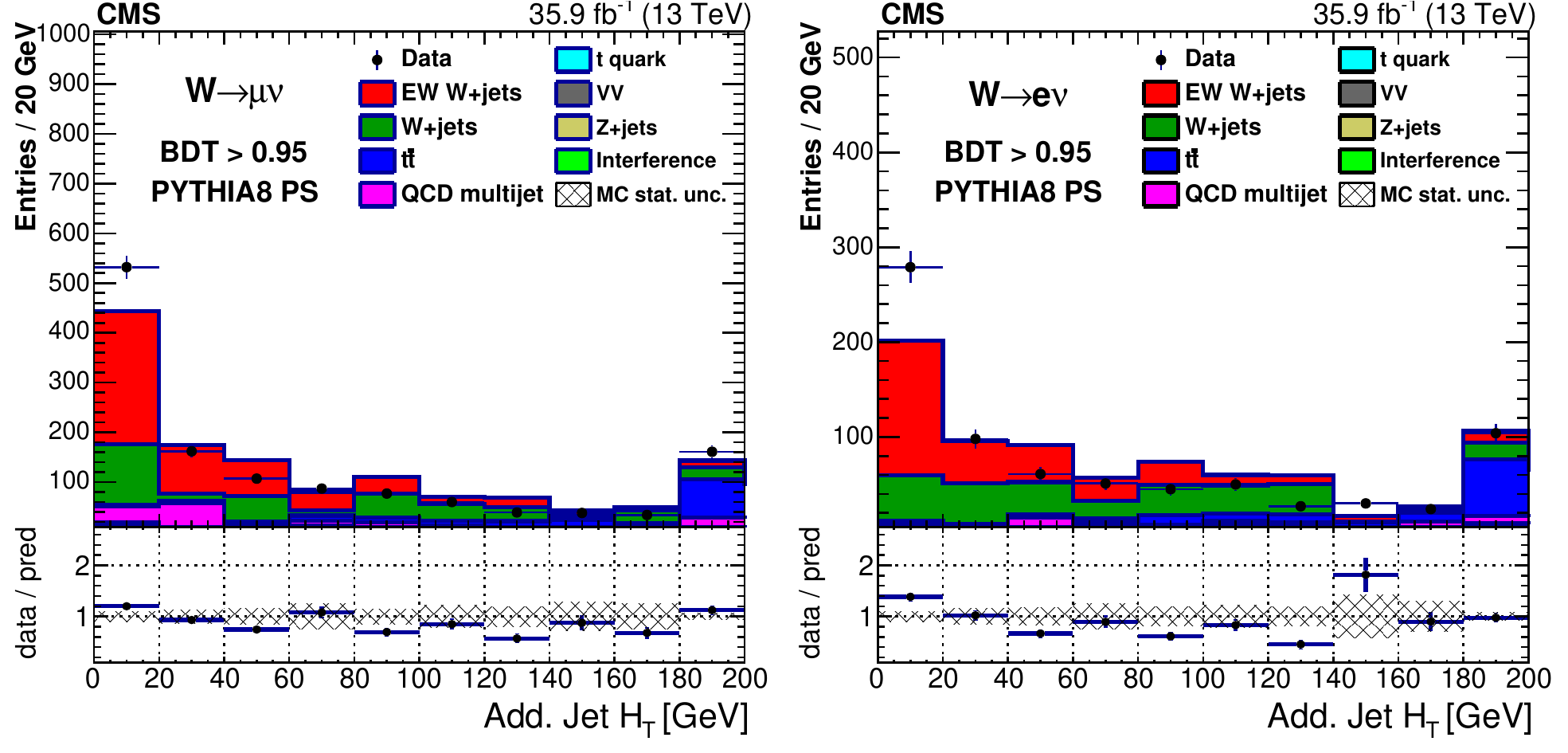}
    \includegraphics[width=0.98\textwidth]{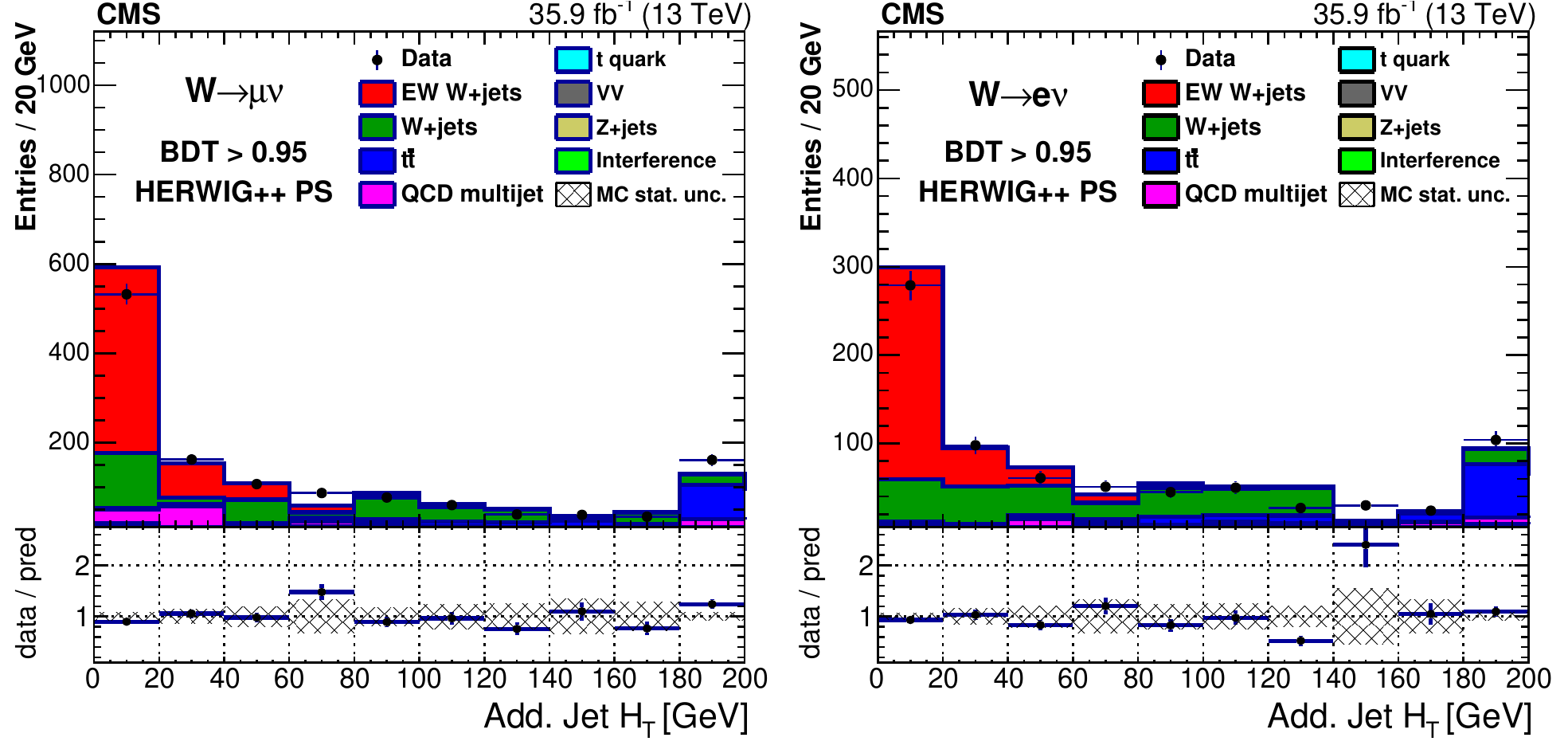}
    \caption{Total \HT of the additional jets for $\mathrm{BDT} > 0.95 $ in the muon (left) and electron (right)
    channels including the signal prediction from \MGvATNLO interfaced with \PYTHIA parton showering (upper) and \HERWIGpp
    parton showering (lower). In all plots the last bin contains overflow events, and the first bin
    contains events where no additional jet with \pt$>15$\GeV is present.}
     \label{fig:addjetht_bdtcut}

\end{figure*}

The \pt of the leading additional jet in \Wjj events,
as well as the scalar \pt sum ($\HT$) of all additional jets,
are shown in Figs.~\ref{fig:jet3pt_bdtcut} and~\ref{fig:addjetht_bdtcut},
comparing data and simulations including the signal prediction from \MGvATNLO interfaced with either \PYTHIA or \HERWIGpp parton showering.
The comparison reveals a deficit in the simulation predictions with
\PYTHIA parton showering
 for the rate of events with lower additional jet activity, whereas the
tail of higher additional activity is generally in better agreement.

A suppression of additional jets is observed in data compared
with the background-only simulation shapes.
In the simulation of the signal, the additional jets
are produced by the PS (see Section~\ref{sec:simulation}),
so studying these distributions provides insight on the PS model
in the rapidity gap region.

\subsection{Study of charged hadron activity}
\label{subsec:soft}

For this study, a collection  is formed
of high-purity tracks \cite{CMS-PAS-TRK-10-005} with $\pt > 0.3\GeV$,
uniquely associated with the main PV in the event.
Tracks associated with the lepton or with the tagging jets are
excluded from the selection.
The association between the selected tracks and the reconstructed PVs
is carried out by minimizing the longitudinal
impact parameter, which is defined
as the $z$-distance
between the PV and the point of closest approach of
the track helix to the PV, labeled $d_z^\mathrm{PV}$.
The association is required to satisfy the conditions $d_z^\mathrm{PV}<2\unit{mm}$ and
 $d_z^\mathrm{PV}<3\delta d_z^\mathrm{PV}$, where $\delta d_z^\mathrm{PV}$ is the uncertainty
in $d_z^\mathrm{PV}$.

A collection of ``soft-track'' jets is defined
by clustering the selected tracks using the anti-\kt clustering algorithm~\cite{Cacciari:2008gp}
with a distance parameter of $R=0.4$. The use of track jets represents a
clean and well-understood
method~\cite{CMS-PAS-JME-10-006} to reconstruct jets with energy
as low as a few \GeV.
These jets are not affected by pileup because of the association
of the constituent tracks with the hard scattering vertex~\cite{CMS-PAS-JME-08-001}.

\begin{figure*}[htb!]
  \centering
    \includegraphics[width=0.98\textwidth]{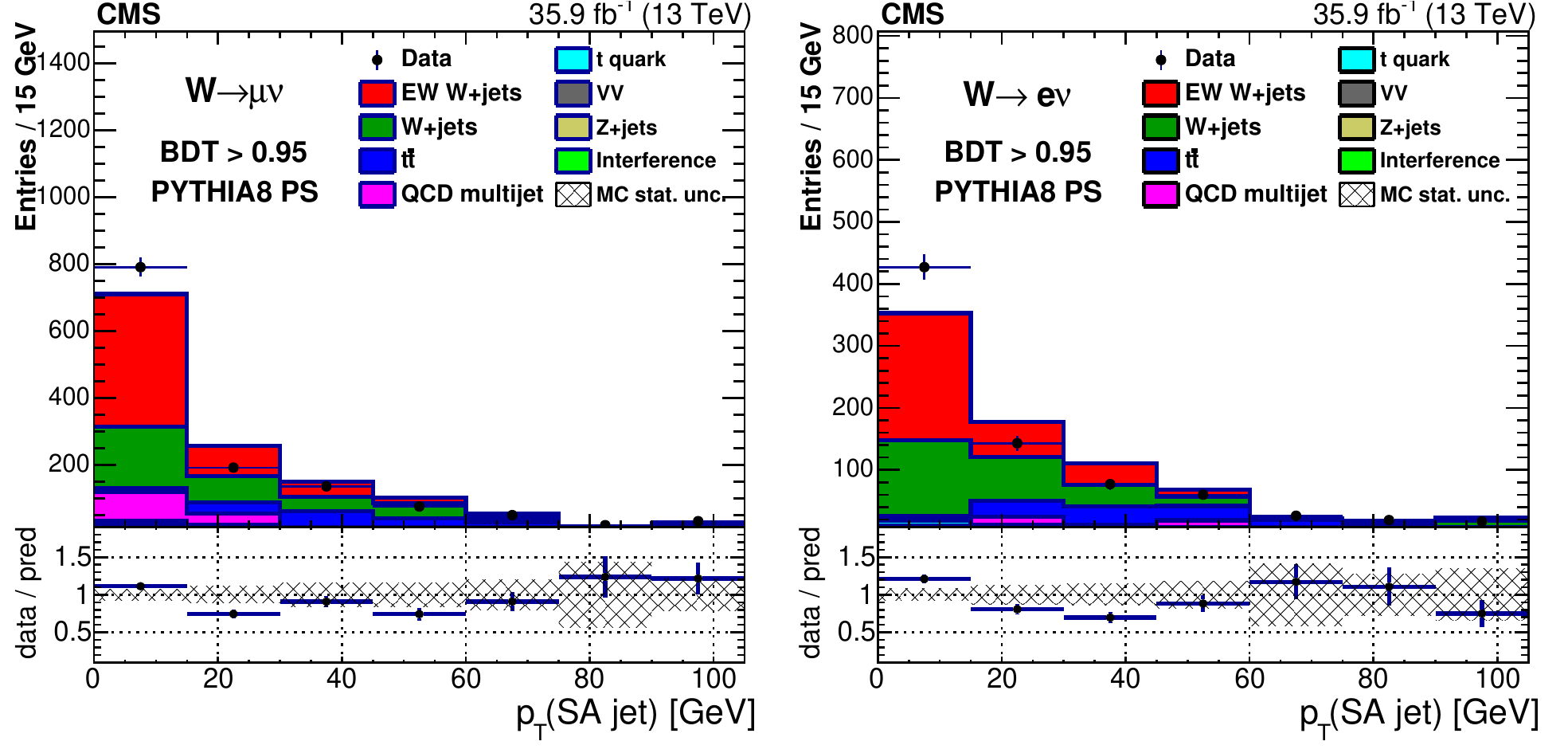}
    \includegraphics[width=0.98\textwidth]{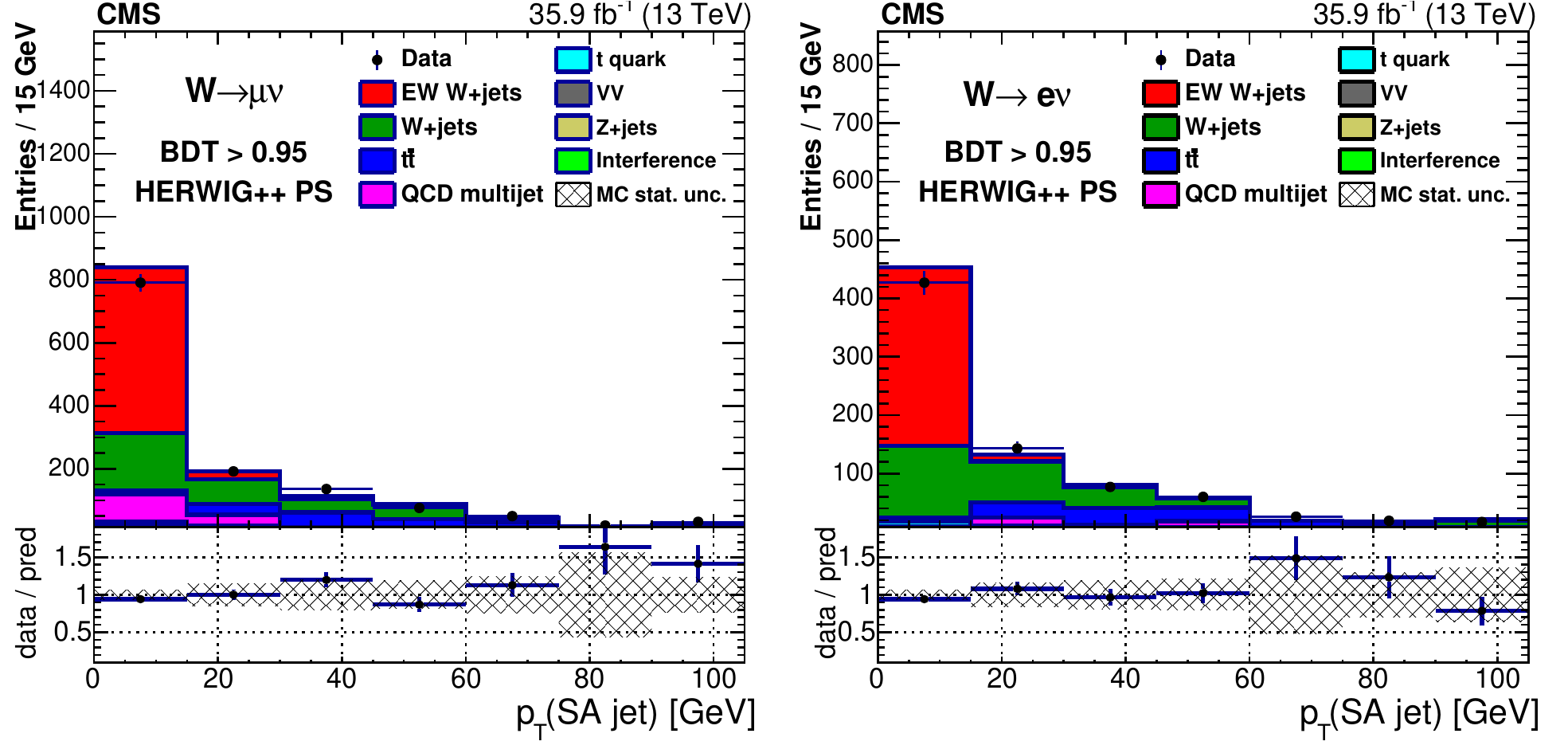}
    \caption{Leading additional soft-activity (SA) jet \pt for BDT~$> 0.95$
    in the muon (left) and electron (right)
    channels including the signal prediction from \MGvATNLO interfaced with \PYTHIA parton showering (upper) and \HERWIGpp parton showering (lower). }
     \label{fig:leadsajetpt_bdtcut}

\end{figure*}

\begin{figure*}[htb!]
  \centering
    \includegraphics[width=0.98\textwidth]{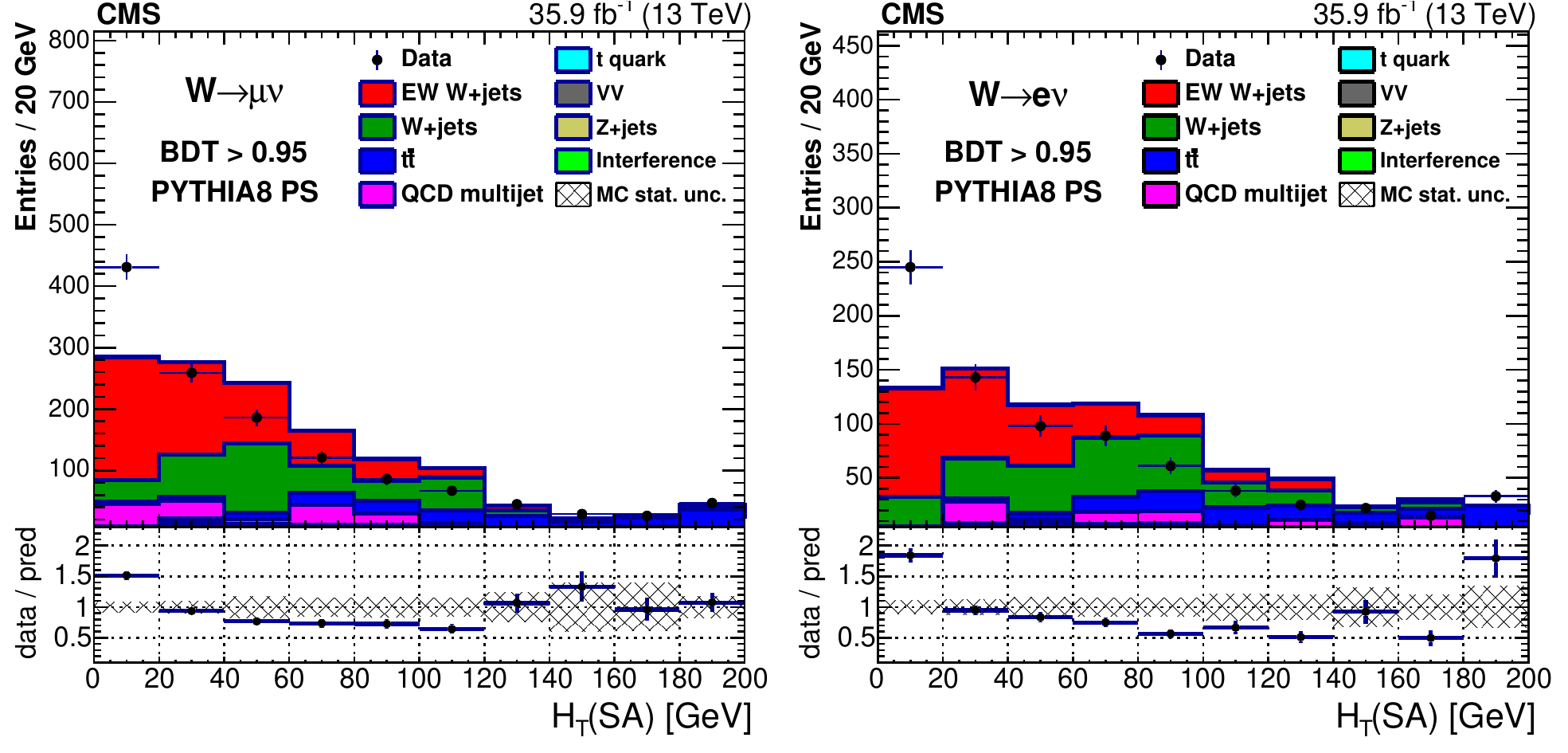}
    \includegraphics[width=0.98\textwidth]{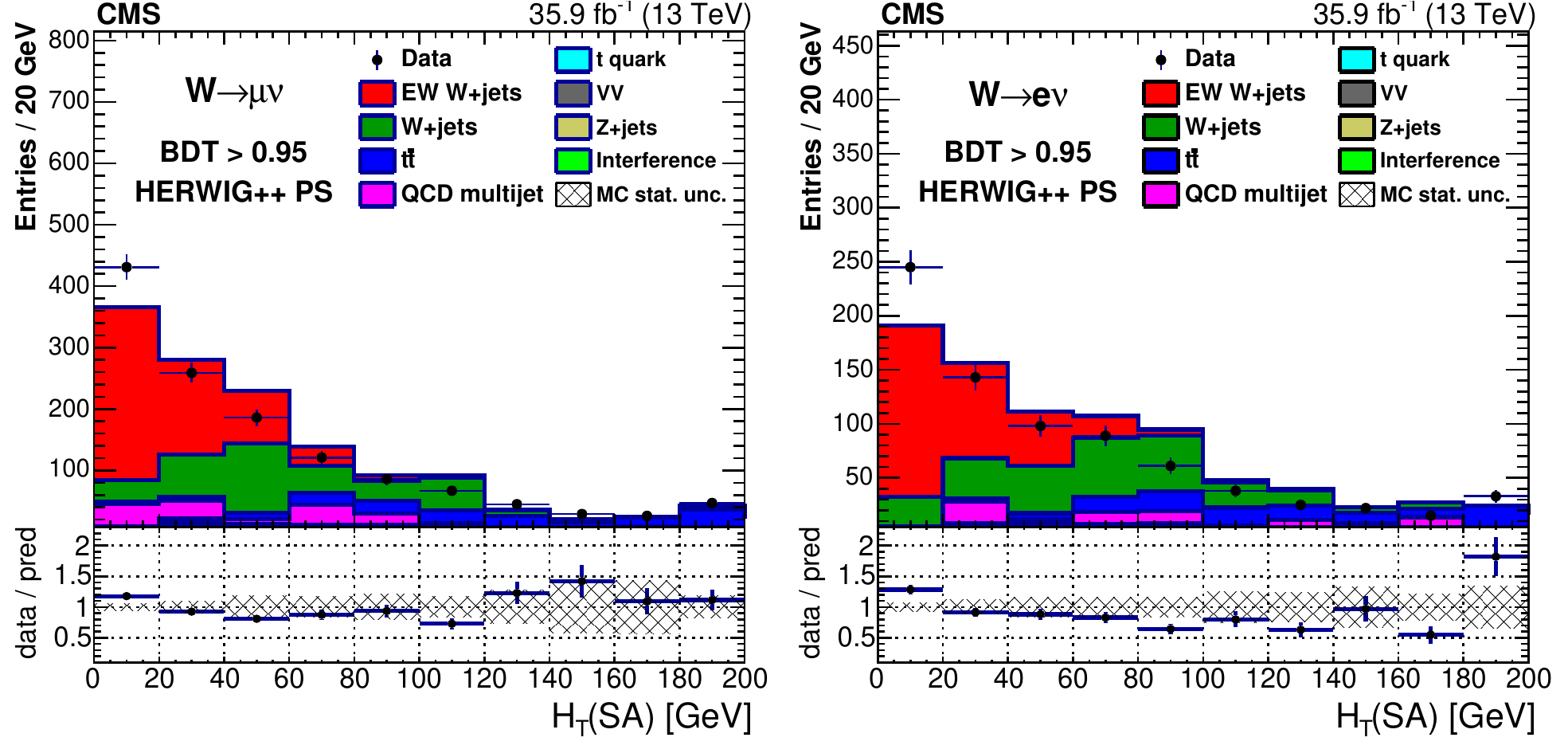}
    \caption{Total soft activity (SA) jet \HT for BDT~$> 0.95$
    in the muon (left) and electron (right)
    channels including the signal prediction from \MGvATNLO interfaced with \PYTHIA parton showering (upper) and \HERWIGpp parton showering (lower).
    In all plots the last bin contains overflow events.}
     \label{fig:sajetht_bdtcut}

\end{figure*}

Track jets of low \pt and within
$\eta^\text{tag jet}_\text{min} < \eta < \eta^\text{tag jet}_\text{max} $ are
considered for the study of the hadronic activity between the tagging jets,
and referred to as ``soft activity'' (SA).
For each event, the scalar  \pt sum of
the soft-track jets with \pt$>1$\GeV is computed, and referred to as
the ``soft $\HT$'' variable.
Figures~\ref{fig:leadsajetpt_bdtcut} and~\ref{fig:sajetht_bdtcut} show the distribution of the
leading soft-track jet \pt and soft \HT
in the signal-enriched region (BDT~$>0.95$), for the electron and muon
channels, compared to predictions from \PYTHIA and \HERWIGpp PS models.
The plots show some disagreement between the data and the predictions, in particular in the
regions of small additional activity, when compared with the \PYTHIA predictions.

\subsection{Study of hadronic activity vetoes}
\label{subsec:gapveto}

The efficiency of a hadronic activity veto corresponds to the fraction of events with a measured gap activity
below a given threshold. This efficiency is studied as a function of the applied threshold
for various gap activity observables.
The veto thresholds studied here start at 15\GeV for gap activities measured with standard PF jets,
while they go down to 1\GeV for gap activities measured with soft-track jets.

\begin{figure*}[htb!]
  \centering
    \includegraphics[width=0.48\textwidth]{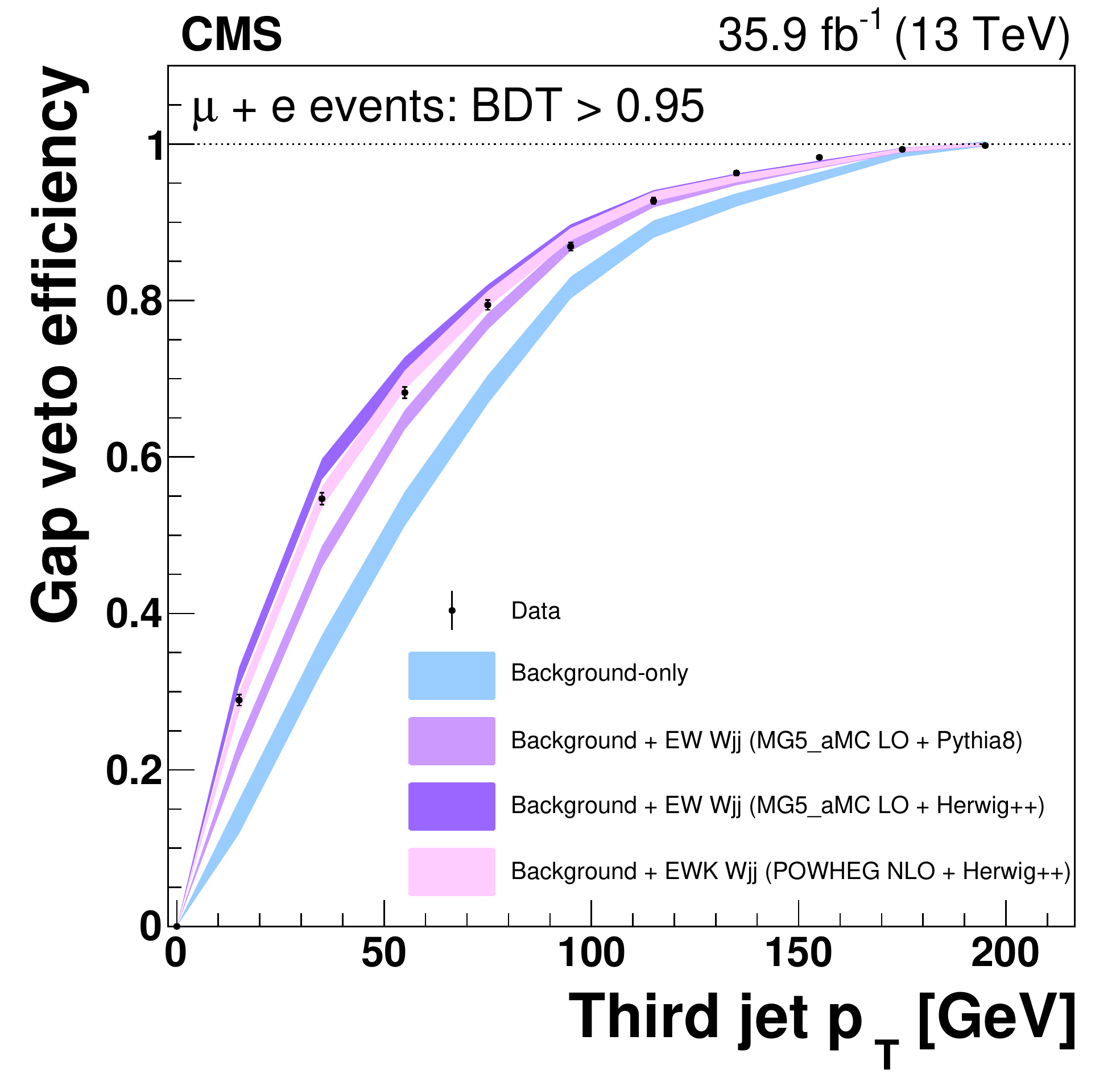}
    \includegraphics[width=0.48\textwidth]{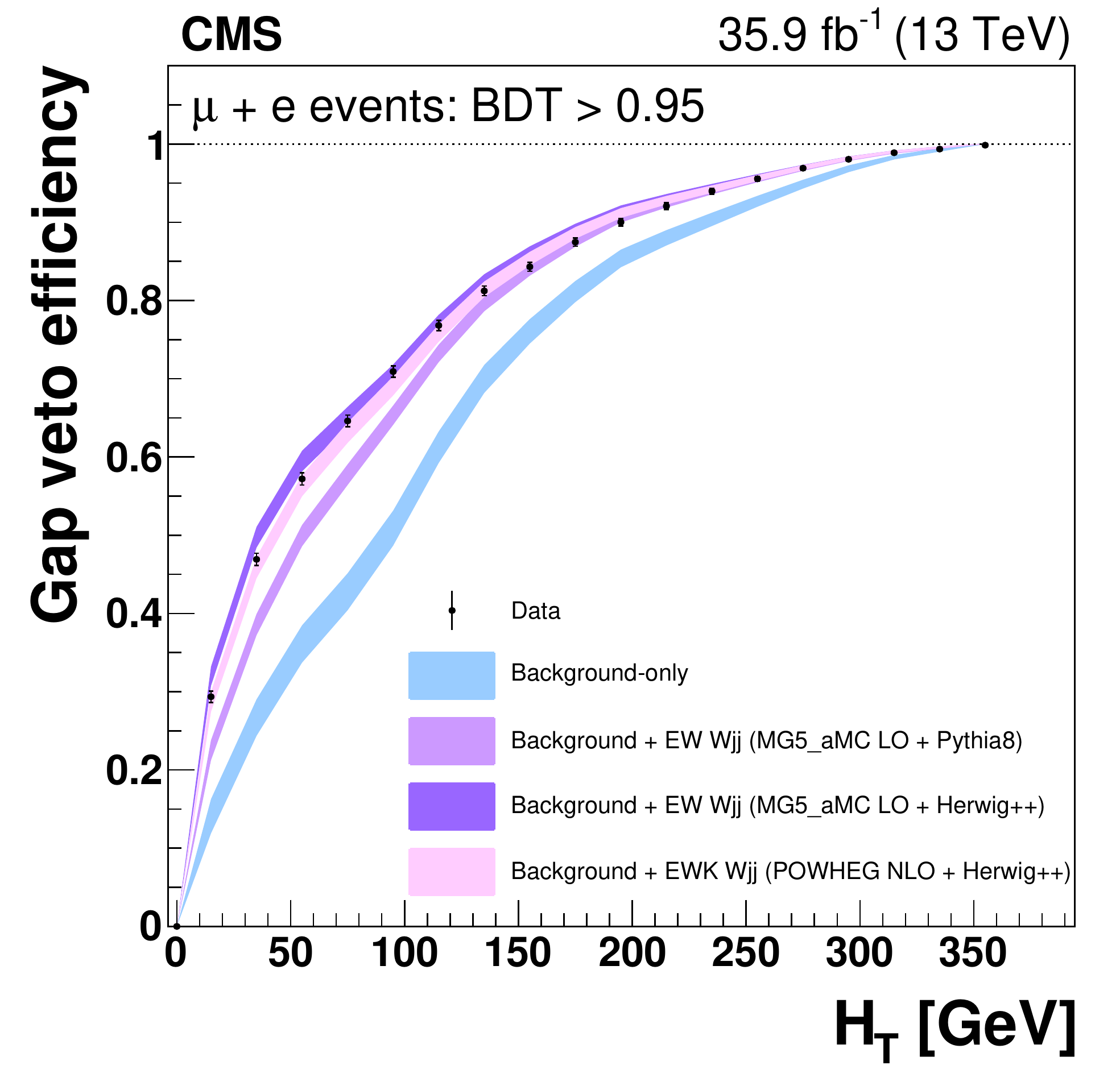}
    \includegraphics[width=0.48\textwidth]{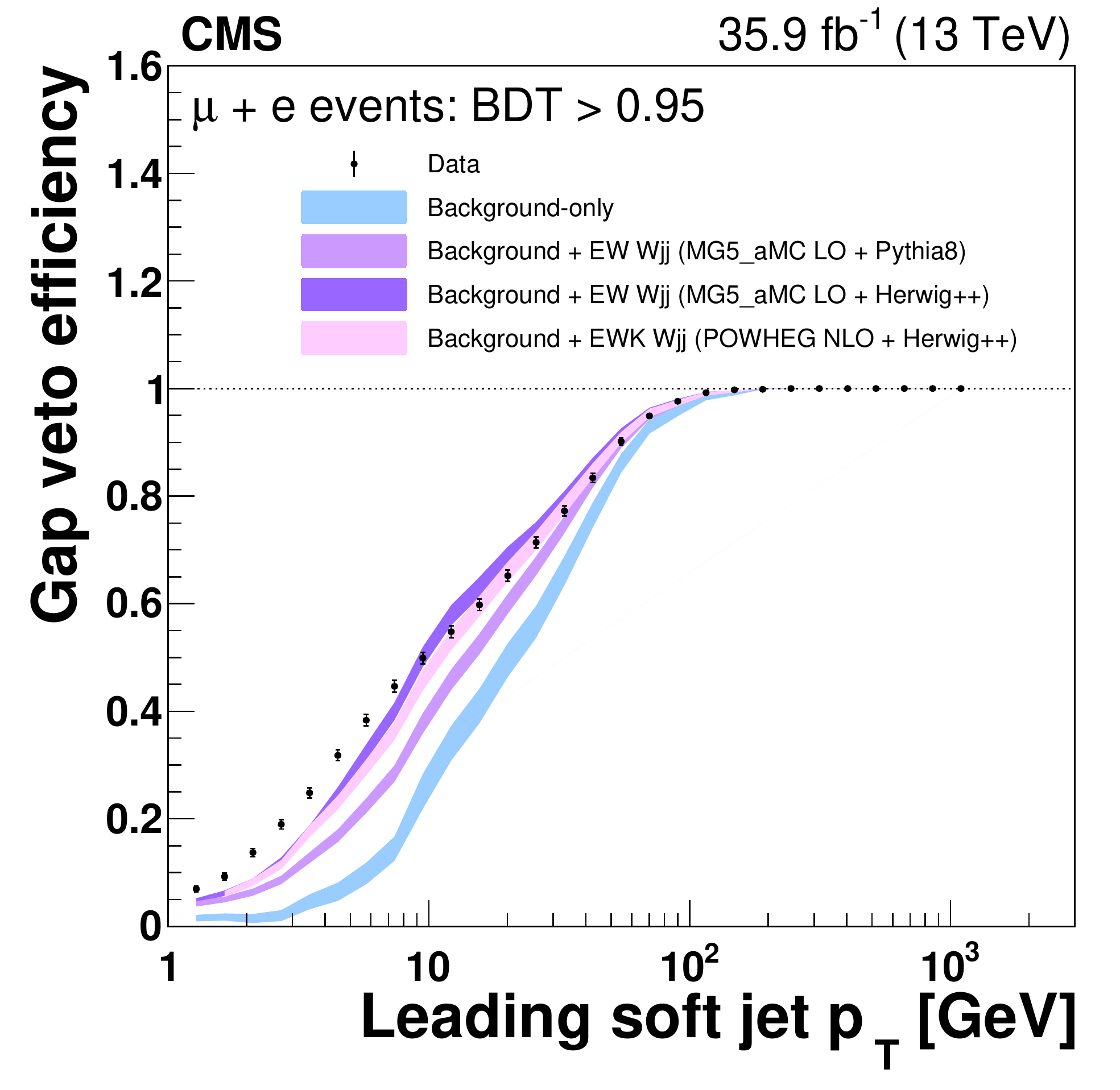}
    \includegraphics[width=0.48\textwidth]{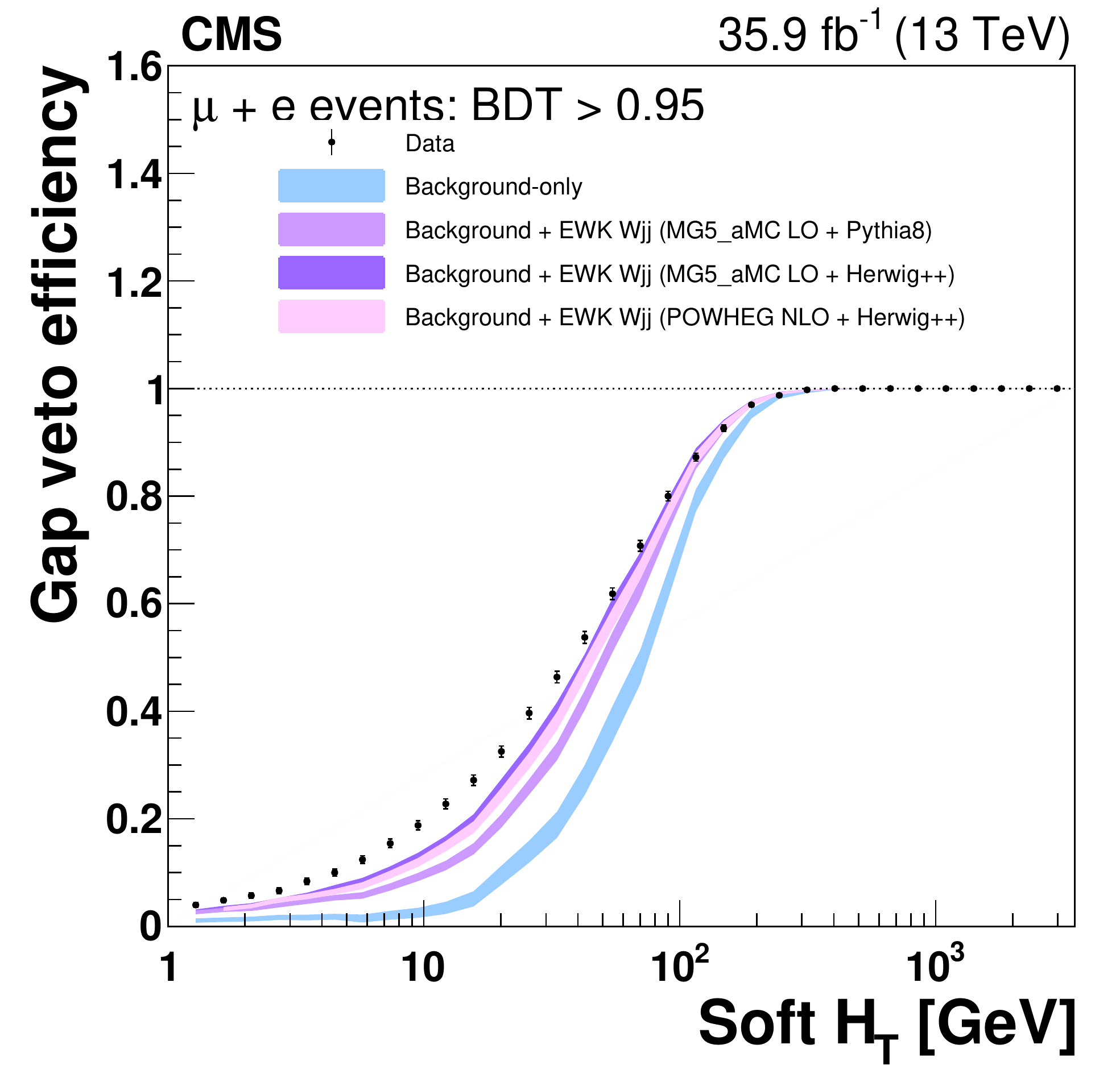}
    \caption{Hadronic activity veto efficiencies in the signal-enriched
    $\mathrm{BDT}>0.95$
    region for the muon and electron channels combined, as a function of the leading
    additional jet \pt (upper left), additional jet \HT (upper right), leading
    soft-activity jet \pt (lower left), and soft-activity jet \HT (lower right). The data
    are compared with the background-only prediction as well as background+signal with \PYTHIA
    parton showering and background+signal with \HERWIGpp parton showering. In addition, the background+signal
    prediction from \POWHEG plus \HERWIGpp parton showering is included. The uncertainty bands
    include only the statistical uncertainty in the prediction from simulation, and the data points include
    only the statistical uncertainty in data.
    }
     \label{fig:gapvetoeff}

\end{figure*}

Figure~\ref{fig:gapvetoeff} shows the gap activity veto efficiency of combined muon and electron events
in the signal-enriched region when placing an upper threshold on
the \pt of the additional third jet, on the \HT of all additional jets,
on the leading soft-activity jet \pt, or on the soft-activity \HT.
The observed efficiency in data is compared to expected efficiencies for
background-only events, and efficiencies for background plus signal events where
the signal is modeled with \PYTHIA or \HERWIGpp.
Data points clearly disfavor the background-only predictions and are in reasonable agreement
with the presence of the signal with the \HERWIGpp PS predictions for gap activities above 20\GeV,
while the signal with \PYTHIA PS seems to generally overestimate the gap activity.
In the events with very low gap activity, in particular below 10\GeV as measured with the soft track jets,
the data indicates gap activities also below the \HERWIGpp PS predictions. In addition, the expected efficiencies
are included for background plus signal events where the signal is modeled with \POWHEG (Sec.~\ref{sec:simulation})
with \HERWIGpp PS. The \POWHEG plus \HERWIGpp prediction is in good agreement with the LO plus \HERWIGpp prediction.

\section{Summary}
\label{sec:summary}

The cross section of the electroweak production of a
\PW\ boson in association with two jets
is measured in the kinematic region defined as invariant mass $m_\mathrm{jj} >120\GeV$
and transverse momenta $p_\mathrm{T j} > 25\GeV$.
The data sample corresponds to
an integrated  luminosity of $35.9~\mathrm{fb}^{-1}$ of proton-proton collisions at centre-of-mass energy
$\sqrt{s}=13\TeV$ recorded by the CMS Collaboration at the LHC.
 The measured cross section
$\sigma_\mathrm{EW}(\PW\mathrm{jj})= 6.23 \pm 0.12 \stat\pm 0.61 \syst\unit{pb}$
agrees with the leading order standard model prediction. This is the first
observation of this process at $\sqrt{s}=13\TeV$.

{\tolerance=4800
A search is performed for anomalous trilinear gauge couplings associated with dimension-six operators
as given in the framework of an effective field theory. No evidence for ATGCs
is found, and the corresponding
95\% confidence level intervals on the dimension-six operators are
$-2.3 < c_{{\PW\PW\PW}}/\Lambda^2 < 2.5\TeV^{-2}$,
$-8.8 < c_{\PW}/\Lambda^2  < 16\TeV^{-2}$, and
$-45 < c_{\mathrm{B}}/\Lambda^2 < 46\TeV^{-2}$.
These results are combined with previous results  on the electroweak production of a Z boson in association with two jets,
yielding the limit on the $c_{{\PW\PW\PW}}$ coupling
$-1.8 <  c_{{\PW\PW\PW}}/\Lambda^2  < 2.0\TeV^{-2}$.
\par}

The additional hadronic activity, as well as the efficiencies for gap activity vetos,
are studied in a signal-enriched region. Generally reasonable agreement is found between the
data and the quantum chromodynamics predictions with the \HERWIGpp parton shower
and hadronization model, while the \PYTHIA model predictions typically show greater activity in the rapidity gap between the two tagging jets.

\begin{acknowledgments}
We congratulate our colleagues in the CERN accelerator departments for the excellent performance of the LHC and thank the technical and administrative staffs at CERN and at other CMS institutes for their contributions to the success of the CMS effort. In addition, we gratefully acknowledge the computing centers and personnel of the Worldwide LHC Computing Grid for delivering so effectively the computing infrastructure essential to our analyses. Finally, we acknowledge the enduring support for the construction and operation of the LHC and the CMS detector provided by the following funding agencies: BMBWF and FWF (Austria); FNRS and FWO (Belgium); CNPq, CAPES, FAPERJ, FAPERGS, and FAPESP (Brazil); MES (Bulgaria); CERN; CAS, MoST, and NSFC (China); COLCIENCIAS (Colombia); MSES and CSF (Croatia); RPF (Cyprus); SENESCYT (Ecuador); MoER, ERC IUT, and ERDF (Estonia); Academy of Finland, MEC, and HIP (Finland); CEA and CNRS/IN2P3 (France); BMBF, DFG, and HGF (Germany); GSRT (Greece); NKFIA (Hungary); DAE and DST (India); IPM (Iran); SFI (Ireland); INFN (Italy); MSIP and NRF (Republic of Korea); MES (Latvia); LAS (Lithuania); MOE and UM (Malaysia); BUAP, CINVESTAV, CONACYT, LNS, SEP, and UASLP-FAI (Mexico); MOS (Montenegro); MBIE (New Zealand); PAEC (Pakistan); MSHE and NSC (Poland); FCT (Portugal); JINR (Dubna); MON, RosAtom, RAS, RFBR, and NRC KI (Russia); MESTD (Serbia); SEIDI, CPAN, PCTI, and FEDER (Spain); MOSTR (Sri Lanka); Swiss Funding Agencies (Switzerland); MST (Taipei); ThEPCenter, IPST, STAR, and NSTDA (Thailand); TUBITAK and TAEK (Turkey); NASU and SFFR (Ukraine); STFC (United Kingdom); DOE and NSF (USA).

\hyphenation{Rachada-pisek} Individuals have received support from the Marie-Curie program and the European Research Council and Horizon 2020 Grant, contract Nos.\ 675440 and 765710 (European Union); the Leventis Foundation; the A.P.\ Sloan Foundation; the Alexander von Humboldt Foundation; the Belgian Federal Science Policy Office; the Fonds pour la Formation \`a la Recherche dans l'Industrie et dans l'Agriculture (FRIA-Belgium); the Agentschap voor Innovatie door Wetenschap en Technologie (IWT-Belgium); the F.R.S.-FNRS and FWO (Belgium) under the ``Excellence of Science -- EOS" -- be.h project n.\ 30820817; the Beijing Municipal Science \& Technology Commission, No. Z181100004218003; the Ministry of Education, Youth and Sports (MEYS) of the Czech Republic; the Lend\"ulet (``Momentum") Program and the J\'anos Bolyai Research Scholarship of the Hungarian Academy of Sciences, the New National Excellence Program \'UNKP, the NKFIA research grants 123842, 123959, 124845, 124850, 125105, 128713, 128786, and 129058 (Hungary); the Council of Science and Industrial Research, India; the HOMING PLUS program of the Foundation for Polish Science, cofinanced from European Union, Regional Development Fund, the Mobility Plus program of the Ministry of Science and Higher Education, the National Science Center (Poland), contracts Harmonia 2014/14/M/ST2/00428, Opus 2014/13/B/ST2/02543, 2014/15/B/ST2/03998, and 2015/19/B/ST2/02861, Sonata-bis 2012/07/E/ST2/01406; the National Priorities Research Program by Qatar National Research Fund; the Programa Estatal de Fomento de la Investigaci{\'o}n Cient{\'i}fica y T{\'e}cnica de Excelencia Mar\'{\i}a de Maeztu, grant MDM-2015-0509 and the Programa Severo Ochoa del Principado de Asturias; the Thalis and Aristeia programs cofinanced by EU-ESF and the Greek NSRF; the Rachadapisek Sompot Fund for Postdoctoral Fellowship, Chulalongkorn University and the Chulalongkorn Academic into Its 2nd Century Project Advancement Project (Thailand); the Welch Foundation, contract C-1845; and the Weston Havens Foundation (USA).
\end{acknowledgments}

\bibliography{auto_generated}

\appendix

\section{Additional rapidity gap observables}

\begin{figure*}[htb!]
  \centering
    \includegraphics[width=0.98\textwidth]{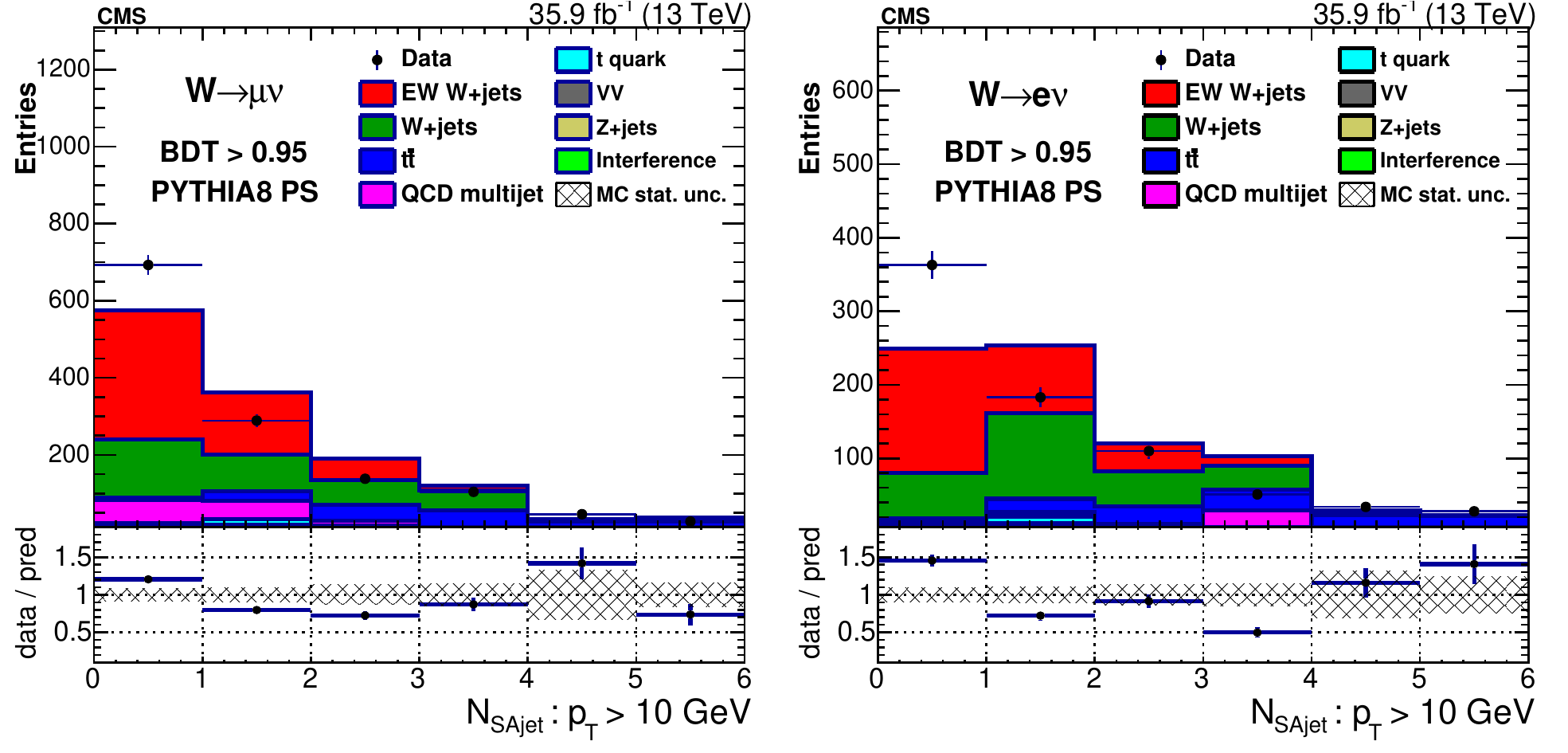}
    \includegraphics[width=0.98\textwidth]{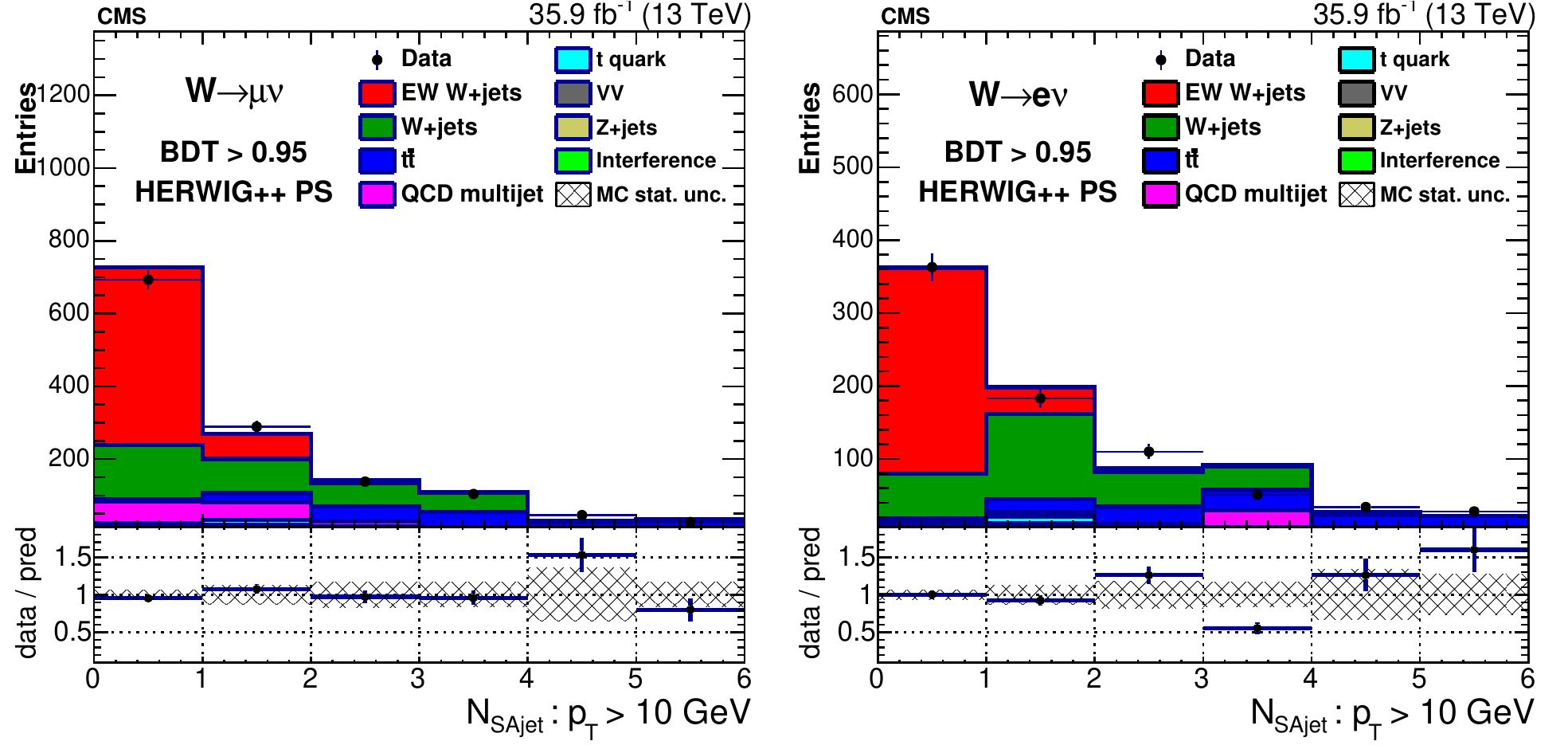}
    \caption{Number of soft activity jets with $\pt> 10\GeV$ in the rapidity gap for $\mathrm{BDT} > 0.95$ in the muon (left) and electron (right)
    channels including signal with \PYTHIA parton showering (upper) and \HERWIGpp
    parton showering (lower). In all plots the last bin contains overflow events.}
     \label{fig:nsa10_bdtcut}

\end{figure*}

\begin{figure*}[htb!]
  \centering
    \includegraphics[width=0.98\textwidth]{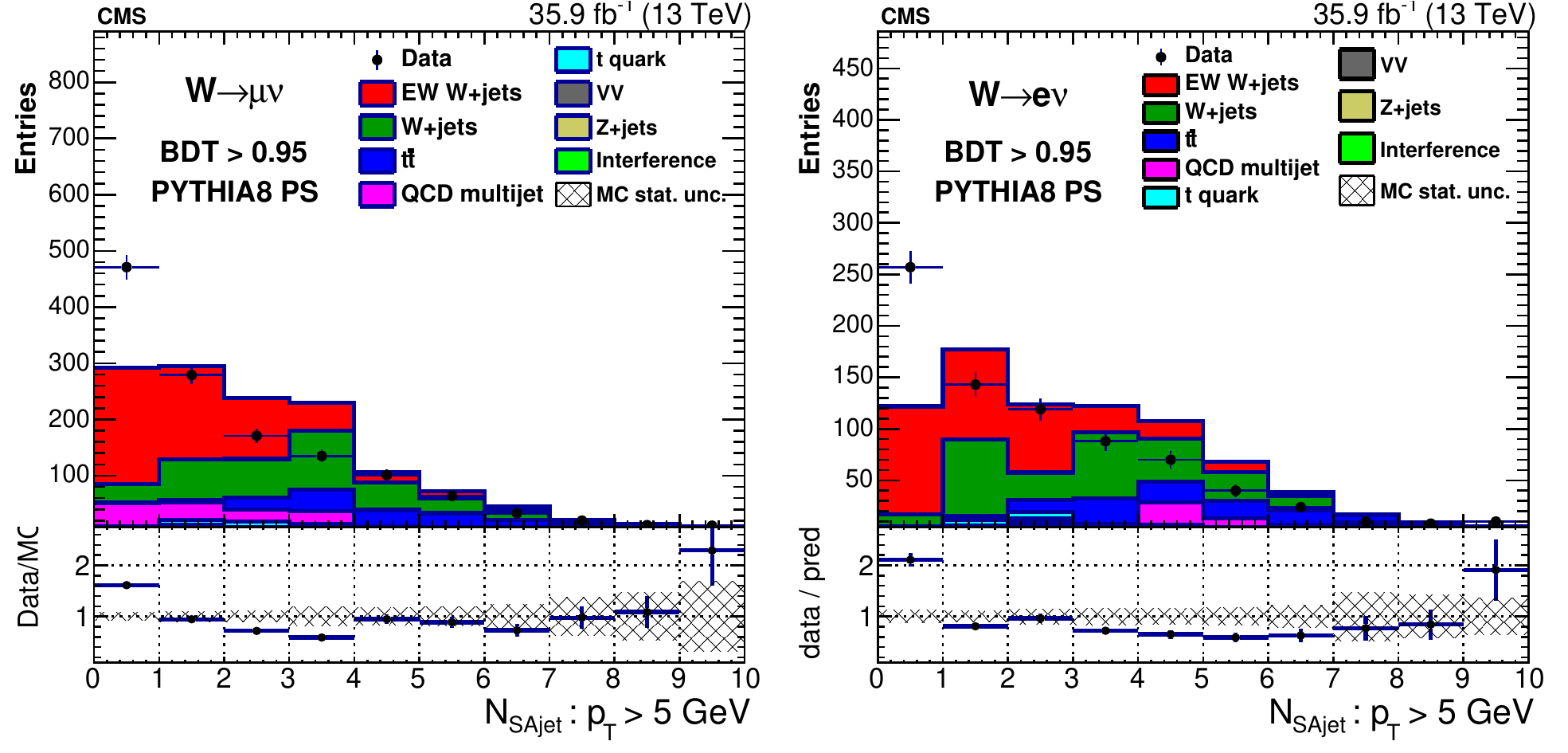}
    \includegraphics[width=0.98\textwidth]{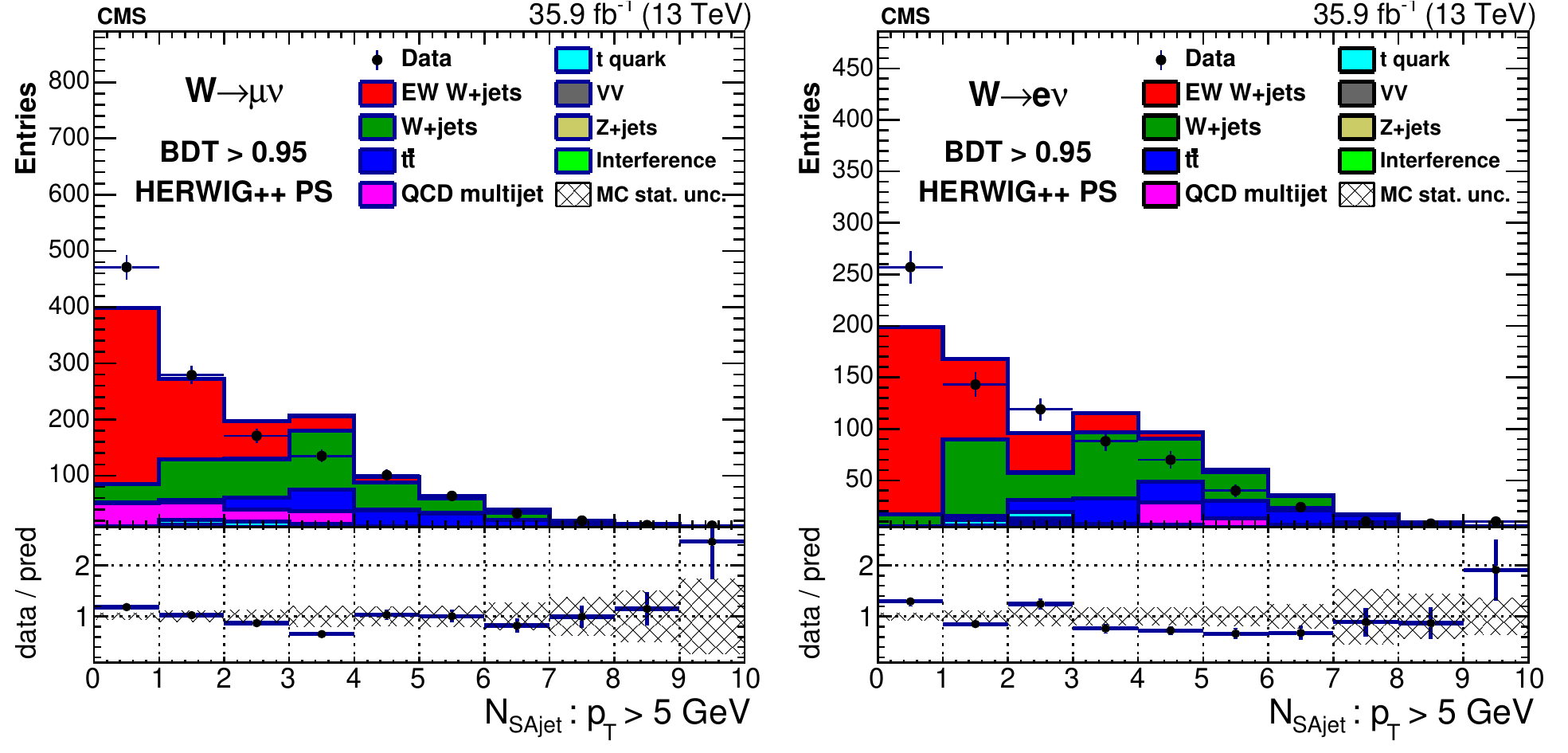}
    \caption{Number of soft activity jets with $\pt>5\GeV$ in the rapidity gap for $\mathrm{BDT} > 0.95 $ in the muon (left) and electron (right)
    channels including signal with \PYTHIA parton showering (upper) and \HERWIGpp
    parton showering (lower).}
     \label{fig:nsa5_bdtcut}

\end{figure*}

\begin{figure*}[htb!]
  \centering
    \includegraphics[width=0.98\textwidth]{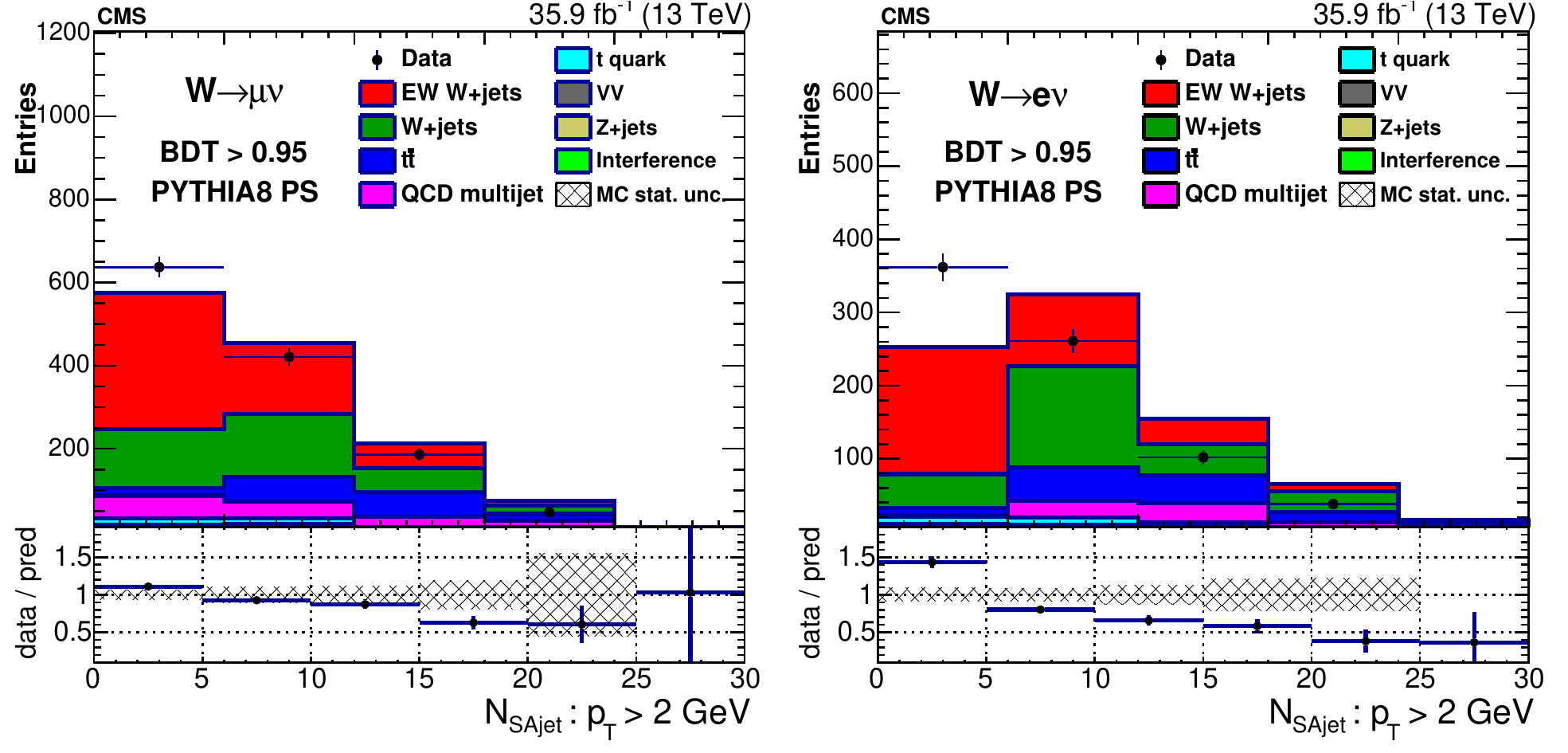}
    \includegraphics[width=0.98\textwidth]{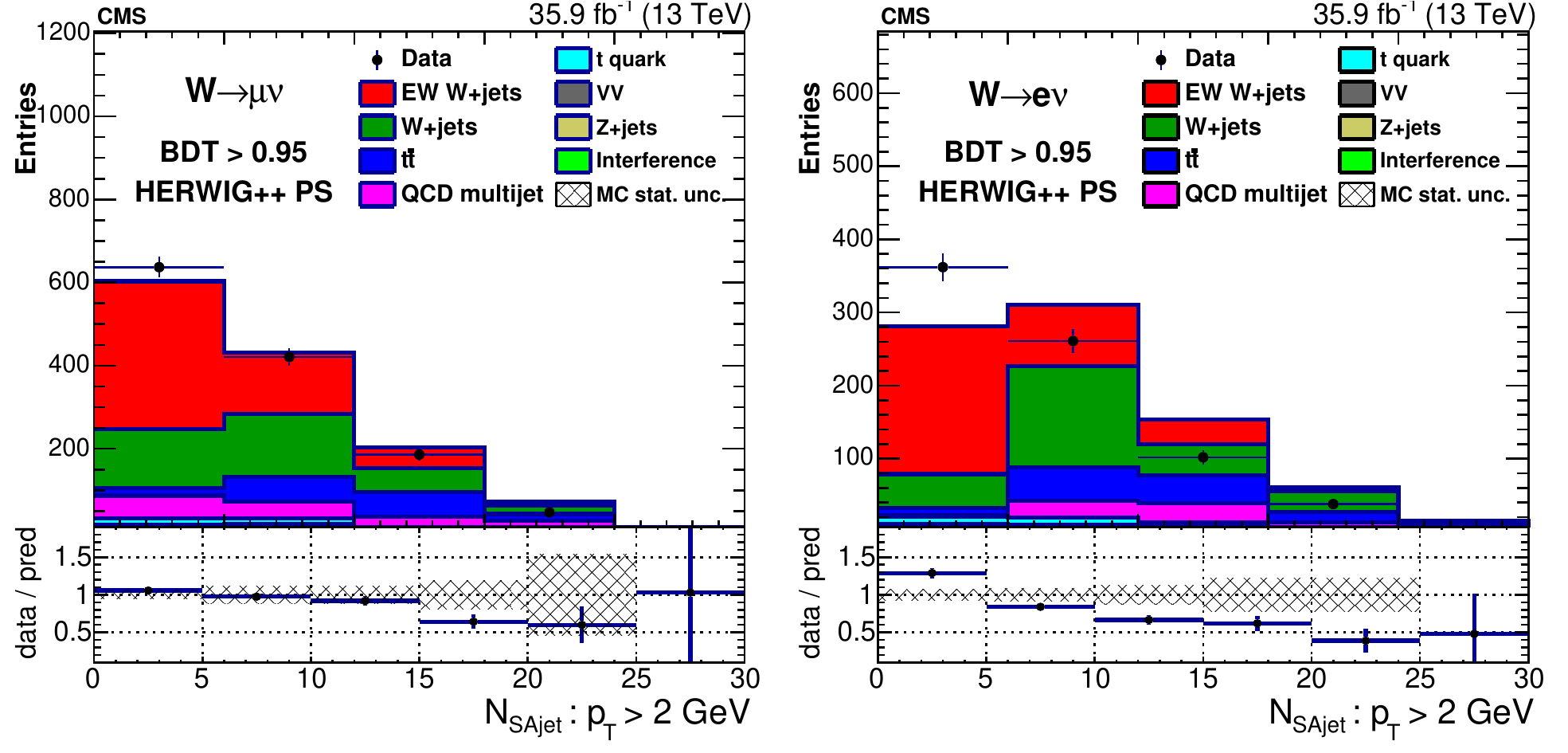}
    \caption{Number of soft activity jets with $\pt>2\GeV$ in the rapidity gap for $\mathrm{BDT} > 0.95 $ in the muon (left) and electron (right)
    channels including signal with \PYTHIA parton showering (upper) and \HERWIGpp
    parton showering (lower).}
     \label{fig:nsa2_bdtcut}

\end{figure*}

A set of rapidity gap observables in the high signal purity region $\mathrm{BDT} > 0.95 $ is studied in addition to the results
described in Section~\ref{sec:hadactivity}. The number of soft activity jets, defined in Section \ref{subsec:soft}, in the rapidity gap between the
two tag jets is shown for soft activity jet $\pt> 10$, 5, and 2\GeV in Figures \ref{fig:nsa10_bdtcut}, \ref{fig:nsa5_bdtcut}, and \ref{fig:nsa2_bdtcut},
respectively. These distributions are consistent with the general underestimation of the simulation with respect to data at low activity values, particularly
for the \PYTHIA parton showering.

\clearpage
\section{Jet activity in signal-depleted region}

\begin{figure*}[htb!]
  \centering
    \includegraphics[width=0.98\textwidth]{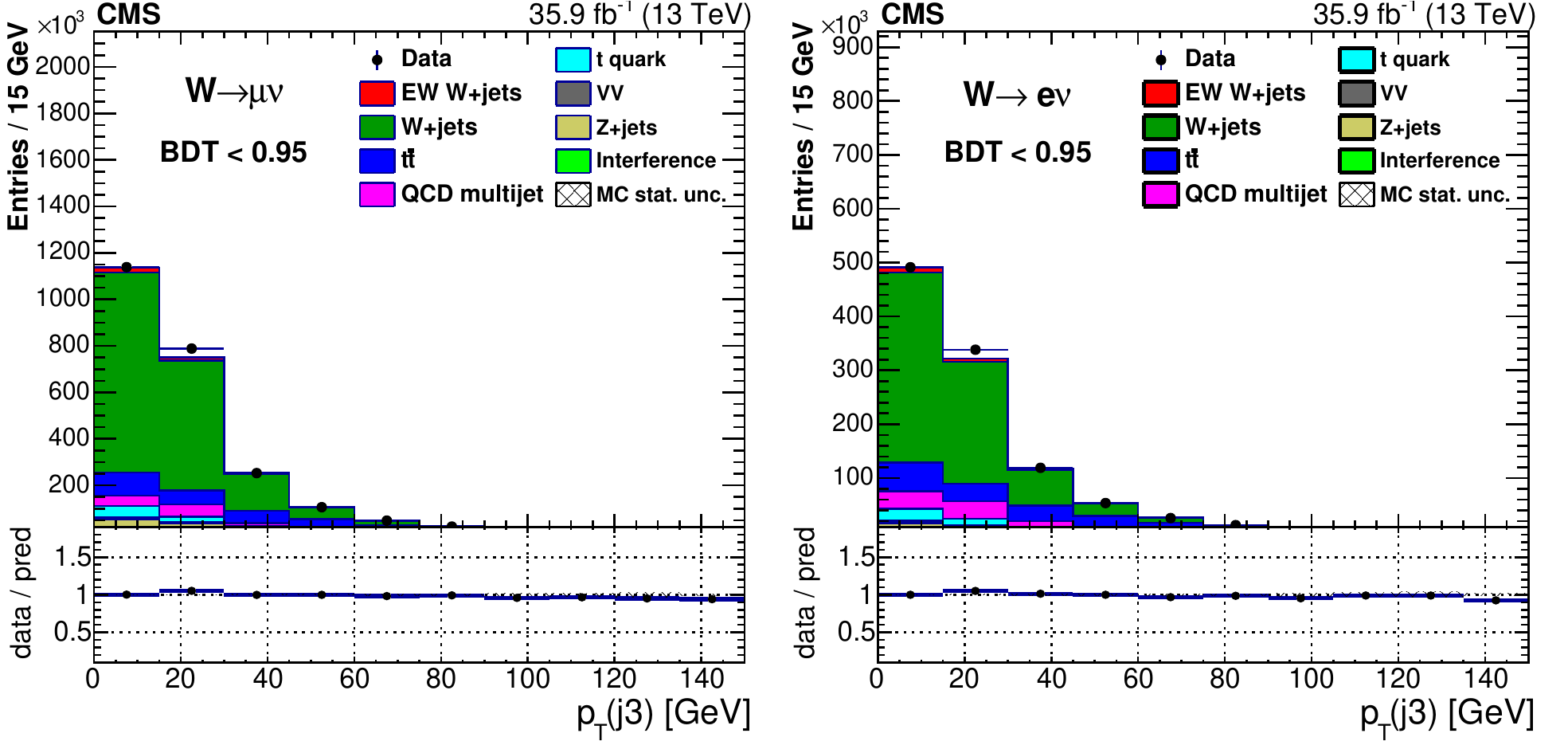}
    \caption{Leading additional jet \pt (\pt(j3)) for BDT $<$ 0.95 in the muon (left) and electron (right)
    channels. In all plots the first bin
    contains events where no additional jet with $\pt>15\GeV$ is present.}
     \label{fig:jet3pt_bdtcutInv}

\end{figure*}

\begin{figure*}[htb!]
  \centering
    \includegraphics[width=0.98\textwidth]{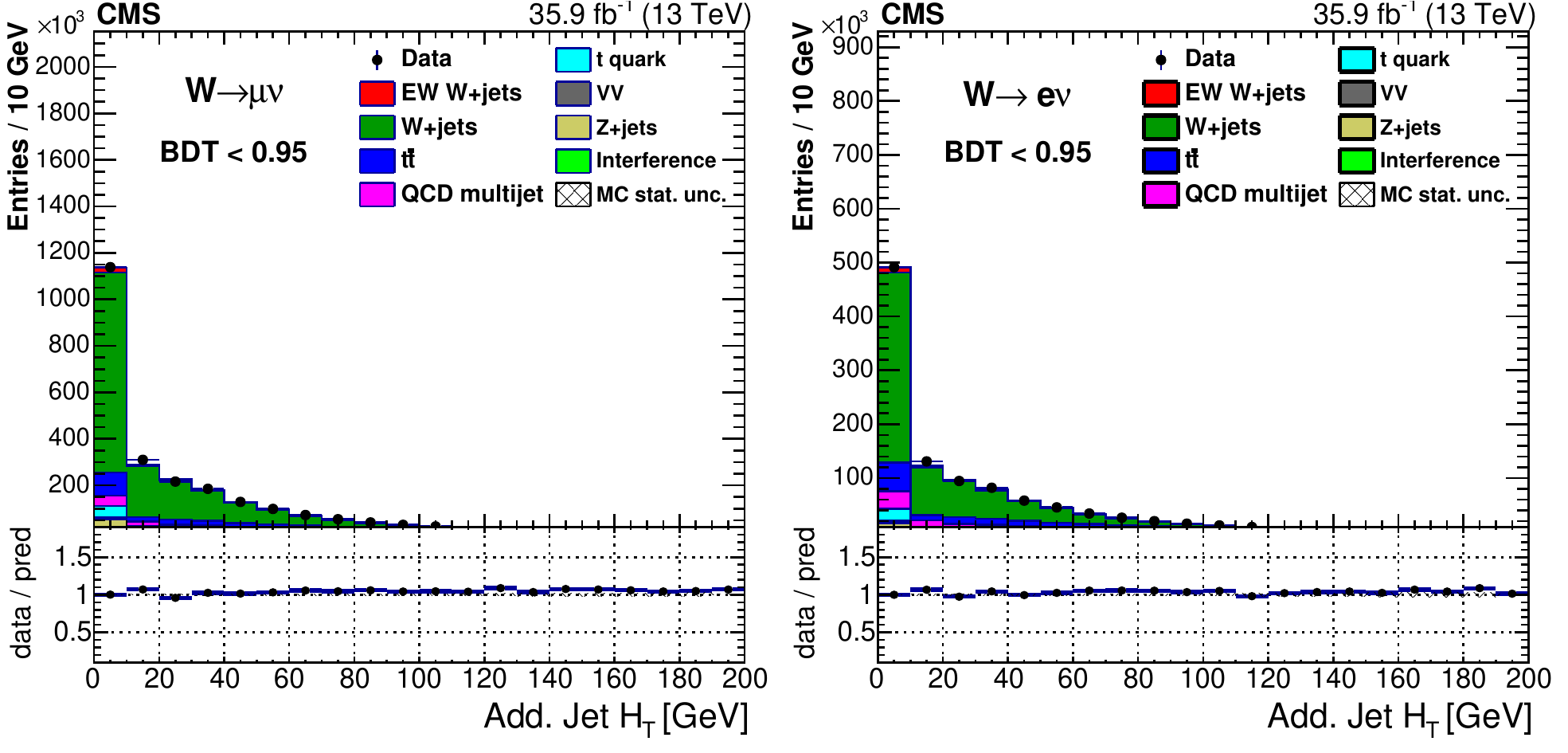}
    \caption{Total \HT of the additional jets for BDT $<$ 0.95 in the muon (left) and electron (right)
    channels. In all plots the fist bin
    contains events where no additional jet with $\pt>15\GeV$ is present.}
     \label{fig:addjetht_bdtcutInv}

\end{figure*}

\begin{figure*}[htb!]
  \centering
    \includegraphics[width=0.98\textwidth]{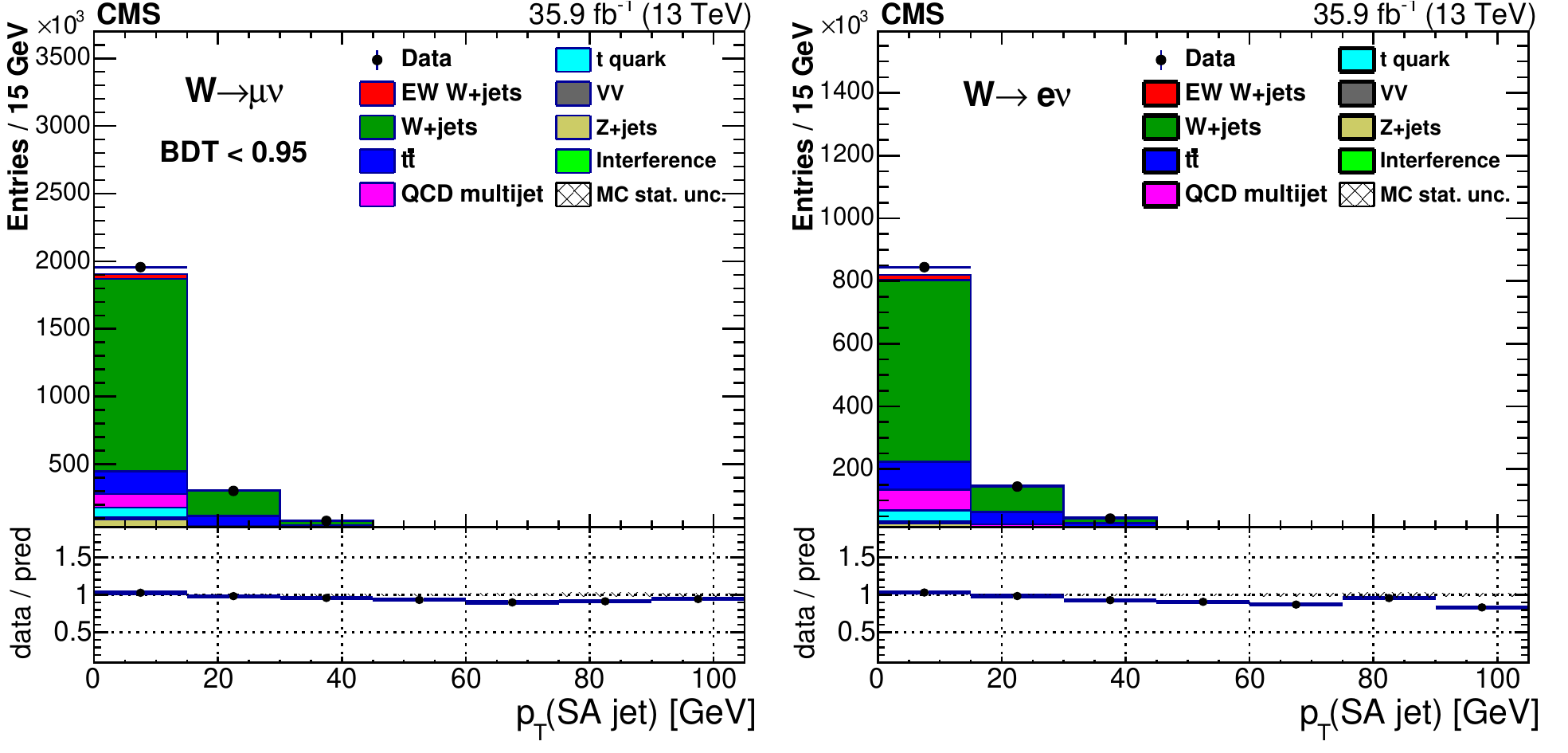}
    \caption{Leading additional soft-activity (SA) jet \pt for BDT $<$
     0.95 in the muon (left) and electron (right) }
     \label{fig:leadsajetpt_bdtcutInv}

\end{figure*}

\begin{figure*}[htb!]
  \centering
    \includegraphics[width=0.98\textwidth]{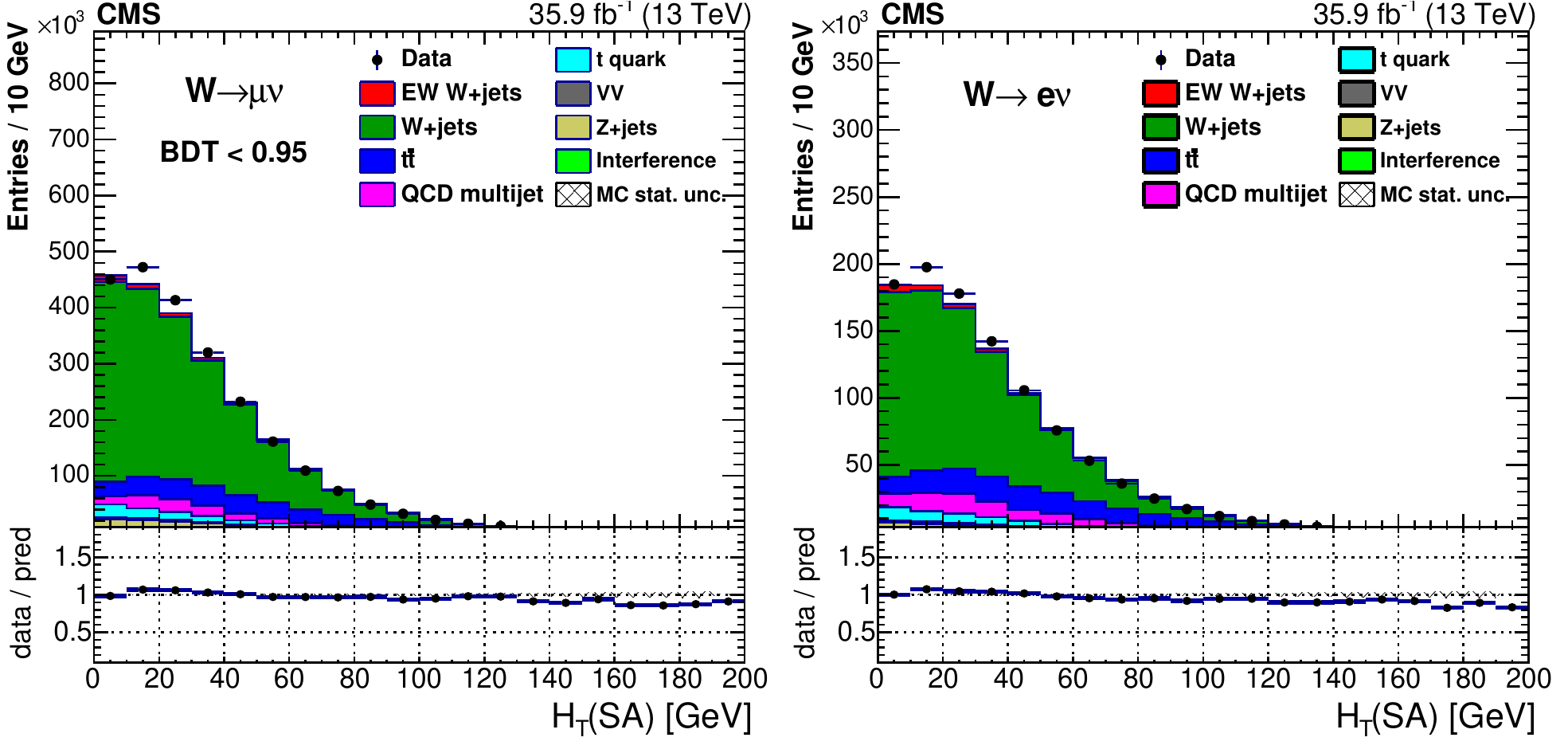}
    \caption{Total soft activity (SA) jet \HT for BDT $<$
     0.95 in the muon (left) and electron (right)
    channels.}
     \label{fig:sajetht_bdtcutInv}

\end{figure*}

Section~\ref{sec:hadactivity} shows a comparison of the data with simulation with \PYTHIA and \HERWIGpp parton showering separately
in a high purity signal region with $\mathrm{BDT} > 0.95. $ The agreement of the simulation with data for the background prediction is validated for
the rapidity gap observables in the signal-depleted region BDT $<$ 0.95, where the signal purity is less than 2\%. Figures~\ref{fig:jet3pt_bdtcutInv},
\ref{fig:addjetht_bdtcutInv}, \ref{fig:leadsajetpt_bdtcutInv}, and \ref{fig:sajetht_bdtcutInv} show the leading additional jet \pt, the total \HT of the additional jets,
the leading soft activity jet \pt, and the total soft activity jet \HT, respectively, in the region BDT $<$ 0.95. Good agreement is observed between the background
prediction and the data for all observables.

\clearpage
\subsection{Hadronic activity vetoes}

\begin{figure*}[htb!]
  \centering
    \includegraphics[width=0.48\textwidth]{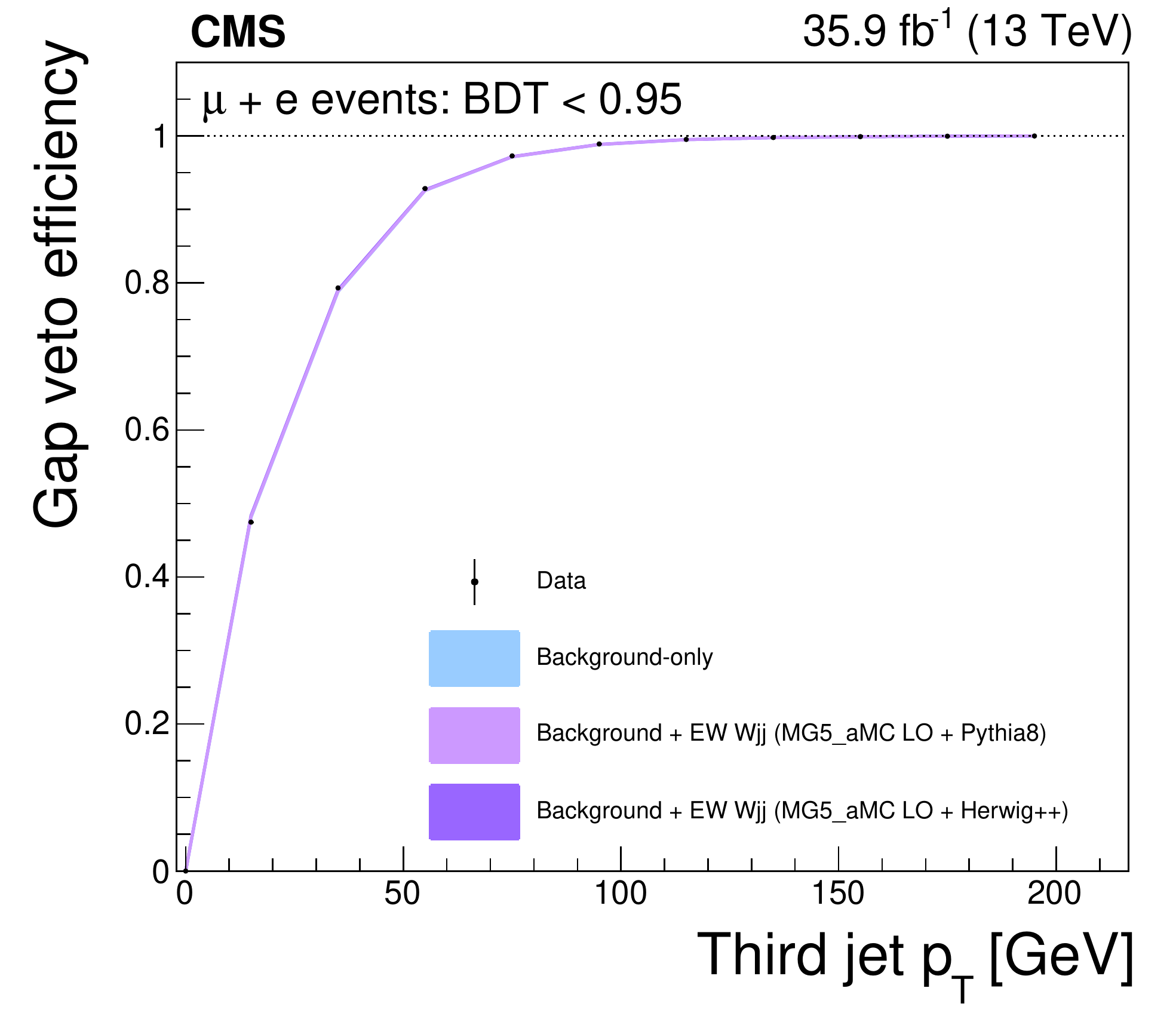}
    \includegraphics[width=0.48\textwidth]{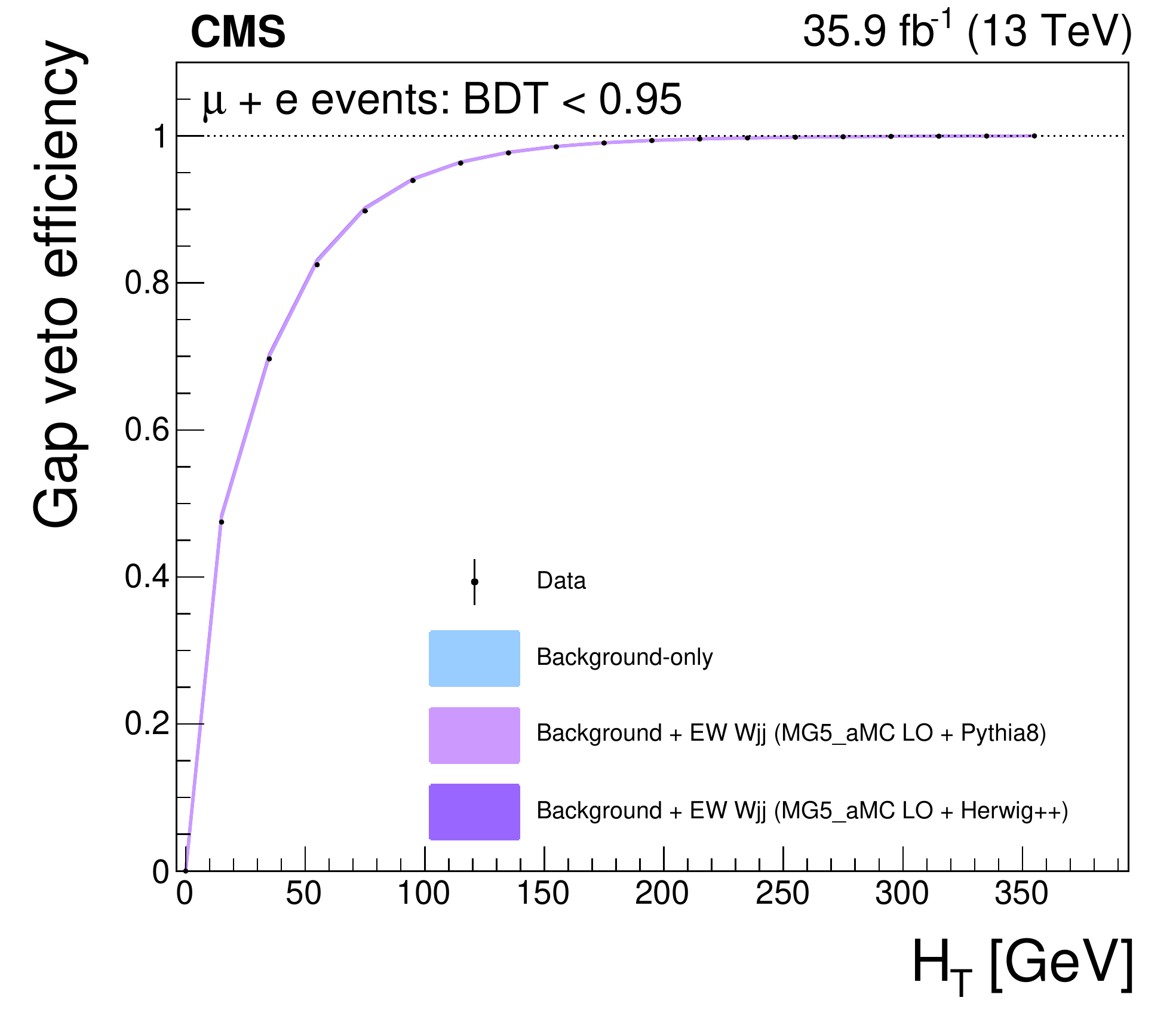}
    \includegraphics[width=0.48\textwidth]{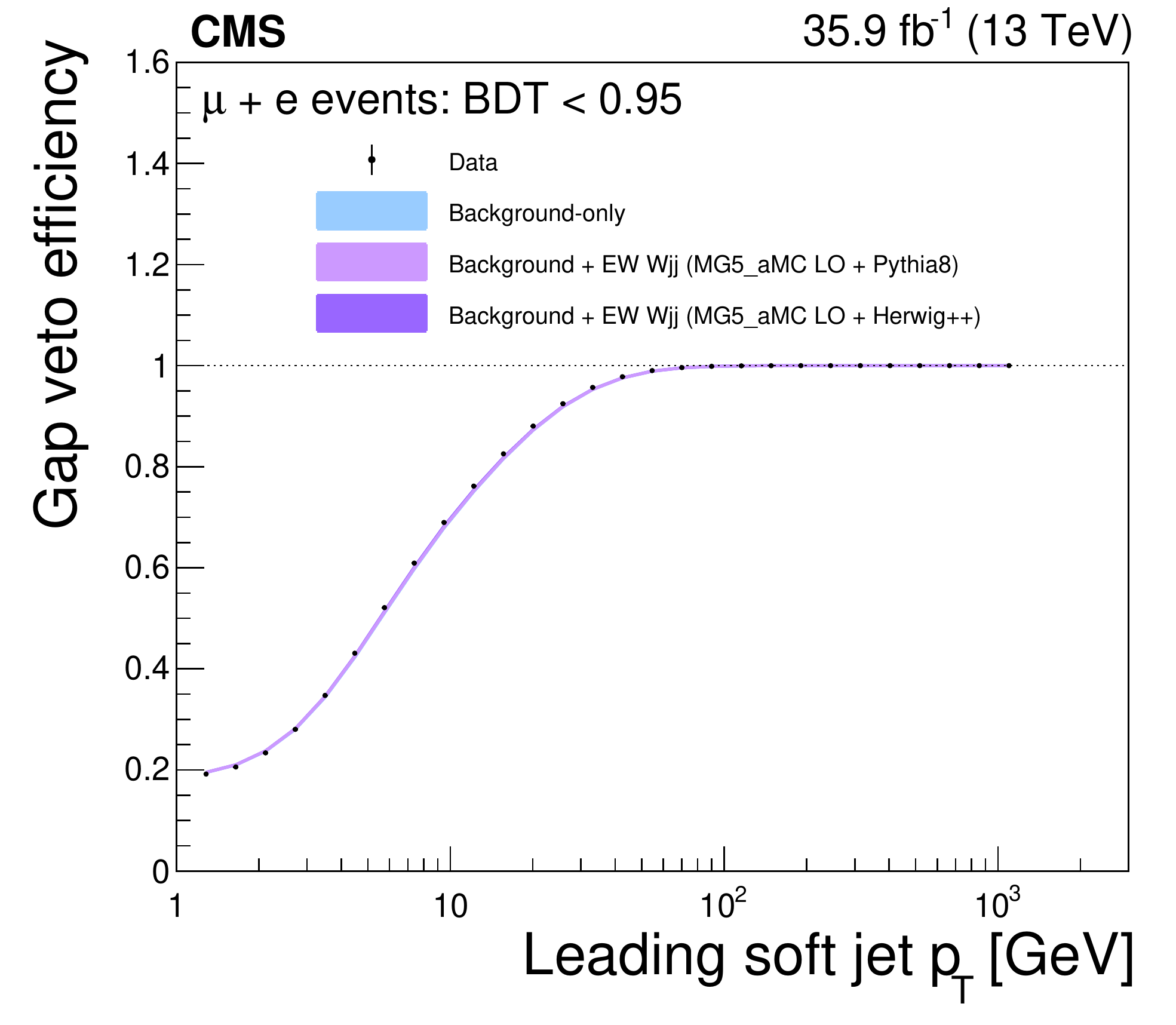}
    \includegraphics[width=0.48\textwidth]{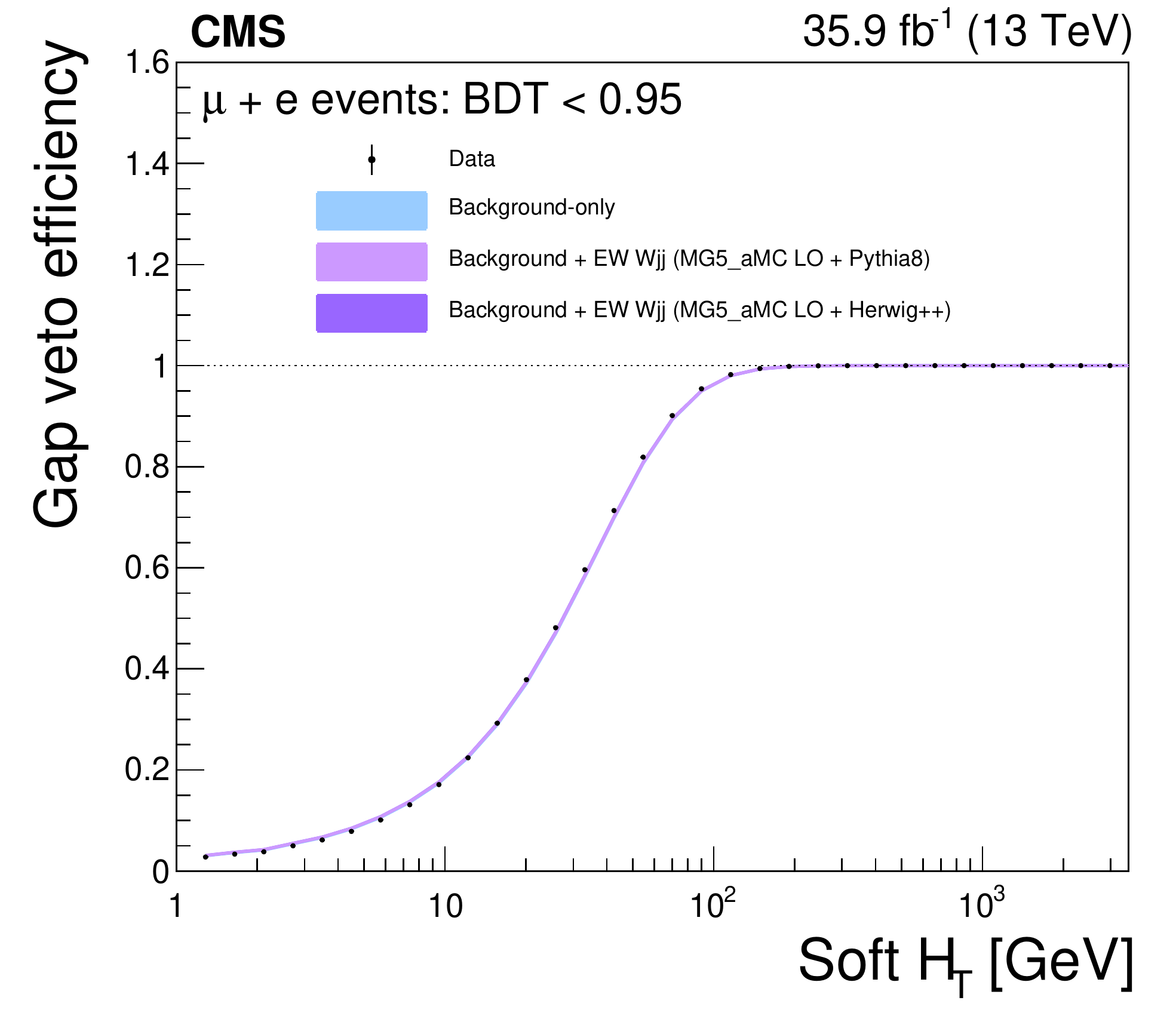}
    \caption{Hadronic activity veto efficiencies in the signal-depleted BDT $<$ 0.95
    region for the muon and electron channels combined, as a function of the leading
    additional jet \pt (upper left), additional jet \HT (upper right), leading
    soft-activity jet \pt (lower left), and soft-activity jet \HT (lower right). The data
    are compared with the background-only prediction as well as background+signal with \PYTHIA
    parton showering and background+signal with \HERWIGpp parton showering. The uncertainty bands
    include only the statistical uncertainty in the prediction from simulation. There is very little
    difference between the predictions due to the small fraction of signal in this region.
    }
     \label{fig:gapvetoeff_bdtInv}
\end{figure*}

The efficiency of a hadronic activity veto, as described in Section~\ref{subsec:gapveto}, is studied in the signal-depleted
$\mathrm{BDT} < 0.95$ region. Figure~\ref{fig:gapvetoeff_bdtInv} shows the gap activity veto efficiency of combined muon and electron events
in the signal-depleted region when placing an upper threshold on
the \pt of the additional third jet, on the \HT of all additional jets,
on the leading soft-activity jet \pt, or on the soft-activity \HT.
There is very little difference between the background-only prediction and the predictions including signal with either \PYTHIA or \HERWIGpp
parton showering due to the very small fraction of signal in this region. Good agreement is observed between the data and the simulation,
 giving further confidence in the modelling of the background observables for the rapidity gap studies.

\cleardoublepage \section{The CMS Collaboration \label{app:collab}}\begin{sloppypar}\hyphenpenalty=5000\widowpenalty=500\clubpenalty=5000\vskip\cmsinstskip
\textbf{Yerevan Physics Institute, Yerevan, Armenia}\\*[0pt]
A.M.~Sirunyan, A.~Tumasyan
\vskip\cmsinstskip
\textbf{Institut f\"{u}r Hochenergiephysik, Wien, Austria}\\*[0pt]
W.~Adam, F.~Ambrogi, E.~Asilar, T.~Bergauer, J.~Brandstetter, M.~Dragicevic, J.~Er\"{o}, A.~Escalante~Del~Valle, M.~Flechl, R.~Fr\"{u}hwirth\cmsAuthorMark{1}, V.M.~Ghete, J.~Hrubec, M.~Jeitler\cmsAuthorMark{1}, N.~Krammer, I.~Kr\"{a}tschmer, D.~Liko, T.~Madlener, I.~Mikulec, N.~Rad, H.~Rohringer, J.~Schieck\cmsAuthorMark{1}, R.~Sch\"{o}fbeck, M.~Spanring, D.~Spitzbart, W.~Waltenberger, J.~Wittmann, C.-E.~Wulz\cmsAuthorMark{1}, M.~Zarucki
% ...
\vskip\cmsinstskip
% ...
1:  Also at Vienna University of Technology, Vienna, Austria\\
% ...
\end{sloppypar}
\end{document}